\documentclass[12pt]{article}
\usepackage{amsmath, amssymb, graphics,epsfig}
\usepackage{bbm}
\usepackage{appendix}

\usepackage{graphics, psfrag}
\usepackage{amsbsy}
\usepackage{amsfonts}
\usepackage{amsmath}
\usepackage{amssymb}
\usepackage{graphicx}
\usepackage{color}
\newcommand{\mathsym}[1]{{}}

\def\id{\protect{{1 \kern-.28em {\rm l}}}}

\def\be{\begin{eqnarray}}
\def\ee{\end{eqnarray}}

\def\nn{\nonumber}
\makeatletter
\renewcommand\section{\@startsection {section}{1}{\z@}%
                                   {-3.5ex \@plus -1ex \@minus -.2ex}%
                                   {2.3ex \@plus.2ex}%
                                   {\normalfont\large\bfseries}}
\renewcommand\subsection{\@startsection{subsection}{2}{\z@}%
                                   {-3.25ex\@plus -1ex \@minus -.2ex}%
                                   {1.5ex \@plus .2ex}%
                                   {\normalfont\normalsize\bfseries}}

\makeatother

\numberwithin{equation}{section}

\def \l {\lambda}

\def \sql {{\sqrt \l}}
\def \adss {$AdS_5 \times S^5~$ }

\newcommand{\Smatrix}{\mathbbm{S}}  
\newcommand{\smatrix}{\mathbf{S}}   
\newcommand{\Tmatrix}{\mathbbm{T}}  
\newcommand{\tmatrix}{\mathrm{T}}   

\newcommand{\ket}[1]{\mathopen{|}#1\mathclose{\rangle}}

\newcommand{\Asmatrix}{\mathbf{A}}
\newcommand{\Bsmatrix}{\mathbf{B}}
\newcommand{\Csmatrix}{\mathbf{C}}
\newcommand{\Dsmatrix}{\mathbf{D}}
\newcommand{\Esmatrix}{\mathbf{E}}
\newcommand{\Fsmatrix}{\mathbf{F}}
\newcommand{\Gsmatrix}{\mathbf{G}}
\newcommand{\Hsmatrix}{\mathbf{H}}
\newcommand{\Ksmatrix}{\mathbf{K}}
\newcommand{\Lsmatrix}{\mathbf{L}}

\newcommand{\Atmatrix}{\mathrm{A}}
\newcommand{\Btmatrix}{\mathrm{B}}
\newcommand{\Ctmatrix}{\mathrm{C}}
\newcommand{\Dtmatrix}{\mathrm{D}}
\newcommand{\Etmatrix}{\mathrm{E}}
\newcommand{\Ftmatrix}{\mathrm{F}}
\newcommand{\Gtmatrix}{\mathrm{G}}
\newcommand{\Htmatrix}{\mathrm{H}}
\newcommand{\Ktmatrix}{\mathrm{K}}
\newcommand{\Ltmatrix}{\mathrm{L}}

\newcommand{\sfrac}[2]{{\textstyle\frac{#1}{#2}}}
\newcommand{\half}{\sfrac{1}{2}}
\newcommand{\ihalf}{\sfrac{i}{2}}

\newcommand{\Half}{\frac{1}{2}}

\newcommand{\Quarter}{\frac{1}{4}}

\newcommand{\grp}[1]{\mathrm{#1}}

\newcommand{\grSU}{\grp{SU}}

\newcommand{\grPSU}{\grp{PSU}}

\newcommand{\hypref}[2]{\ifx\href\asklfhas #2\else\href{#1}{#2}\fi}

\newcommand{\brk}[1]{(#1)}
\newcommand{\lrbrk}[1]{\left(#1\right)}

\newcommand{\biggsbrk}[1]{\biggl[#1\biggr]}

\newcommand{\abs}[1]{{|#1|}}

\newcommand{\lAA}{{a}}

\newcommand{\laa}{{\alpha}}

\newcommand{\lBB}{{b}}

\newcommand{\lbb}{{\beta}}

\newcommand{\lCC}{{c}}

\newcommand{\lcc}{{\gamma}}

\newcommand{\lDD}{{d}}

\newcommand{\ldd}{{\delta}}

\newcommand{\levi}{\epsilon}
\newcommand{\energy}{\varepsilon}

\newcommand{\unit}{\mathbbm{1}}
\newcommand{\order}{\mathcal{O}}

\newcommand{\cpp}{\energy' p - \energy p'}
\def\hg{{\hat g}}

\def\adss{AdS$_5\times$S$^5$}
\def\adssss{AdS$_3\times$S$^3\times$S$^3\times$S$^1$}
\def\adsst{AdS$_3\times$S$^3\times$T$^4$}

\def\adscp{AdS$_4\times$CP$^3$}

\def\NeqFour{{\cal N}=4}

\begin{document}


\overfullrule=0pt
\parskip=2pt
\parindent=12pt
\headheight=0in \headsep=0in \topmargin=0in \oddsidemargin=0in

\vspace{ -3cm}
\thispagestyle{empty}
\vspace{-1cm}


\

\

\begin{center}
\vspace{1cm}
{\Large\bf  
%
Generalized unitarity and 
 \\
 \vspace {.2cm}
 the worldsheet S matrix in AdS$_n\times$S$^n\times$M$^{10-2n}$
}

\end{center}

\vspace{.2cm}



\begin{center}
 
Oluf Tang Engelund, Ryan W. McKeown  and   Radu Roiban

\end{center}

\begin{center}
{
\em 
\vskip 0.08cm
\vskip 0.08cm
Department of Physics, The Pennsylvania  State University,\\
University Park, PA 16802 , USA
\vskip 0.08cm
\vskip 0.08cm 

}
 \end{center}



\vspace{1.5cm}

\vspace{.2cm}

\begin{abstract}

\noindent

The integrability-based solution of string theories related to AdS$_n$/CFT$_{n-1}$ dualities relies on 
the worldsheet S matrix. 
Using generalized unitarity we construct the terms with logarithmic dependence on external
momenta at one- and two-loop order in the worldsheet S matrix for strings in a general integrable 
worldsheet theory. We also discuss aspects of calculations at higher orders.
The S-matrix elements are expressed as sums of integrals with coefficients given in terms of  
tree-level worldsheet four-point scattering amplitudes. Off-diagonal one-loop rational functions, not determined 
by two-dimensional unitarity cuts, are fixed by symmetry considerations. They play an important role in the 
determination of the two-loop logarithmic contributions.
We illustrate the general analysis by computing the logarithmic terms in the one- and two-loop 
four-particle S-matrix elements in the massive worldsheet sectors of string theory in AdS$_5\times$S$^5$, 
AdS$_4\times$CP$^3$, AdS$_3\times$S$^3\times$S$^3\times$S$^1$ and AdS$_3\times$S$^3\times$T$^4$.
We explore the structure of the S matrices and provide explicit evidence for the 
absence of higher-order logarithms and for the exponentiation of the one-loop dressing phase.

\end{abstract}

\newpage

\tableofcontents


\section{Introduction}

The integrability of quantum string theory in \adss~ has led to remarkable progress in our understanding of the 
spectrum of anomalous dimensions in the dual $\NeqFour$ super-Yang-Mills theory. Similarly, the
integrability of the spin chain with the dilatation operator of $\NeqFour$ super-Yang-Mills theory as Hamiltonian
offers important insight into the spectrum of string theory in \adss~and also provides tools for the construction of other
important quantities, such as the correlation functions of gauge-invariant operators and four-dimensional scattering amplitudes. 
This remarkable success \cite{Beisert:2010jr} raises the question of applying similar methods, such as algebraic 
curve techniques \cite{Beisert:2005bm} or the Asymptotic Bethe Ansatz (ABA) \cite{BeSt_aba}, to other string theories 
that exhibit integrable structures, such as strings in \adscp~or in AdS$_{n}\times$S$^{n}\times$M$^{10-2{n}}$ supported 
by either an RR flux or a mixture of RR and NSNS fluxes, and shed light on the dual conformal field theories which, 
for $n=3$ are not understood beyond their BPS sector.
The essential ingredient in such an approach is the scattering matrix of worldsheet excitations around a 
suitably-chosen vacuum state or, alternatively, the scattering of spin-chain excitations. 

The assumption of quantum integrability and the symmetries of the theory go a long ways in the construction of worldsheet
S matrices. The former implies that the S matrices obey a form of the Yang-Baxter equation and that higher-point 
S-matrix elements can be constructed by multiplying together  four-point S-matrix elements. The latter 
implies the factorization of the S matrix into factors invariant under each of the symmetry groups of the gauge-fixed 
theory.
All in all, the S matrices are uniquely determined up to overall phases -- known as the dressing phases -- and a function
of the string tension. There is one such phase for each part of an S~matrix that is invariant under the symmetry group.

Green-Schwarz-type supercoset sigma models for AdS$_3\times$S$^3$, AdS$_3\times$S$^3\times$S$^3$ 
and  AdS$_2\times$S$^2$ can be constructed based on 
$D(2,1;\alpha)\times D(2,1;\alpha)/(SU(1,1)\times SU(2)\times SU(2))$,
$PSU(1,1|2)\times PSU(1,1|2)/(SU(1,1)\times SU(2))$
and
$PSU(1,1|2)/(SO(1,1)\times U(1))$, respectively. 
As their dimension is smaller than $d=10$, additional bosonic directions are required for a critical string theory. 
Unlike the NSR string, the worldsheet theory of the Green-Schwarz  string is interacting even when the bosonic part 
of the target space is a product space with interactions induced by fermions which are representations of the 
ten-dimensional Lorenz group.
Thus, these supercoset sigma models can be related to the Green-Schwarz string on 
AdS$_{n}\times$S$^{n}\times$M$^{10-2{n}}$ with M$=$S$^3\times$S$^1$, T$^4$ and T$^6$ only if there 
exists a non-degenerate $\kappa$-symmetry gauge that decouples these additional degrees of freedom.

With worldsheet diffeomorphism invariance fixed to conformal gauge, a $\kappa$-symmetry gauge decoupling the lone 
S$^1$ and the T$^4$ excitations\footnote{String theory 
with M$=$T$^4$ and no excitations along M may be interpreted as the limit $\alpha\to 1$ of string theory with 
M$=$S$^3\times$S$^1$. In this limit M decompactifies to $R^3\times$S$^1$; since worldsheet 
masses depend on $\alpha$ this limit is rather subtle and from the standpoint of the S matrix it involves a nontrivial 
rearrangement of states. The difference in the topology of M is not observable in the absence of excitations on M.}, 
was found for  M$=$S$^3\times$S$^1$ and M$=$T$^4$ \cite{StefanskyZarembo}. An analogous gauge does not 
appear to exist for M$=$T$^6$ \cite{Sorokin:2011rr}. 
If instead worldsheet diffeomorphisms are fixed to a gauge in which all fields are physical -- such as the static gauge 
or the light-cone gauge -- all worldsheet excitations are coupled to each other; on shell one may nevertheless expect 
a decoupling similar to that seen {\em off-shell} in conformal gauge.  
At the classical level, it is nevertheless possible to consistently truncate \cite{Sorokin:2011rr} all the fields orthogonal 
to the supercoset (which are massless). 
While superficially this truncation may seem inconsistent at loop level, we shall argue that the integrability of the 
theory implies that, through two loops, the truncated states affect only the part of the S-matrix terms whose dependence
on external momenta is completely rational.

Similarly to strings in \adss, these theories were quantized around a BMN-type point-like string and 
Asymptotic Bethe Ansatz equations were proposed. An important departure from the structure of the \adss~ string
is that the free worldsheet spectrum is more complex, exhibiting excitations of different masses.
Moreover, masses are 
conserved in the scattering process which, under certain circumstances, also forbids the exchange of 
momentum between particles ({\it i.e.} it is reflectionless).
The same steps as for string theory in \adss~ \cite{AFS} led to the construction of finite-gap equations \cite{Zarembo:2010yz}
and led to constraints on dressing phases.
\footnote{String theory in \adscp, dual to the ABJM theory \cite{ABJM}, was constructed in supercoset form in 
\cite{Arutyunov:2008if, StefanskiADSCP, Fre:2008qc}, finite-gap equations were constructed in \cite{Gromov:2008bz} and all-loop 
Asymptotic Bethe Ansatz equations were proposed \cite{Gromov:2008qe} with 
specific assumptions on the dressing phase of the S matrix. As was discussed in \cite{Arutyunov:2008if} and in more detail 
in  \cite{Gomis:2008jt}, the supercoset action does not describe string configurations with excitations only in the AdS$_4$.}

The S matrices of string theory in \adssss~ and \adsst~ and of the corresponding spin chain were further analyzed from 
an algebraic standpoint  in \cite{David:2010yg, Abbott:2012dd, Borsato:2012ud, Borsato:2013qpa, Borsato:2012ss, Ahn:2012hw}
where it was emphasized that, unlike the case of strings in \adss~ \cite{JanikCrossing, BHL, Volin:2009uv}, the 
dressing phases are not determined by crossing symmetry constraints.
The one-loop (worldsheet) correction to the dressing phases in \adsst~ were found in \cite{Beccaria:2012kb} (see also 
\cite{David:2010yg})
through a comparison of the one-loop correction to the energy of certain extended string configuration and the 
Asymptotic Bethe Ansatz
predictions to these quantities; the result exposes the fact that the dressing phases are different from the original  conjectures
\cite{Zarembo:2010yz, OhlssonSax:2011ms}. 
A direct Feynman graph calculation \cite{Sundin:2013ypa} of the one-loop worldsheet S matrix in the near-flat space limit 
of \adsst~ confirms this conclusion.

These developments underscore the importance of a direct construction of the worldsheet S matrix in these and 
related integrable worldsheet theories. 
In \adss~ such a construction would further test the crossing equation \cite{JanikCrossing} as well as the assumption that 
the relevant dressing phase is the minimal one ({\it i.e.} that no solution of the homogeneous crossing equation needs 
to be included) \cite{BeSt, Volin:2009uv}. 
By testing factorization of the S matrix at higher loops and higher points, a direct evaluation of such  S-matrix elements 
would provide a powerful test of integrability and of the existence of an integrability-preserving regularization.
Moreover, in \adss, tree-level worldsheet calculations \cite{KMRZ} expose the fact that the S-matrix elements depend on
the choice of worldsheet (diffeomorphism and $\kappa$-symmetry) gauges. By directly constructing the S matrix 
in a general gauge one may demonstrate explicitly the independence of the target space spectrum on the 
gauge-choice parameter.

In independent developments, new and powerful methods -- the generalized unitarity method 
\cite{UnitarityMethod, BCFUnitarity} and its refinement, the 
method of maximal cuts \cite{FiveLoop, CompactThree} and its further generalization to certain massive
cases~\cite{Kosower:2011ty, Johansson:2012zv}  -- have been developed for the calculation 
of scattering amplitudes in various quantum field theories.
They have led to the construction with relative ease of a whole host of scattering amplitudes in three-, four- and 
higher-dimensional supersymmetric and non-supersymmetric gauge and gravity theories and to the hope that, 
in the planar limit, the entire S matrix can be found.
The power of the generalized unitarity approach stems from the fact that loop-level amplitudes are constructed
in terms of tree-level amplitudes, which depend only on the physical degrees of freedom. They also manifest 
all the symmetries of the theory, including those that exist only on shell, such as integrability.
Quite generally, this method
suggests that, up to potential anomalies, the symmetries of the tree-level amplitudes are inherited by loop level 
amplitudes. The potential anomalies can also be efficiently identified in this approach.

The direct calculation of the tree-level S matrix in AdS$_5\times$S$^5$ was carried out in \cite{KMRZ}.\footnote{A 
direct one-loop calculation was previously attempted unsuccessfully, using worldsheet Feynman graphs, 
in \cite{MRunpublished}. The main problem that was encountered was non-cancellation of UV divergences.
While such divergences have polynomial momentum dependence and thus can always be eliminated by a local counterterm, 
their existence goes against the expectation that, at least for supersymmetric ground states, the worldsheet theory does not 
exhibit infinite renormalization. The  surviving UV divergences were also present 
in the resulting Bethe equations. The reason for the remaining divergences was never clarified. Similar 
divergences have been reported in Feynman graph calculations in AdS$_3\times$S$^3\times$T$^4$ in \cite{Sundin:2013ypa}.}
Direct tree-level calculations have been carried out for a certain massive subsector of string theory in 
\adssss~ for general $\alpha$
in \cite{Sundin:2013ypa} and in \adsst~ with mixed RR and NSNS fluxes in  \cite{Hoare:2013pma}.\footnote{The bosonic S matrix 
in \adscp~ was found in \cite{Kalousios:2009ey}.}
From a spin chain perspective, the all-loop symmetry-determined parts of the  S matrix in these cases were 
discussed in \cite{Borsato:2012ss, Borsato:2013qpa}.

Here we will use generalized unitarity  to find the two-dimensional cut-constructible part -- that is the terms 
with logarithmic dependence on external momenta -- of the one- and two-loop four-point S-matrices for the Green-Schwarz 
string in \adss, \adssss~ and \adsst~ and also to comment on the relation between the \adscp~ and \adss~ S-matrices. 
The main ingredients of this construction are the tree-level worldsheet scattering amplitudes.
As we shall see, the S matrix can be expected to obey quite generally the factorized structure following from the 
tree-level symmetry group and integrability.
As in all higher-order calculations, regularization is necessary and, ideally, we should use regulated 
tree amplitudes\footnote{For example, in dimensional regularization they are the tree-level amplitudes 
of the $d$-dimensional theory.}. 

The issue of ultraviolet (UV)  regularization of 
the worldsheet theory in Green-Schwarz form is a thorny one. As a conformal field theory (perhaps with spontaneously 
broken conformal invariance) the theory is expected be finite to all orders in perturbation theory; an all-order argument
for finiteness however relies on its symmetries, in particular on $\kappa$-symmetry. 
This symmetry is chiral  (and has a self-dual parameter) and thus exists only in two dimensions making 
dimensional regularization unsuitable. This is also related to the presence of the parity-odd Wess-Zumino term 
in the Green-Schwarz action, which also does not exist\footnote{Dimensional 
regularization can be suitably modified to be applicable in the presence of a Wess-Zumino term. For example,  
may prescribe some analytic continuation of the two-dimensional Levi-Civita symbol (see  e.g. \cite{deWit:1993qv}); 
alternatively, one may do all algebra in two dimensions and continue only the final integrals to $d \ne 2$ (see {\it e.g.} 
\cite{RTT} for a discussion related to the Green-Schwarz string). } in dimensions other that $d=2$.

It is not clear what is an example of regularization that preserves all symmetries of the worldsheet theory. 
In its absence, there are (at least) two possible approaches which, ultimately, need similar additional input to 
yield complete S-matrix elements.  
On general grounds, for any regularization, 
integrability as well as other classical symmetries broken by the regulator can be restored 
by the addition to the S matrix of matrix elements of finite local counterterms in the effective action. Their determination 
relies on the requirement that symmetries be realized at the quantum level.
Since these counterterms contribute only rational terms to the S matrix, one may simply determine their off-diagonal 
components from symmetry considerations (their diagonal components affect the rational part of the dressing phase 
and thus are not determined by symmetries).
Alternatively, two-loop energy calculations \cite{GRRTV, RTT} suggest that finiteness of the theory is observed if in 
loop calculations one does all numerator algebra in $d=2$; this manifestly organizes the result in terms of finite 
combinations of integrals which may then be evaluated in any ({\it e.g.} dimensional) regularization.
Following this lead we will use  two-dimensional tree amplitudes to construct generalized cuts, 
carry out all numerator algebra in two dimensions and regularize the resulting integrals. 
While this approach guarantees that the terms exhibiting imaginary parts in $d=2$ are correctly 
identified, it leaves open the possibility that terms 
with no imaginary parts are missed\footnote{
In particular, in the absence of a complete determination of terms with rational momentum dependence 
we cannot shed light on the fate of the divergences found in earlier attempts \cite{MRunpublished}. 
For the same reason we cannot address the issue pointed out in 
\cite{Borsato:2013qpa} that the \adssss~dressing phases of \cite{Beccaria:2012kb} do not obey the 
generalized crossing equations constructed there, since the offending terms have a completely rational 
dependence on external momenta.}.
As in the previous approach, off-diagonal rational terms are determined by symmetry considerations.
We will follow this approach here.

It is important to note that the specifics of two-dimensional kinematics allow for the existence of four-point loop integrals 
with no net momentum flowing through them\footnote{In higher dimensions they are regular integrals in the forward limit.}. 
One might expect that the interpretation of generalized unitarity as a specific 
organization of the Feynman diagram expansion will capture at least these  (cut-less) terms. Their cuts appear to be 
singular, however,  suggesting that in the absence of a suitable IR regularization their coefficients can at best be determined 
by a prescription.

This paper is organized as follows. 
We begin in sec.~\ref{worldsheetPT} with a discussion of the  structure of the S matrix of a general integrable 
worldsheet theory with a product-group symmetry and of the relation  between the exact (spin chain) and the worldsheet 
S-matrices. 
We shall also discuss the structure of the two-dimensional unitarity cuts to all-loop orders, regularization issues, as well 
as identify the two-dimensional cut-constructible parts of the S matrix. To this end the rational part 
of the symmetry-determined S matrix will play an important role. 

Sec.~\ref{GeneralExpressions} contains the general form of our results;
we collect here in a compact form one- and two-loop amplitudes constructed through 
the generalized unitarity method and extract their logarithmic dependence on external momenta. 
We shall illustrate the application of these general expressions to string theory in \adss, \adscp, \adssss~ and \adsst.

In sec.~\ref{AdS5} we use generalized unitarity to determine the two-dimensional cut-constructible part of the 
worldsheet one- and two-loop S matrix in \adss. Our results may be found in eqs.~\eqref{final_ads5_1loop} and 
\eqref{final_ads5_2loop}; we will explicitly demonstrate the exponentiation of the (logarithmic part of the) 
one-loop dressing phase and thus provide direct evidence for the integrability of the theory at this loop order. 

Using the details of the calculations in sec.~\ref{AdS5},  we briefly comment in sec.~\ref{AdS4CP3} on the 
worldsheet S matrix in \adscp. In particular, we recover the expected result that only a reflectionless S matrix
is consistent with worldsheet perturbation theory if the heavy modes of the worldsheet theory are truncated away
at the classical level. We also confirm the proposed dressing phase through two-loop order.

In sec.~\ref{adss} we determine the one-loop worldsheet 
S matrix in \adssss~ in the massive sector  for general $\alpha$ parameter
and confirm the (logarithmic part of the) dressing phases found in \cite{Beccaria:2012kb} -- see 
eqs.~\eqref{resultat LL} and \eqref{resultat LR}. 

In sec.~\ref{adst} we find the one-loop S matrix in \adsst~ in the massive sector  for a background supported by
a mixture of NSNS and RR fluxes parameterized by the $q$-parameter of  \cite{Hoare:2013pma}; our result may 
be found in eqs.~\eqref{++phase} and \eqref{+-phase}. In the $q\to 0$ limit we find the $m,m'\to 1$ limit of the 
\adssss~phase of \cite{Beccaria:2012kb}. A further near-flat space limit  reproduces the one-loop S-matrix of 
\cite{Sundin:2013ypa}. In the $q\rightarrow 0$ limit we also construct the two-loop S-matrix and demonstrate 
the exponentiation of 
the one-loop dressing phase -- see eqs.~\eqref{ADST_LL_final} and \eqref{ADST_LR_final}. 

We close in sec.~\ref{close} with remarks on the implications and possible extensions of our results.

The Appendices review the conventions used for the calculation of tree-level worldsheet S-matrices pointing out the
differences from the usual four-dimensional ones and collect the details 
of the exact spin-chain S matrices in the various theories discussed in the paper, the general expression for the 
two-particle cuts in \adss, one- and two-loop integrals and their (generalized) cuts.

\section{Worldsheet perturbation theory for the S matrix \label{worldsheetPT}}

Let us consider a general integrable two-dimensional quantum field theory with a 
factorized symmetry group $G_1\otimes G_2$. Examples are the supercoset part of all case for all 
AdS$_{\rm x}\times$S$^{\rm x}\times$M$^{10-2{\rm x}}$ theories, where $G_i$ is either $PSU(2|2)$ or 
$PSU(1|1)^2$.
For such a theory integrability suggests that the four-point S matrix, $\Smatrix$, can be decomposed as 
\be
\Smatrix = 
\smatrix_{G_1}\otimes \smatrix_{G_2} \ ,
\label{ws_smatrix_decomposition_txt}
\ee
where $\smatrix_{G_1}$ and $\smatrix_{G_2}$ are S-matrices invariant under the groups $G_1$ and $G_2$, respectively.
The excitations scattered by $\smatrix_{G_i}$ are 
not natural perturbative excitations of the theory; rather, the latter are bilinear\footnote{While 
of course physically different, this is formally reminiscent of the KLT relations of standard flat space string theory, in which 
tree-level supergravity scattering amplitudes are bilinears of tree-level gauge theory scattering amplitudes. 
The essential difference is that while the latter hold at tree level, eq.~\eqref{ws_smatrix_decomposition_txt} holds to 
all orders.}  in the former.
One may choose to extract an overall phase in eq.~\eqref{ws_smatrix_decomposition_txt}; we will not do this but instead
assign all overall phases to the two $\smatrix_{G_i}$-matrix factors.
We shall denote the worldsheet coupling constant by ${\hat g}= \sqrt{\lambda}/(2\pi)$.

It is possible that the symmetry group of the S matrix has further abelian factors, apart from $G_1\times G_2$, which assign 
nontrivial charges to representations of $G_1\times G_2$. An example in this direction is the case for 
M$=$S$^3\times$S$^1$ where $G_1=G_2=PSU(1|1)$ and the additional symmetry is a $U(1)$ factor\footnote{We 
thank B.~Hoare for emphasizing this out to us.}. 
In such cases integrability no longer requires a factorized 
$\Smatrix$-matrix of the form \eqref{ws_smatrix_decomposition_txt}. Since we will reduce the computation of an 
$\Smatrix$-matrix of the form \eqref{ws_smatrix_decomposition_txt} to a separate computation of the two factors 
$\smatrix_{G_i}$, the discussion in this section applies unmodified to the construction of a non-factorized S matrix as well.

\subsection{Generalities, parametrization, symmetry restrictions}

As usual, the expansion of the worldsheet $\Smatrix$-matrix in the worldsheet coupling constant ${\hat g}^{-1}$ defines the 
$\Tmatrix$ matrix
\be 
\label{asmatrix_expansion}
 \Smatrix = \unit + \frac{1}{\hat g} \, i\Tmatrix^{(0)} +  \frac{1}{{\hat g}^2} \, i\Tmatrix^{(1)}  
 + \order\lrbrk{ \frac{1}{{\hat g}^3} }     \equiv \unit+ i\Tmatrix \ ,
\ee
which contains all scattering amplitudes.  Each factor $\smatrix_{G_i}$ has a similar expansion:
\be
\label{Tmatrix}
 \smatrix_{G_i} = \unit + \frac{1}{\hat g} \, i\tmatrix_{G_i}^{(0)} +  \frac{1}{{\hat g}^2} \, i\tmatrix_{G_i}^{(1)}  
 + \order\lrbrk{ \frac{1}{{\hat g}^3} }  \equiv \unit+ i\tmatrix_{G_i} \ .
\ee
We may also formally refer to the entries of $i\tmatrix_{G_i}$ as "scattering amplitudes".
Integrability of the theory implies that both the $\Tmatrix^{(0)}$-matrix and the $\tmatrix_{G_i}^{(0)}$-matrix 
satisfy the classical limit of the YBE.  

The factorization equation \eqref{ws_smatrix_decomposition_txt} implies a close relation between the 
$L$-loop entries of $i\Tmatrix$ and  the $l\le L$-loop entries of $i\tmatrix_{G_i}$: 
\be
i\Tmatrix^{(L)}=\sum_{l=0}^{L+1}\,(i\tmatrix_{G_1}^{(l-1)})\otimes(i\tmatrix_{G_2}^{(L-l)}) \quad,\qquad 
i\tmatrix_{G_i}^{(-1)}=\unit \ .
\label{tmatrix_decomposition}
\ee
Consequently, if integrability is preserved, to find the $L$-loop $\Tmatrix$ matrix it suffices to find the $L$-loop
$\tmatrix$ matrix, as all the other terms in \eqref{tmatrix_decomposition} are already determined at lower loops.

If we denote by capital and dotted capital letters the indices acted upon by $G_1$ and $G_2$, respectively, 
the tree-level factorized $\Tmatrix$ matrix is given in terms of the $\tmatrix_{G_i}$-matrices as \cite{KMRZ}:
\be
\Tmatrix |\Phi_{A{\dot A}} \Phi'_{B{\dot B}}\rangle = 
(-)^{[{\dot A}]([B]+[D])} |\Phi_{C{\dot A}} \Phi'_{D{\dot B}}\rangle \tmatrix_{AB}^{CD}
+
(-)^{[{B}]([{\dot A}]+[{\dot C}])} |\Phi_{A{\dot C}} \Phi'_{B{\dot D}}\rangle \tmatrix_{{\dot A}{\dot B}}^{{\dot C}{\dot D}}  \ ,
\label{sign_def_Tmatrix_elements}
\ee
where $[\bullet]$ represents the grade of the argument, which is zero for a bosonic index and unity for a fermionic index.
Similarly, the matrix elements of a generic term $\tmatrix_i\otimes \tmatrix_j\subset \Tmatrix_{i+j+1}$ between 
worldsheet states $\Phi_{A{\dot A}} \Phi'_{B{\dot B}}$ are related to the matrix elements of $\tmatrix_i$ and $\tmatrix_j$ 
between (fictitious) two-particle states $|AB'\rangle$ and $|{\dot A}{\dot B}'\rangle$
\be
\tmatrix |AB'\rangle = |CD'\rangle (\tmatrix_i)^{CD}_{AB}
\qquad 
\tmatrix_{AB'}^{CD'} = \langle D'C|\tmatrix |AB'\rangle
\label{def_halfSmatrix}
\ee
by
\be
\Tmatrix  |\Phi_{A{\dot A}} \Phi'_{B{\dot B}}\rangle\supset \tmatrix_{i}\otimes \tmatrix_j  |\Phi_{A{\dot A}} \Phi'_{B{\dot B}}\rangle
= (-)^{[{\dot A}] [B]+[{\dot C}] [D]}|\Phi_{C{\dot C}} \Phi'_{D{\dot D}}\rangle
(\tmatrix_i)^{CD}_{AB}(\tmatrix_j)^{{\dot C}{\dot D}}_{{\dot A}{\dot B}}\ ;
\ee
momenta of the primed states are different from momenta of the other states. 
If either $\tmatrix_i$ or $\tmatrix_j$ are the identity matrix, {\it i.e.} if $(\tmatrix_i)^{{C}{D}}_{{A}{B}}
=\delta^{{C}}_{{A}}\delta^{{D}}_{{B}}$ or $(\tmatrix_j)^{{\dot C}{\dot D}}_{{\dot A}{\dot B}}
=\delta^{{\dot C}}_{{\dot A}}\delta^{{\dot D}}_{{\dot B}}$, we recover the two terms in~\eqref{sign_def_Tmatrix_elements}.

Since the world sheet theory is not Lorentz-invariant (by the existence of a fixed vector related to the choice of 
vacuum), neither the $\Smatrix$-matrix nor the $\smatrix$-matrix is invariant under crossing transformations
\be
\Smatrix^\text{cross} = {\cal C}^{-1}\Smatrix^\text{st}{\cal C} 
\qquad\qquad
\smatrix^\text{cross} = {\cal C}^{-1}\smatrix^\text{st}{\cal C}  \ .
\label{crossing_transformation}
\ee
Here $\Smatrix^\text{st}$ is the super-transpose of the $\Smatrix$-matrix in the two labels corresponding to the 
crossed particles,
\be
(M^\text{st})_{AB}=(-)^{[A][B]+[B]}M_{BA} \ ,
\label{crossing_rel}
\ee
%
%
${\cal C}$ is the charge conjugate matrix and one is also to change the sign of the energy and 
momentum of the particles that are crossed. The super-transpose is necessary if hermitian conjugation 
and complex conjugation are defined in the same way as for regular matrices. 
The standard (relativistic) crossing symmetry is expected to be replaced by the generalized 
crossing equations suggested in \cite{JanikCrossing}, which  relate in a nontrivial way 
$\Smatrix$ and $\Smatrix^\text{cross}$.
It is not difficult to see that the two transformations \eqref{crossing_transformation} are consistent with the factorization
\eqref{ws_smatrix_decomposition_txt}.
Through a sequence of crossing transformations  \eqref{crossing_transformation} we shall consistently construct the 
$u$-channel cuts by relating them to  $s$-channel without using the explicit form of the crossed S matrix.

\subsection{The perturbative expansion of the worldsheet S matrix}

As mentioned previously, the worldsheet S matrix is determined by symmetries up to an overall phase denoted by 
$\theta_{12}$ whose general structure in terms of spin-chain variables was discussed in~\cite{Arutyunov:2004vx}. 
Its strong coupling expansion at fixed spin-chain momenta is reviewed in Appendix~\ref{exp_phase}. 

Contact with worldsheet perturbation theory is however made \cite{Roiban:2006yc, KMRZ} in the strong 
coupling expansion at fixed worldsheet momenta (the "small momentum expansion")
\be
p_{\rm chain}=\frac{2\pi}{\sqrt{\lambda }}p_{\rm ws}=\frac{1}{\hg}p_\text{ws} \ .
\label{small_p_exp}
\ee
In \adssss, \adsst~ and \adscp~ the role of the coupling constant is played by a nontrivial function $h$ whose 
relation to the naive worldsheet coupling $\hg$ in these cases with less-than-maximal supersymmetry 
is subject to finite renormalization.
Taking this limit is straightforward for the state-dependent part of the S matrices and the result has the same 
structure as \eqref{Tmatrix}. 
We focus here only the features added by the dressing phase to the perturbative expansion of the 
worldsheet S matrix. Depending on the theory one may have different dressing phases in different sectors 
describing the scattering of different multiplets of the symmetry group; the 
same discussion applies separately for each of them.

The general form of the dressing phase is included in eq.~(\ref{theta_ito_chi}) with the Zhukowsky variables 
$x^\pm$ defined in \eqref{x+-} and expanded in \eqref{expanded_x+-}. It is not difficult to see that the leading 
term in the expansion of $x^\pm$,  which is independent of the indices $\pm$, cancels out in eq.~(\ref{theta_ito_chi}). 
It is also not difficult to construct the next order in the expansion  of the right-hand side 
of eq.~(\ref{theta12n}) for any $\chi^{(n)}$; thus, in the small momentum expansion, $\theta_{12}^{(n)}$ 
is ${\cal O}(\hg^{-2})$. Extracting the leading $\hg^2$ factor, ${\hat\theta}_{12}^{(n)}=\hg^2 \theta_{12}^{(n)}$, 
we have (see Appendix~\ref{ws_from_sc})
\be
\theta_{12}&=&\frac{1}{\hg}\sum_{n=0}^\infty\,\frac{1}{\hg^{n}}{\hat \theta}_{12}^{(n)} \ ,
\label{theta12_general_ws_expansion}
\ee
with ${\hat \theta}^{(1)}$ contributing at ${\cal O}({\hat g}^{-2})$, {\it i.e.} at one-loop order. 
For string theory in \adss, the leading term ${\hat\theta}_{12}^{(0)}$ is also the leading term in the small momentum 
expansion of the AFS phase~\cite{AFS}. While the AFS phase contains logarithms of worldsheet momenta, each term 
in its small momentum expansion \eqref{small_p_exp} is a rational function of worldsheet momenta and energies.

Defining ${\hat \smatrix}$ to be the symmetry-determined part of the S matrix dressed with ${\hat\theta}_{12}^{(0)}$,
the $\smatrix$-matrix can be written as 
\be
\smatrix = e^{\frac{i}{2}(\theta_{12}-\frac{1}{\hg}{\hat \theta}_{12}^{(0)})}{\hat\smatrix} \ ;
\label{def_S_phase}
\ee
Its large $\hg$ expansion identifies the entries of $\smatrix$ containing information on the loop corrections
to the dressing phase:
\be
\smatrix 
=\unit+\frac{1}{{\hat g}}i{\tmatrix}^{(0)}+\frac{1}{{\hat g}^2}i\left({\hat\tmatrix}^{(1)}+\frac{1}{2}{\hat\theta}^{(1)}_{12}\unit\right)
+\frac{1}{{\hat g}^3}i\left({\hat\tmatrix}^{(2)}+\frac{i}{2} {{\hat\theta}^{(1)}}_{12}{\tmatrix}^{(0)}
+\frac{1}{2}{{\hat\theta}^{(2)}}_{12}\unit \right)+ \order\lrbrk{ \frac{1}{{\hat g}^4} } \ .
\label{SmatExpansion}
\ee
We see in particular that the one-loop correction to the dressing phase affects only the 
diagonal entries of the one-loop S matrix.\footnote{
We shall use the notation ${\tmatrix}^{(0)}$ rather than ${\hat\tmatrix}^{(0)}$ for the leading term in the small 
momentum expansion of $\smatrix$ because this term is unaffected by the corrections to the dressing phase.
}
Moreover, since the part of the S matrix that is determined by symmetries 
has rational dependence on momenta and so does the classical phase ${\hat\theta}_{12}^{(0)}$, 
the only transcendental dependence on external momenta comes from $\theta_{12}^{(i)}$ 
with $i\ge 1$. 
A direct demonstration of the structure of the ${\cal O}({\hg}^{-3})$ term would give strong indication of 
the exponentiation of the one-loop phase. 

For example, for string theory in AdS$_5\times$S$^5$, the expansion of the first (loop correction to the) 
dressing phase is \cite{Arutyunov:2006iu, Roiban:2006yc, KZ} 
\be
{\hat\theta}^{(1)}_{12}
=-\frac{1}{\pi}\frac{p^2p'{}^2(\energy\energy'{}-pp'{})}{(\energy'{}p-\energy p'{})^2}
\ln\Big|\frac{p'_{-}}{p_{-}}\Big| +\text{rational}\ ,
\label{1loopphase}
\ee
where 
\be
p_{\pm} = \frac{1}{2}(\energy \pm p) \  , 
\ee
the mass of the worldsheet excitations is set to $m=1$ and $p_+$ and $p'_+$ were eliminated 
from the argument of the logarithm through the on-shell condition
\be
4p_+p_- = \energy^2 - p^2 = 1
\ .
\ee
In the sec.~\ref{AdS5} we will find this expression for ${\hat\theta}_{12}^{(1)}$ from a direct calculation 
through the generalized unitarity method; we will also demonstrate that the ${\cal O}({\hg}^{-3})$ term 
in the perturbative expansion of the worldsheet S matrix has the form given by \eqref{SmatExpansion}.

\subsection{Generalized unitarity and the worldsheet S matrix}

The ($d$-dimensional) generalized unitarity method \cite{UnitarityMethod, BCFUnitarity}, its implementation 
in the method of maximal cuts \cite{FiveLoop, CompactThree} and in cases massive of massive particles 
\cite{Kosower:2011ty, Johansson:2012zv} provide powerful tools for the construction of one- and higher-loop 
quantum field theory scattering amplitudes. 
In the presence of a suitable regularization, such as dimensional regularization, complete amplitudes can 
be constructed. Terms with rational dependence on momenta (and thus with no cuts) are related to terms 
that exhibit cuts in the presence of the regulator but disappear as the regulator is removed. We refer to them 
as "rational terms" or $d$-dimensional cut constructible terms. The rest, which can be determined from 
unregularized unitarity cuts, are referred to as two-dimensional cut-constructible terms.

Here we will use the generalized unitarity method to find the two-dimensional cut-constructible parts of the 
one- and two-loop corrections to the worldsheet S matrix. It is usually the case that rational terms undetermined 
at some loop level lead to missing terms with logarithmic (or, in general, transcendental) dependence on momenta 
at the next loop level. While at one loop we do not determine explicitly
rational terms, we bypass this issue by making use of the rational terms that are determined by symmetries 
at one loop; the rational terms proportional to the identity matrix -- and thus undetermined -- turn out to be irrelevant 
for finding all logarithms at two loops.
\footnote{\label{notadpoles}
Since we do not evaluate directly all rational terms we will not be able to completely 
address questions regarding the fate of divergences that appeared in previous attempts to 
compute the one-loop worldsheet S matrix in \adss~\cite{MRunpublished}. In two dimensions, the only one-loop 
divergent rational terms are related to tadpole integrals; following the four-dimensional construction of
\cite{Britto:2012mm}, in general theories may be determined from one-particle cuts. As we shall discuss in more detail in 
sec.~\ref{higher_loops_reg}, 
as a consequence of the factorization of six-point tree-level amplitudes in integrable theories into a product of two 
four-point amplitudes, single-particle cuts of one-loop four-point amplitudes have an additional hidden cut condition 
and thus are in fact two-particle cuts. This suggests that there are no tadpole integrals in one-loop amplitudes apart 
from those related to wave-function renormalization. 
Since the string tension should not receive infinite renormalization (which is supported by the symmetry-based determination 
of the off-diagonal terms with completely rational momentum dependence employed here), we may therefore say that, 
to some extent, generalized unitarity provides a divergence-free construction of the one-loop S matrix.
}

Quite generally, one may use the generalized unitarity method to determine directly the $\Smatrix$-matrix elements, 
which are the scattering amplitudes of the worldsheet theory. The factorization of the $\Tmatrix$ matrix suggests 
however that we ought to construct the $\smatrix$-matrix instead. Indeed, at any loop order $L$, cuts
not already computed at lower loops contribute only to  
$\tmatrix_{G_1}^{(L)}\otimes \unit +\unit\otimes \tmatrix_{G_2}^{(L)}$. 
Since the $\tmatrix$-matrix elements are substantially simpler than the entries of the $\Tmatrix$ matrix, we shall focus 
on the former. 

The structure of the tree-level S matrix is tightly constrained by the (assumed) integrability of the theory which, 
in particular, implies absence of particle production and thus that the number of incoming particles is 
the same as the number of outgoing ones. Consequently, all components of generalized cuts should 
also obey this constraint. For example, for the $2\rightarrow 2$ S matrix at two loops, the three-particle 
cut with two external particles on each side of the cut  vanishes identically. Of course, not all three-particle cuts of 
two-loop four-point amplitudes vanish identically; an example is the cut with one external line on one side and the other 
three on the other side of the cut. 

At one-loop level, two-particle cuts in two dimensions play a role analogous 
to that of quadruple cuts in four dimensions. Indeed, since in two dimensions
momenta have two components, cutting two internal lines of a one-loop amplitude 
completely constrains the loop momentum.
Thus, two-particle cuts are maximal cuts for one-loop amplitudes; similarly, four-particle cuts are maximal cuts at two loops. 
Consequently, the coefficients of one-loop bubble integrals are simply given by products 
of tree-level amplitudes appropriately summed over all possible internal states, in close 
analogy to the coefficients of box integrals in four-dimensional field theories.

To construct either the $\Tmatrix$ matrix or, separately, the $\tmatrix_{G_i}$-matrix factors  it is necessary 
to have a spectral decomposition of the identity operator in the Hilbert space of states. We will focus 
here on the two-particle identity operator, for which we can write:
\be
\unit = |\Phi_{E{\dot E}}\Phi'_{F{\dot F}}\rangle   \langle \Phi'{}^{{\dot F}F}\Phi^{{\dot E}E}|
~~~\text{and}~~~
 \langle \Phi'{}^{{\dot A}A}\Phi^{{\dot B}B}|\Phi_{E{\dot E}}\Phi'_{F{\dot F}}\rangle  
 =\delta_E^B\delta_{\dot E}^{\dot B}\delta_F^A\delta_{\dot F}^{\dot A} \ .
\label{unit_full_space}
\ee
Here $|\Phi_{E{\dot E}}\rangle$ corresponds to the state created by the field $\Phi_{E{\dot E}}$
and $\langle \Phi_{{\dot E}E}|$ is the conjugate of that state.  
We may further split the spectral decomposition (\ref{unit_full_space}) into $\unit_{G_1}\otimes\unit_{G_2}$ identity operators:
\be
\unit_{G_1} \otimes\unit_{G_2} = (|EF'\rangle\langle F' E|)\otimes (|{\dot E}{\dot F}'\rangle\langle {\dot F}' {\dot E}|)
~~~\text{with scalar product:}~~~
 \langle A'  B|E F' \rangle  
 =\delta_E^B\delta_F^A \ ,
\ee
where the primes denote the fact that two excitations carry momenta different from the other two; excitations
with different momenta are orthogonal, $\langle A'| E\rangle = 0$.

As usual, we interpret a generalized unitarity cut as selecting from an amplitude the parts that have a certain 
set of propagators present. Using the spectral decomposition in one of the two group factors and the definition \eqref{def_halfSmatrix} of the matrix elements of $\tmatrix$, it is easy to see 
that the $(L_1+L_2+1)\rightarrow L_1\times L_2$ $s$-channel cut of the $(L_1+L_2+1)$-loop component of the 
$\tmatrix$ matrix  is:
\be
\label{schannelcut}
(i\tmatrix^{(L_1+L_2+1)})_{AB'}^{CD'} \Big|^{L_1\times L_2}_{s-\text{cut}}&=&
(i)^2\langle D' C|(i\tmatrix^{(L_1)}) |EF'\rangle\langle F' E| (i\tmatrix^{(L_2)})|AB\rangle
\cr
&=& 
(i)^2 (i\tmatrix^{(L_1)})_{EF'}^{CD'} (i\tmatrix^{(L_2)})_{AB'}^{EF'} \ ,
\ee
where the $(i)^2$ factor originate from the two cut propagators.

To construct the $u$-channel cut we use the crossing transformation \eqref{crossing_rel} 
to relate it to an $s$-channel cut. The crossing transformation (\ref{crossing_rel}) acts consistently 
on both matrices $\Smatrix$ and $\smatrix$. Here we will use the transformation on the latter one:  
\be
&&(-)^{[B][D]+[B]}\langle D' C |(i\tmatrix^{(L_1+L_2+1)}) |AB'\rangle\Big|^{L_1\times L_2}_{u-\text{cut}}  
= \langle B' C|(i\tmatrix^\text{st})|AD'\rangle\Big|^{L_1\times L_2}_{s-\text{cut}}
\nonumber\\[1pt]
&=& (i)^2\langle B' C|(i\tmatrix^{(L_1),\text{st }})|EF'\rangle \langle F' E|(i\tmatrix^{(L_2),\text{st}})|AD'\rangle
\nonumber\\[1pt]
&=& (i)^2(-)^{[B][F]+[B]}\langle F' C|(i\tmatrix^{(L_1)}) |EB'\rangle\;(-)^{[F][D]+[F]}\langle D' E|(i\tmatrix^{(L_2)}) |AF'\rangle \ .
\label{uchannel_details}
\ee
In terms of matrix elements this becomes 
\be
\label{uchannelcut}
(i\tmatrix^{(L_1+L_2+1)})_{AB'}^{CD'} \Big|^{L_1\times L_2}_{u-\text{cut}} = 
(i)^2(-)^{([B]+[F])([D]+[F])}
(i\tmatrix^{(L_1)})_{EB'}^{CF'} (i\tmatrix^{(L_2)})_{AF'}^{ED'} \ .
\ee
One may also understand the sign factors by {\em formally} permuting indices and bringing them to the same order 
as in the $s$-channel cut. 
We will use these relations repeatedly in the following sections to construct loop-level worldsheet S-matrix elements
in \adss, \adscp, \adssss~ and \adsst.

The $t$-channel cut is structurally different from the other two: 
there is no net momentum flow across the cut.
Because of this, in the product of two scattering amplitudes in this channel one encounters a kinematic singularity
in the form of a factor of $\delta(0)$ or perhaps as the square of a delta function, {\it e.g.} $\delta(p_1-p_3)^2$. This 
singular momentum configuration is forced on us by two-dimensional kinematics and the integrability of the theory.
Clearly, some form of IR regularization is necessary to extract reliable information from this cut. 

It is not obvious what such a regularization might be. Continuing external momenta to $d=2-2\epsilon$ 
would forbid the appearance $\delta(0)$ by allowing {\em individual} momenta to not be conserved 
in the scattering process by an ${\cal O}(\epsilon)$ amount. As discussed previously however, this regularization
cannot preserve all the classical symmetries of the theory. Moreover, since integrability also requires that individual 
momenta be conserved in a scattering process (for any number of external lines and independently of 
two-dimensional kinematics), regulating $\delta(0)$ as above would also break integrability. 
Since the $t$-channel bubble integral is in fact a constant, we will simply ignore it and determine it together with
all the other rational terms from symmetry considerations.

The description of unitarity cuts earlier in this section assumed that scattering amplitudes are normalized in the standard 
Lorentz-invariant way. As reviewed in Appendix~\ref{recall_trees}, worldsheet tree-level S-matrix elements are normalized 
slightly differently both because of the mode expansion of fields and because the on-shell conditions and momentum 
conservation are solved using special properties of two-dimensional kinematics.
Thus, to apply the usual rules we need to compensate for the solution of the momentum conservation constraint and also  
adjust the normalization of the creation operators to the relativistic one for each cut leg for both the left and right side of
the cut, {\it i.e.} we need to multiply by  $(\sqrt{2\energy})^2$ for each cut leg with energy $\energy$.
For example, for each two-particle cut obtained by multiplying  two S-matrix elements in the 
standard worldsheet normalization we need to supply an additional factor of 
\be
J=(\sqrt{2\energy}\sqrt{2\energy'{}})^2 \left(\frac{d\energy}{dp} - \frac{d\energy'}{dp'}\right) \ ;
\label{Jac0}
\ee
For a standard dispersion relation this is:
\be
(\sqrt{2\energy}\sqrt{2\energy'{}})^2
\frac{\energy'{}p-\energy p'{}}{\energy\energy'{}} = 
4(\energy'{}p-\energy p'{}) = \frac{2(m^2 p'_{-}{}^2-m'{}^2p_{-}^2)}{p_{-}p'_{-}} \ .
\label{Jacobian}
\ee
We shall denote this expression by $J_{s, u}$, depending on the channel in which it appears. It is
not difficult to see however that $J_s=J_u$ so we will at times also simply denote it by $J$.

\subsection{On higher loops, regularization, factorization and related issues \label{higher_loops_reg}}

Two-particle cuts are sufficient for one-loop calculations. In general, the $2$-, $3$-, $\dots$, $2L$-particle 
generalized cuts of an $L$-loop amplitude determine its cut-constructible part. It is interesting to examine 
the cuts  needed for higher-loop calculations and understand their structure in view of 
the special properties of the theory. 

Let us assume that we focus on the massive sector of the theory and that we choose a regularization that 
preserves the integrability; for example, a possibility is to assume that, since the theory should be 
finite both in the IR (being massive) and in the UV (being a theory with spontaneously broken conformal invariance) no regularization is necessary. 
A regularization preserving the integrability of the theory will also preserve the absence of particle production. 
In particular, all nonzero amplitudes have an even number of external legs and an equal number of incoming 
and outgoing particles; $2k$-point amplitudes are a sequence of four-point scattering events. 
This structure places a number of constraints on the cuts that determine the higher-loop S-matrix elements. 

We first notice that any cut that is a product of tree-level amplitudes is also a maximal cut.
Indeed, in an integrable quantum field theory  the higher-point S-matrix elements are given by 
the sum of products of four-point S-matrix elements; the internal lines connecting them are on shell and, as 
such, may be interpreted as cut propagators. Thus, a generalized cut of an $L$-loop amplitude that is a product 
of tree-level amplitudes is also naturally the cut of the same $L$-loop amplitude in which all tree-level factors are 
four-point tree-level amplitudes -- {\it i.e.} the maximal cut of the amplitude.  The coefficient of an 
$L$-loop integral that has a non-zero maximal cut is simply given -- for any $L$ -- by the product of tree-level 
amplitudes appropriately summed over all the possible states crossing the generalized cut. 
For example, the two-loop six-point tree-level amplitude the two-loop factorizes as a sum of products of two 
four-point amplitudes;  Then, the three-particle cut with one external leg on one side and three on the other side 
of the cut may be interpreted as contains a hidden on-shell condition for a propagator internal to the six-point 
tree-level amplitude \cite{KMRZ} and thus it is in fact a four-particle cut.

In the simplest choice of regularization described above such cuts do not determine all
terms with logarithmic dependence beyond one loop. For this purpose, and to be able to include the contribution 
of symmetry-determined rational terms at lower-loop orders, it is useful to consider cuts that break up an amplitude
into a product of an $L_1$-loop and an $L_2$-loop amplitude. 
Integrability then implies that each such amplitudes is a sum of products of $l_i\le L_1$ and $l'_j\le L_2$ 
four-point amplitudes with $\sum l_i=L_1$ and $\sum l'_j=L_2$. 
In non-integrable theories, focusing on a single term in this sum requires imposing additional cut conditions.
In contrast, in an integrable theory these conditions are naturally present and do not impose new 
restrictions on the amplitudes building up the generalized cut.
In each amplitude factor we may recursively include the symmetry-determined rational terms at $l_i$-loop order. 
Based on the structure of one-loop integrals and on the structure of two-particle cuts it is possible to argue\footnote{
A property of the two-dimensional one-loop bubble integrals ${\tilde I}_s$ and ${\tilde I}_u$ listed in Appendix~\ref{multimass}, 
is that only their difference contains logarithms; thus, the difference of the coefficients of these integrals determines the 
coefficient of the logarithm of external momenta.
}
quite generally that $(L-1)$-loop rational terms proportional to the identity matrix in field space -- and thus 
undetermined by symmetries --  do not contribute to $L$-loop terms with  logarithmic dependence on momenta.
One may check on a case by case basis whether rational terms which are not determined by symmetries at 
$l\le (L-2)$-loop order make any contributions to the logarithmic terms at $L$-loops. The argument here guarantees 
that no such contributions exist at two loops.

An argument similar in spirit to the one in the previous paragraph suggests that logarithmic terms at $L$-loops 
preserve the factorization \eqref{tmatrix_decomposition} of the $\Tmatrix$ matrix. Indeed, let us consider the cut in which 
one side of the cut is a four-point tree-level amplitude. Since
\be
i\Tmatrix^{(0)} = (i\tmatrix^{(0)}_{G_1})\otimes \unit + \unit\otimes (i\tmatrix^{(0)}_{G_1}) \ ,
\ee 
it follows that the product $(i\Tmatrix^{(L)})(i\Tmatrix^{(0)})$ will exhibit the factorization \eqref{tmatrix_decomposition} 
after the cut conditions are released and the integrals are evaluated.
This argument is sufficient to guarantee factorization through two-loop order, where only cuts of the type above 
are important. While at higher loops the factorization \eqref{tmatrix_decomposition} clearly holds at the level of cuts,
it is less straightforward to see it in general once cut conditions are removed. Nevertheless, consistency of cuts in all channels 
will relate all contributions of $\Tmatrix^{(L)}$ to the ones that are factorized following the argument above, suggesting 
that it is plausible that eq.~\eqref{tmatrix_decomposition} indeed holds to all orders in perturbation theory.  

A class of theories that feature in the context of the gauge/string duality are those containing multiplets of the symmetry 
group that have different masses. An interesting question which arose in the comparison of worldsheet and spin chain 
S matrices is whether calculations in the theory truncated to some 
subset of fields ({\it e.g.} all fields with some subset of masses) yield the same result as calculations in the complete theory.
From a Lagrangian point of view this is clearly impossible unless the desired subset is decoupled from 
the other fields.  Generalized unitarity provides more structure since it makes use of properties of the S matrix 
not immediately visible in the Lagrangian.
To understand the terms that are missed by restricting ourselves to a subset of fields, let us start with an 
integrable quantum field theory with particles of different masses and ignore one of them, denoted by $\varphi$.
We aim to construct the scattering amplitudes of the remaining fields from their generalized cuts; ignoring $\varphi$ 
means that in generalized unitarity cuts we sum over all states except $\varphi$.
It is not difficult to see that, in a generic quantum field theory, these steps result in an incorrect S matrix. 

\begin{figure}[ht]
\begin{center}
\includegraphics[height=32mm]{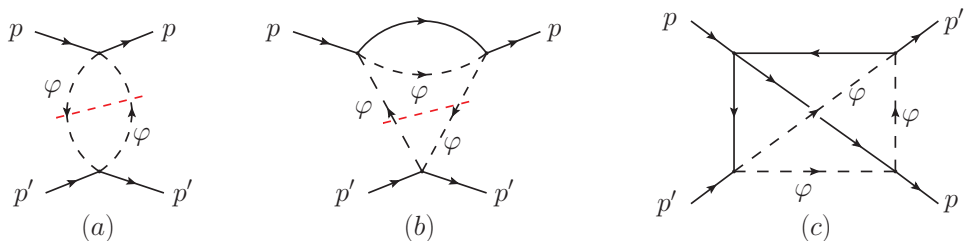}
\caption{Integrals with fields that are truncated away at the classical level. The external momentum configuration 
guarantees that in (a) and (b) there are $\varphi$ states crossing the cut. The one- and two-loop integrals are 
constants, independent of external momenta.
The three-loop integral depends on both external momenta and therefore need not be a rational function.
\label{truncated_fields}}
\end{center}
\end{figure}

If the tree-level S matrix has further special properties, then the two-dimensional cut-constructible 
terms may be reproduced correctly. Indeed, let us assume that masses of individual particles are conserved and that 
the scattering of $\varphi$ (or, more generally, the scattering of fields we want to truncate away) is {\em reflectionless}, 
{\it i.e.} that there is no momentum exchanged  between the two scattered particles. All theories with 
different worldsheet masses that we will  discuss in later sections are of this type\footnote{This includes the massless 
particles that do not decouple from the supercoset sigma model for strings in \adsst~ but can be consistently truncated 
away at the classical level \cite{Hoare:2013pma}.}.
With such assumptions we see that there is no net momentum flow across any cut putting on shell 
only $\varphi$ fields, such as the cuts in fig.~\ref{truncated_fields}$(a)$ and  \ref{truncated_fields}$(b)$. 
Indeed, since integrability implies that momenta of $\varphi$ fields are separately conserved, 
$\varphi$ fields run in the loop only if the external momenta are chosen as shown.
The  two integrals are momentum-independent because the integrand depends on at most one external momentum 
and thus disappears upon use of the on-shell condition.
It is not difficult to argue in a similar way that all integrals having this property --~that they have no momentum flow 
across some generalized cut that sets on shell only $\varphi$ fields~-- are independent of external momenta and 
thus contribute only to the rational part of the S matrix.

This argument guarantees that, at one and two loops, all logarithmic terms are correctly reproduced by 
calculations in the theory  obtained by truncating away fields that scatter reflectionlessly off the remaining ones. 
At three loops and beyond it is possible construct integrals --~such as the one in fig.~\ref{truncated_fields}$(c)$~--  
that do not have any generalized cut crossed only by $\varphi$-type fields and thus the argument here does 
not immediately imply that such integrals are rational functions. 
Thus, at three loops and beyond quantum calculations in the truncated theory do not necessarily yield the complete 
result.

In a theory exhibiting nontrivial two-point functions the construction of the S matrix through the LSZ reduction
requires that the physical pole of the two-point function of fields be identified and that its residue be correctly 
included in the reduction of Green's functions to S-matrix elements. In general, this implies that the naive $L$-loop 
amplitudes are corrected by the addition of lower-loop amplitudes multiplied by the residue of the two-point function.
In all theories we will discuss in later sections it is expected that the first correction to the dispersion relation 
(and hence to the two-point function) is at two loops (see {\it e.g.} \cite{Klose:2007rz} for a calculation in the 
near-flat space limit of \adss). Thus, through two loops, only the rational terms in the S matrix will be affected;
since we determine the (off-diagonal) rational terms from symmetry considerations we will ignore the corrections 
to the propagator.
At higher loops it is, of course, important to include such contributions.

One way to construct scattering amplitudes through the generalized unitary method is to begin with an ansatz  
for amplitudes in terms of Feynman-like integrals whose coefficients are subsequently determined by comparing 
the generalized cuts of the ansatz with the generalized cuts of amplitudes constructed in terms of lower-order amplitudes.
The ansatz is based on the structure of the Feynman graphs of the theory.
It is interesting to note that, in all cases we are interested in, all cuts that break up an $L$-loop amplitude into a 
product of tree-level amplitudes completely freeze the loop momenta and therefore cannot distinguish 
between scalar integrals and tensor integrals. Since rational terms are supplied separately, one may thus attempt 
to construct ans\"atze in terms of only scalar integrals; such ans\"atze turn out to be sufficient at one and two loops. 

\section{General expressions for one- and two-loop amplitudes \label{GeneralExpressions}}

Due to the special properties of S-matrices of the Green-Schwarz string in \adss, \adscp, \adssss~ and \adsst, 
generalized unitarity allows us to derive compact general expressions for all logarithmic terms of the one- and 
two-loop amplitudes in these theories.  We shall outline the derivation in this section; we will then proceed in 
later sections to discuss each of these theories separately, pointing out the features specific to each of them.

For a compact notation we will interpret all indices used in sec.~\ref{worldsheetPT} 
as multi-indices while keeping their capital letter appellation, $A, B, C,\dots$. Each multi-index stands 
for the set $(\text{field label}, \text{mass of field}, \text{sector})$.\footnote{\label{footnoteLnR}
The last entry of the multi-index refers to transformation of the excitation under some representation ($L$) or the 
conjugate representation ($R$) of the symmetry group, such as $PSU(1|1)^2$ for strings in \adssss.   They are 
also referred to as left- and right-moving excitations in \cite{OhlssonSax:2011ms}.  In integrable models, the concept 
of left- and right-excitations was originally introduced in the context of massless factorized 
scattering \cite{Zamolodchikov:1992zr}.  
In that case the  $LL$ and $RR$ scattering is non-perturbative and $LR$ and $RL$ is perturbative. While we use the 
original notation, perhaps a more natural one is  $L\leftrightarrow +$, $R\leftrightarrow -$ used in \cite{Hoare:2013pma}  
in a related context.
For us all states are massive and therefore a left-motion can be transformed into right-motion by a change of 
two-dimensional frame. Because of this scattering in all four sectors is perturbative.
The notation $RR$ (sector of the S matrix) should not be confused with that for the RR flux. 
}
Not all entries 
are relevant in all theories. For example, in \adss~ all fields have the same mass and 
there are no left and right excitations; thus, in this case only the field label is relevant. In \adssss~ however all entries 
are important.
The grade $[A]$ of a multi-index is the grade of the field label ($0/1$ for bosons/fermions). 

\begin{figure}[ht]
\begin{center}
\includegraphics[height=32mm]{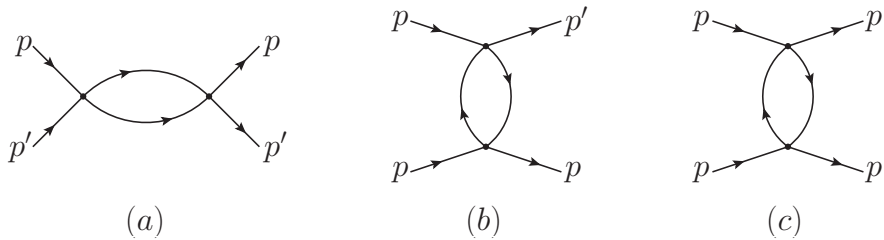}
\caption{The integrals appearing in the one-loop four-point amplitudes.  Tensor integrals can be reduced to them 
as well as to tadpole integrals, which are momentum-independent. \label{1loop_integrals}}
\end{center}
\end{figure}

\begin{figure}[ht]
\begin{center}
\includegraphics[height=32mm]{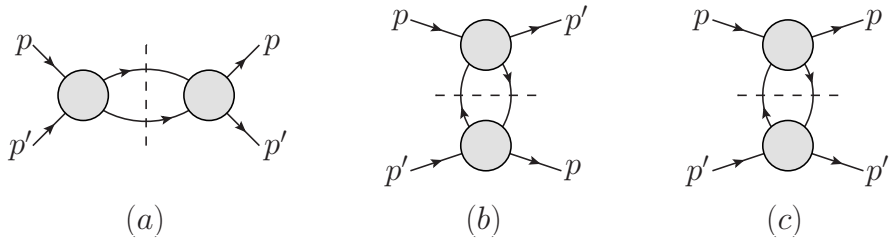}
\caption{Two-particle cuts of the one-loop four-point amplitudes \label{1loop_2p_cut}}
\end{center}
\end{figure}

\subsection{The general expression for one-loop amplitudes}

The integrals that can appear in one-loop amplitudes are shown in fig.~\ref{1loop_integrals}; their details 
depend on the worldsheet spectrum. The general structure of one-loop amplitudes is
\be
i\tmatrix^{(1)}&=&\frac{1}{2} C_s\,{\tilde I}_s
+\frac{1}{2}C_u\, {\tilde I}_u +\frac{1}{2}C_t\, {\tilde I}_t +\text{rational}\ . 
\label{ansatz_1loop}
\ee
with the factors of $1/2$ being the symmetry factors of  bubble integrals; 
the expression for these integrals  ${\tilde I}$ as well as their equal-mass versions are collected in 
Appendix~\ref{1loopint}. 
The coefficients $C_{u,s}$ are tensors in field space. As discussed in the previous section, we cannot reliably 
determine the coefficient $C_t$ due to the kinematic singularity of the $t$-channel cut.
As emphasized there, at one-loop, two-particle cuts are also maximal; this simply implies that the two coefficients, 
$C_s$ and $C_u$, have simple expressions in terms of the tree-level S matrix $i\tmatrix^{(0)}$:
\be
(C_s)_{AB'}^{CD'}  &=&(i)^2 J_s \sum_{E, F'}(i\tmatrix^{(0)})_{EF'}^{CD'} (i\tmatrix^{(0)})_{AB'}^{EF'}
\nonumber\\[2pt]
(C_u)_{AB'}^{CD'}  &=& 
(i)^2J_u\sum_{E, F'}(-)^{([B]+[F])([D]+[F])}
(i\tmatrix^{(0)})_{EB'}^{CF'} (i\tmatrix^{(0)})_{AF'}^{ED'} \ .
\label{generalC1loop}
\ee
The Jacobians $J_{s}$ and $J_{u}$ are the adjustment factors 
eq.~(\ref{Jacobian}) needed to transform the S-matrix elements to the relativistic normalization.
We recall that the precise form of these factors depends on the dispersion relation.

The explicit form of the one-loop bubble integrals from Appendix~\ref{1loopint} implies that the difference
\be
\frac{C_s}{J_s}-\frac{C_u}{J_u}
\label{difference}
\ee
governs the logarithmic dependence of one-loop amplitudes on external momenta. Noticing that only ${\tilde I}_s$
has a rational component, we can cleanly separate all logarithmic dependence on momenta by organizing 
$i\tmatrix^{(1)}$ as
\be
i\tmatrix^{(1)}=\frac{1}{2} \frac{C_s}{J_s}\,(J_s{\tilde I}_s+1 )
+\frac{1}{2}C_u\, {\tilde I}_u  +i{\tilde \tmatrix}^{(1)} \ ,
\ee
where 
\be
i{\tilde \tmatrix}^{(1)}=i{\hat\tmatrix}^{(1)}+i\Phi \unit \ .
\label{tildeT}
\ee
$i{\hat\tmatrix}^{(1)}$ was introduced in eq.~\eqref{SmatExpansion} as the ${\cal O}({\hat g}^{-2})$ term in 
the small momentum expansion of the symmetry-determined part of the S matrix dressed with the classical part 
of the dressing phase and $\Phi$ is the contribution of rational terms in the dressing phase which are not 
determined by symmetries. We will notice that, in all theories we analyze, the off-diagonal entries of 
$i{\hat\tmatrix}^{(1)}$ are proportional to the corresponding tree-level amplitudes.

\subsection{The general expression for two-loop amplitudes}

\begin{figure}[ht]
\begin{center}
\includegraphics[height=68mm]{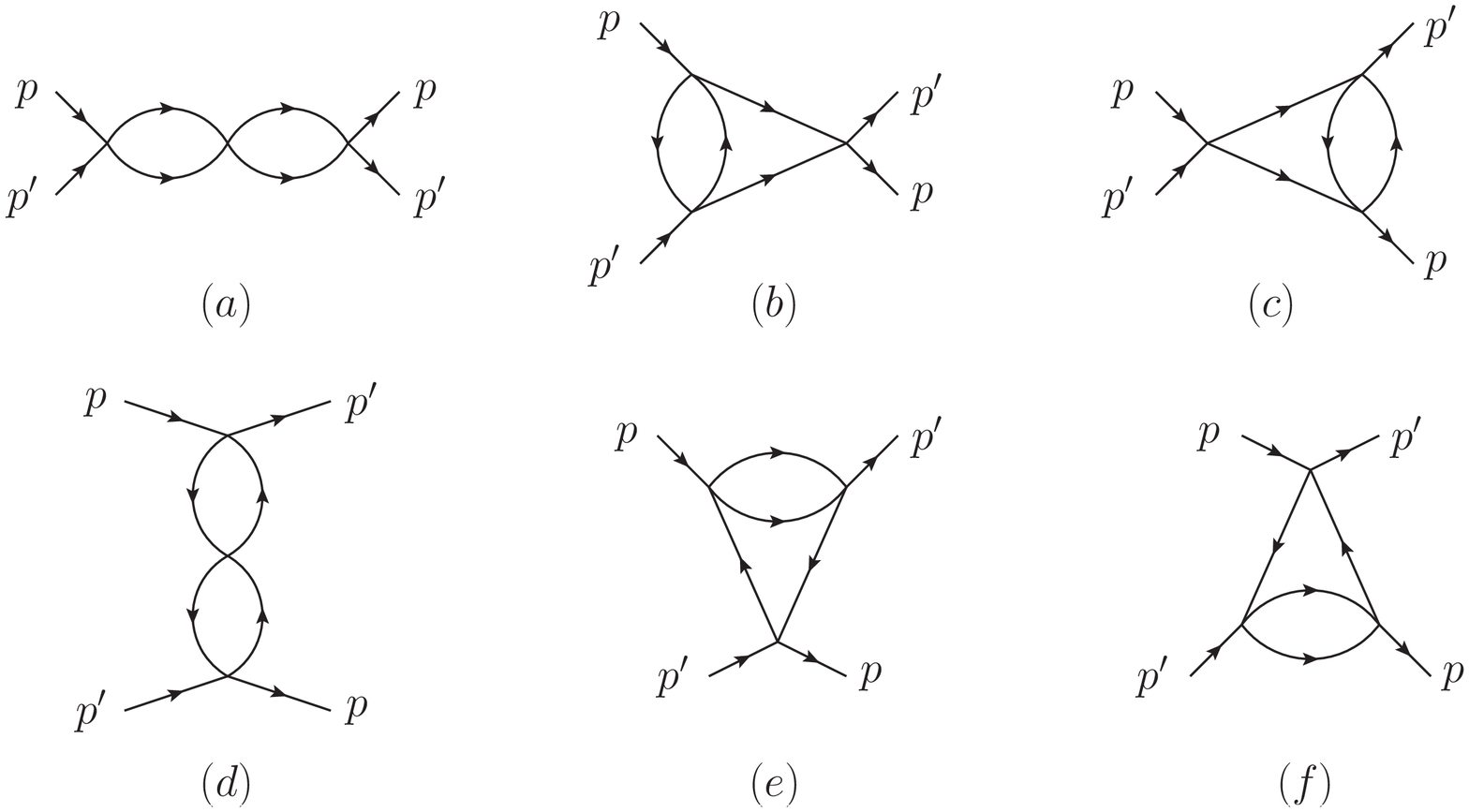}
\caption{The integrals appearing in the two-loop four-point amplitudes. Each cut in fig.~\ref{2loop_i2p_cut}
determines the coefficient of one of these integrals.  There exist, of course, other two-loop four-point integrals; 
the structure of the Lagrangian suggests that integrals with vertices with an odd number of edges cannot 
appear while the integral with a six-point vertex is momentum-independent and thus it can contribute only to 
terms with rational momentum dependence.\label{2loop_integrals}}
\end{center}
\end{figure}

\begin{figure}[t]
\begin{center}
\includegraphics[height=60mm]{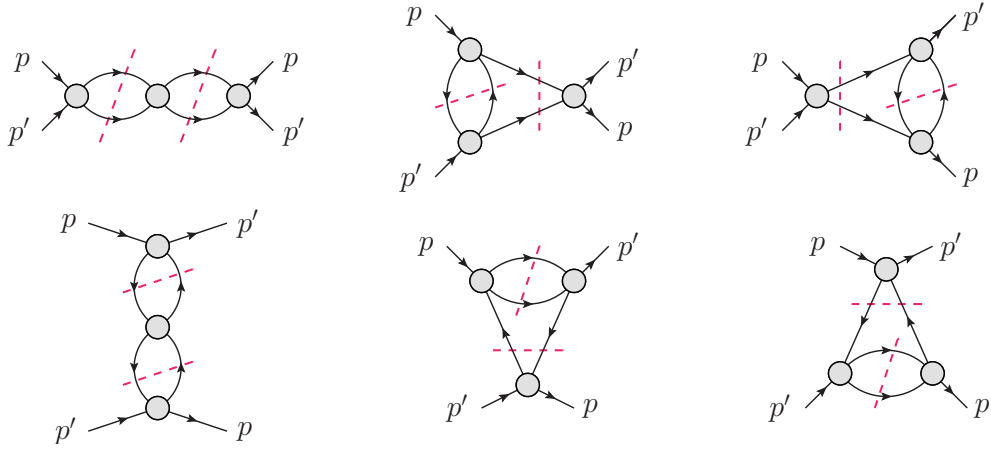}
\caption{Iterated two-particle cuts of two-loop four-point amplitudes. They are all maximal cuts (in two dimensions). 
It is not possible to relax the cut condition on any propagator either because the corresponding tree-level amplitude 
does not exist or because the resulting higher-point tree amplitude has an on-shell propagator as a consequence of 
integrability and S-matrix factorization. As discussed in sec.~\ref{worldsheetPT} all cuts of a four-point two-loop 
amplitude which is a product of tree amplitudes is equivalent to a sum of the cuts shown here.\label{2loop_i2p_cut}}
\end{center}
\end{figure}

To construct the two-loop correction to the worldsheet S matrix,
we begin by constructing an ansatz that contains all two-loop integrals 
that can appear and have a logarithmic dependence on momenta.  
We list in fig.~\ref{2loop_integrals} all integrals that have maximal cuts that are kinematically 
non-singular. The cuts determining their coefficients are shown in fig.~\ref{2loop_i2p_cut}.
In addition, we will include integrals that do not have non-singular maximal cuts but have single two-particle cuts;
they are just products of the $s$- or $u$-channel bubble integrals with the $t$-channel integral ${\tilde I}_t$. The latter 
factor is constant and can thus be absorbed in the coefficient of the one-loop integrals.
The ansatz is therefore:
\be
i\tmatrix^{(2)}&=&\frac{1}{4}C_a{\tilde I}_a+\frac{1}{2}C_b{\tilde I}_b+\frac{1}{2}C_c{\tilde I}_c
+\frac{1}{4}C_d{\tilde I}_d+\frac{1}{2}C_e\tilde{I}_e+\frac{1}{2}C_f{\tilde I}_f
\nonumber\\[1pt]
&+&\frac{1}{2} C_{s,\text{extra}}\,{\tilde I}_s+\frac{1}{2}C_{u,\text{extra}}\,{\tilde I}_u
\nonumber\\[2pt]
&+&\text{rational} \ ,
\label{twoloop_tmatrix_improved}
\ee
where we explicitly included the symmetry factors of integrals. A sum over possible distributions of internal masses is 
assumed. While here we are keeping the setup general, in all our explicit two-loop 
calculations we shall have all masses equal. The 
relevant integrals are listed in Appendix~\ref{one_n_twoloop_ints}.
This ansatz manifestly satisfies the vanishing of the three-particle cut containing five-point tree-level amplitudes.

As at one loop, the coefficients $C$ 
are tensors in field space. Maximal cuts determine them in terms of the tree-level S-matrix 
elements or, alternatively, in terms of the tree-level S-matrix elements and one-loop 
integral coefficients \eqref{generalC1loop}:
\be
(C_a){}_{AB'}^{CD'}&=&(i)^2J_s\sum_{G,H'}(i\tmatrix^{(0)})_{GH'}^{CD'}(C_s)_{AB'}^{GH'}
\cr
(C_b){}_{AB'}^{CD'}&=&(i)^2J_s \sum_{G,H'}(i\tmatrix^{(0)})_{GH'}^{CD'}
(C_u)_{AB'}^{GH'}  
\cr
(C_c){}_{AB'}^{CD'}&=&(i)^2J_s \sum_{G,H'}(C_s){}_{GH'}^{CD'} (i\tmatrix^{(0)})_{AB'}^{GH'}
\cr
(C_d){}_{AB'}^{CD'}&=&(i)^2J_u \sum_{G,H'}   
(-)^{([B]+[H])([D]+[H])} (i\tmatrix^{(0)})_{GB'}^{CH'} (C_u){}_{AH'}^{GD'}
\cr
(C_e){}_{AB'}^{CD'}&=&(i)^2J_u\sum_{G,H'} (-)^{([B]+[H])([D]+[H])} (i\tmatrix^{(0)}){}_{GB'}^{CH'} 
(C_s){}_{A H'}^{G D'}
\cr
(C_f){}_{AB'}^{CD'}&=&(i)^2J_u\sum_{G,H'}   (-)^{([B]+[H])([D]+[H])}(C_s){}_{GB'}^{CH'}
(i\tmatrix^{(0)}){}_{AH'}^{GD'} \ .
\label{general_C2loop}
\ee


\begin{figure}[t]
\begin{center}
\includegraphics[height=30mm]{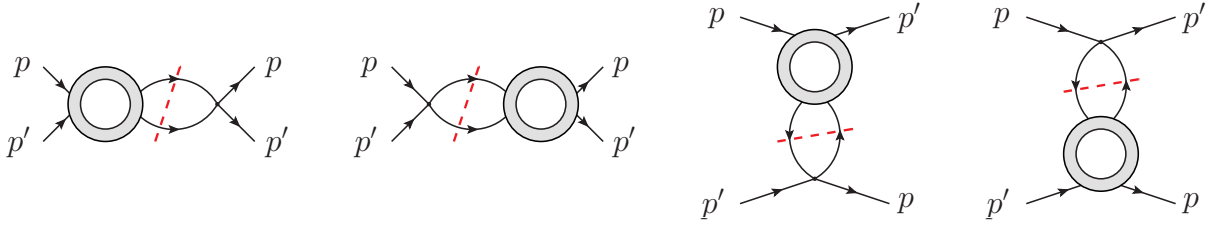}
\caption{The single two-particle cuts of two-loop four-point amplitudes.
They are used to determine the subleading logarithms not  
captured by maximal cuts. \label{2loop_single2p_cut}}
\end{center}
\end{figure}

With these coefficients the three-particle cuts of  two-loop amplitudes are also reproduced.

The two remaining coefficients, $C_{s,\text{extra}}$ and $C_{u,\text{extra}}$, are found by comparing the 
single two-particle cuts of the ansatz with the single two-particle cut of the two-loop amplitude 
(see fig.~\ref{2loop_single2p_cut}). 
The former are given in terms of complete one-loop amplitudes $i\tmatrix^{(1)}$ 
\be
i\tmatrix^{(2)}{}_{AB'}^{CD'}\Big|_{s-\text{cut}}\!\!\!\!\!\!&=&\!\!\!
(i)^2J_s\sum_{G,H'}
\left((i\tmatrix^{(0)})_{GH'}^{CD'}(i\tmatrix^{(1)})_{AB'}^{GH'}+(i\tmatrix^{(1)})_{GH'}^{CD'}(i\tmatrix^{(0)})_{AB'}^{GH'}\right)
\label{sdirect}
\\
i\tmatrix^{(2)}{}_{AB'}^{CD'}\Big|_{s-\text{cut}}\!\!\!\!\!\!&=&\!\!\!
(i)^2J_u\sum_{G,H'} 
(-)^{([B]+[H])([D]+[H])} \left( (i\tmatrix^{(0)}){}_{GB'}^{CH'} 
(i\tmatrix^{(1)}){}_{A H'}^{G D'} + (i\tmatrix^{(1)}){}_{GB'}^{CH'} 
(i\tmatrix^{(0)}){}_{A H'}^{G D'}  \right)~~~~~~
\label{udirect}
\ee
while the latter are given by
\be
i\tmatrix^{(2)}\big|^{\text{eq.}\;\ref{twoloop_tmatrix_improved}}_{s-\text{cut}} &=& 
   \frac{C_a}{J^2_s} \, ((J_s {\tilde I}_s+1)-1) + \frac{1}{2}\left(\frac{C_b}{J_s J_u}
   +\frac{C_c}{J_uJ_s}\right) J_u {\tilde I}_u
   \cr
   &&+\frac{1}{2}\left(\frac{C_b}{J_s}+\frac{C_c}{J_s}\right){\tilde I}_t
   +\frac{C_{s,\text{extra}}}{J_s} \ ,
\label{sansatz}
\\
i\tmatrix^{(2)}\big|^{\text{eq.}\;\ref{twoloop_tmatrix_improved}}_{u-\text{cut}} &=& 
\frac{C_d}{J^2_u}\,J_u {\tilde I}_u + \frac{1}{2}\left(\frac{C_e}{J_s J_u}
+\frac{C_f}{J_uJ_s}\right) ((J_s {\tilde I}_s+1)-1)
\cr
&&
+\frac{1}{2}\left(\frac{C_e}{J_s}+\frac{C_f}{J_s}\right){\tilde I}_t
+\frac{C_{u,\text{extra}}}{J_u} \ .
\label{uansatz}
\ee 
Using eq.~\eqref{general_C2loop} it is not difficult to see that the coefficients of $(J_s {\tilde I}_s+1)$ and $J_u {\tilde I}_u$ 
are the same in eqs.~\eqref{sdirect} and \eqref{sansatz} as well as in eqs.~\eqref{udirect} and \eqref{uansatz}.
This is a consequence of identities between the one-loop and the two-loop maximal cuts stemming from the fact 
that both of them are given as sums of products of tree-level amplitudes; it is also a manifestation of the consistency
of the generalized unitarity method.

We can therefore immediately read off the remaining coefficients; from eqs.~\eqref{sdirect} and \eqref{udirect}
only the terms in which $\tmatrix^{(1)}$ is replaced with its rational part ${\tilde\tmatrix}^{(1)}$ contribute:
\be
\label{Cextra}
\frac{1}{J_s}(C_{s,\text{extra}}){}_{AB'}^{CD'}&=&(i)^2J_s\sum_{G,H'}
\left((i\tmatrix^{(0)})_{GH'}^{CD'}(i{\tilde \tmatrix}^{(1)})_{AB'}^{GH'}+(i{\tilde \tmatrix}^{(1)})_{GH'}^{CD'}(i\tmatrix^{(0)})_{AB'}^{GH'}\right)
\cr
&&+\frac{(C_a){}_{AB'}^{CD'}}{J^2_s} - \frac{1}{2}\left(\frac{(C_e){}_{AB'}^{CD'}}{J_s}+\frac{(C_f){}_{AB'}^{CD'}}{J_s}\right)I_t \ ,
\\
\frac{1}{J_u}(C_{u,\text{extra}}){}_{AB'}^{CD'}&=&(i)^2J_u\sum_{G,H'} 
(-)^{([B]+[H])([D]+[H])} \left( (i\tmatrix^{(0)}){}_{GB'}^{CH'} 
(i{\tilde \tmatrix}^{(1)}){}_{A H'}^{G D'} + (i{\tilde \tmatrix}^{(1)}){}_{GB'}^{CH'} 
(i\tmatrix^{(0)}){}_{A H'}^{G D'}  \right)
\cr
&&+\frac{1}{2}\left(\frac{(C_e){}_{AB'}^{CD'}}{J_s J_u}+\frac{(C_f){}_{AB'}^{CD'}}{J_uJ_s}\right) 
-\frac{1}{2}\left(\frac{(C_e){}_{AB'}^{CD'}}{J_s}+\frac{(C_f){}_{AB'}^{CD'}}{J_s}\right)I_t \ .
\ee
The structure of one-loop integrals implies again that any logarithmic dependence 
on momenta that is not fixed by two-loop maximal cuts depends only on the difference 
\be
\frac{C_{s,\text{extra}}}{J_s}-\frac{C_{u,\text{extra}}}{J_u} \ .
\ee
One can check  that  all terms proportional to the identity matrix in ${\tilde\tmatrix}^{(1)}$, 
${\tilde\tmatrix}^{(1)}{}_{AB'}^{CD'}\propto\delta_A^C\delta_{B'}^{D'}$, 
cancel out in this difference and thus all two-loop logarithmic terms can be determined unambiguously. If desired 
({\it e.g.} for a three-loop calculation),
the two-loop rational terms $i{\hat\tmatrix}^{(2)}$ can be supplied separately, by expanding the symmetry-determined 
part of the S matrix.

Using the explicit expressions for the integrals in Appendix~\ref{one_n_twoloop_ints} one can see that in 
theories in which all worldsheet masses are equal the coefficient of the double logarithm
is given by the combination
\be
{\cal C}_{\text{ln}^2} = \frac{1}{8\pi^2J^2}\big(-2C_a+C_b+C_c-2C_d+C_e+C_f  \big)\ ,
\label{coef_2log}
\ee
while the coefficient of the simple logarithm $\ln \frac{p'_{-}}{p_{-}}$ is given by
\be
{\cal C}_{\text{ln}^1}
&=&\frac{i}{2\pi}\Big[\frac{1}{2 J^2}(2C_a -C_b -C_c)
\nonumber\\[1pt]
&&
-\frac{1}{ J}(C_{s,\text{extra}}-C_{u,\text{extra}})
     -\frac{i}{8\pi J}(C_b +C_c-C_e -C_f)\Big]  \ .
\label{simple_logs}
\ee
Here we used the fact that $J_s=J_u$.

In the following sections we shall compute the one-loop integral coefficients for \adss, \adscp~ 
and \adsst~ with NSNS and RR fluxes. We shall find that the difference
\be
\frac{C_s}{J_s}-\frac{C_u}{J_u}
\ee
is proportional to the identity matrix in each of the different sectors of the S matrix as labeled by the third
entry of the multi-index ({\it e.g.} $LL$ and $LR$ scattering, see footnote~\ref{footnoteLnR}); 
the proportionality coefficient can be identified with the one-loop dressing phase in each 
sector, cf. eq.~\eqref{SmatExpansion}. 
We shall compute the two-loop integral coefficients for \adss, \adscp~ 
and \adsst~ with RR flux and find that 
\be
{\cal C}_{\text{ln}^2} &=& 0
\cr
{\cal C}_{\text{ln}^1} &=&\frac{i}{4\pi^2 J^2} (2C_a-C_b-C_c) \propto -\frac{1}{2} \tmatrix^{(0)}  \ . 
\ee
The proportionality coefficient in each sector is given by the coefficient of the identity matrix in $C_s/J_s-C_u/J_u$.
This demonstrates the exponentiation of the one-loop dressing phase in each sector and thus provides 
support for two-loop integrability in all sectors.

\section{The S matrix for strings in AdS$_5\times$ S$^5$ \label{AdS5}}

In this section we will use generalized unitarity and the special properties of two-dimensional integrable quantum 
field theories discussed in sec.~\ref{worldsheetPT} to recover the known S matrix for string theory in \adss. In this 
case $G_1=G_2=PSU(2|2)$. The worldsheet theory contains eight bosons and eight fermions of equal mass which 
we shall normalize to $m=1$. Therefore, the multi-indices relabeling the S-matrix have a single entry -- the field label, which 
is just the fundamental representation of $PSU(2|2)$. Each of them is represented by a pair of a 
two-component bosonic index and a two-component fermionic index, such as $(a,\alpha)$ {\it etc.}, each acted upon by 
an $SU(2)\subset PSU(2|2)$.
The symmetry-determined part of the S matrix was found in \cite{BeSmatrix} and the 
dressing phase conjectured in \cite{BeSt} was tested through two loops in the $SL(2)$ sector. 
The $\tmatrix$ matrix is parametrized as
\begin{align}
\tmatrix_{\lAA\lBB}^{\lCC\lDD} & = \Atmatrix \,\delta_\lAA^\lCC \delta_\lBB^\lDD + \Btmatrix \,\delta_\lAA^\lDD \delta_\lBB^\lCC \; , &
\tmatrix_{\lAA\lBB}^{\lcc\ldd} & = \Ctmatrix \,\levi_{\lAA\lBB} \levi^{\lcc\ldd} \; , \nn \\
\tmatrix_{\laa\lbb}^{\lcc\ldd} & = \Dtmatrix \,\delta_\laa^\lcc \delta_\lbb^\ldd + \Etmatrix \,\delta_\laa^\ldd \delta_\lbb^\lcc \; , &
\tmatrix_{\laa\lbb}^{\lCC\lDD} & = \Ftmatrix \,\levi_{\laa\lbb} \levi^{\lCC\lDD} \; , \label{eqn:tmatrix-coeff} \\
\tmatrix_{\lAA\lbb}^{\lCC\ldd} & = \Gtmatrix \,\delta_\lAA^\lCC \delta_\lbb^\ldd \; , &
\tmatrix_{\laa\lBB}^{\lcc\lDD} & = \Ltmatrix \,\delta_\laa^\lcc \delta_\lBB^\lDD \; , \nn \\
\tmatrix_{\lAA\lbb}^{\lcc\lDD} & = \Htmatrix \,\delta_\lAA^\lDD \delta_\lbb^\lcc \; , &
\tmatrix_{\laa\lBB}^{\lCC\ldd} & = \Ktmatrix \,\delta_\laa^\ldd \delta_\lBB^\lCC \; . \nn
\end{align}
Each of the coefficients $\Atmatrix \dots \Ktmatrix$ has an inverse-$\hg$ expansion, {\it e.g.}
\be
\Atmatrix=\frac{1}{\hg}\Atmatrix^{(0)}+\frac{1}{\hg^2} \Atmatrix^{(1)}+\dots \ ,
\label{egExpansionA}
\ee
and similarly for all other coefficients. Here $\Atmatrix^{(0)}$, {\it etc.} are tree-level S-matrix elements, {\it i.e.} the 
entries of $\tmatrix^{(0)}$ introduced in eq.~\eqref{SmatExpansion}. 

%
The action has standard Lorentz-invariant quadratic terms  but interactions break this symmetry.
The tree-level S-matrix elements were found in \cite{KMRZ}:
\begin{align}
\Atmatrix^{(0)}(p,p') & = \Quarter \biggsbrk{ (1-2a)\lrbrk{\cpp} + \frac{\lrbrk{p-p'}^2}{\cpp} } \; , \nn \\
\Btmatrix^{(0)}(p,p') & = -\Etmatrix^{(0)}(p,p')= \frac{pp'}{\cpp} \; , \nn \\
\Ctmatrix^{(0)}(p,p') & = \Ftmatrix^{(0)}(p,p') = \Half \frac{\sqrt{\left(\energy+1\right)\left(\energy'+1\right)}
\lrbrk{\cpp+p'-p}}{\cpp} \; , \label{ourfinalequation} \\
\Dtmatrix^{(0)}(p,p') & = \Quarter \biggsbrk{ (1-2a)\lrbrk{\cpp} - \frac{\lrbrk{p-p'}^2}{\cpp} } \; , \nn \\
\Gtmatrix^{(0)}(p,p') & = -\Ltmatrix^{(0)}(p',p) = \Quarter \biggsbrk{ (1-2a)\lrbrk{\cpp} - \frac{p^2-p'^2}{\cpp} } \; , \nn \\
\Htmatrix^{(0)}(p,p') & = \Ktmatrix^{(0)}(p,p') = \Half \, \frac{pp'}{\cpp} \, \frac{\left(\energy+1\right)\left(\energy'+1\right)-pp'}{\sqrt{\left(\energy +1\right)\left(\energy '+1\right)}} \; . \nn
\end{align}
Here $\energy = \sqrt{1+p^2}$ denotes the relativistic energy.  The parameter $a$ reflects the dependence of 
the S matrix on the choice of physical states selected by the gauge-fixing of two-dimensional diffeomorphism invariance. 
We shall see that  it does not affect the logarithmic part of the dressing phase.

As noted in \cite{KMRZ}, the tree-level S matrix determined by the coefficients (\ref{ourfinalequation}) differs 
from the one obtained by expanding the one in \cite{BeSmatrix} by terms linear in the particle's momenta. These 
terms may be accounted for by a suitable rephasing of the S matrix in \cite{BeSmatrix}  (included for convenience in 
Appendix~\ref{sec:results}) as 
\be
\begin{array}{lcl}
{\hat \Asmatrix}^{B} = \Asmatrix^B e^{i(1-2a)(p-p')}
&~~~~~~&
{\hat \Bsmatrix}^{B} = \Bsmatrix^B e^{i(1-2a)(p-p')}
\cr
{\hat \Csmatrix}^{B} = \Csmatrix^B e^{i((\frac{5}{4}+b-2a)p-(\frac{1}{4}-b-2a)p')}
&&
{\hat \Dsmatrix}^{B} = \Dsmatrix^B e^{i((\frac{1}{2}-2a)p-(\frac{1}{2}-2a)p')}
\cr
{\hat \Esmatrix}^{B} = \Esmatrix^B e^{i((\frac{1}{2}-2a)p-(\frac{1}{2}-2a)p')}
&&
{\hat \Fsmatrix}^{B} = \Fsmatrix^B e^{i((\frac{1}{4}-b-2a)p-(\frac{5}{4}+b-2a)p')}
\cr
{\hat \Gsmatrix}^{B} = \Gsmatrix^B e^{i(-\frac{1}{2}p+(1-2a)(p-p'))}
&&
{\hat \Hsmatrix}^{B} = \Hsmatrix^B  e^{i((\frac{3}{4}+b-2a)p-(\frac{3}{4}+b-2a)p')}
\cr
{\hat \Ksmatrix}^{B} = \Ksmatrix^B e^{i((\frac{3}{4}-b-2a)p-(\frac{3}{4}-b-2a)p')}
&&
{\hat \Lsmatrix}^{B} = \Lsmatrix^B e^{i(\frac{1}{2}p'+(1-2a)(p-p'))} \ .
\end{array}
\ee
One may check that the $\smatrix$-matrix with these coefficients obeys the graded untwisted Yang-Baxter equation;
it is therefore a particular case of the $\smatrix$-matrix constructed in  \cite{Arutyunov:2006yd}.
The phases added to $\Asmatrix^B,\Bsmatrix^B, \Dsmatrix^B,\Esmatrix^B,\Gsmatrix^B, \Lsmatrix^B$ eliminate
the terms linear in momenta that are different between the spin chain and the world sheet calculations; the other 
phases, depending on the free parameter $b$ may be adjusted (or eliminated) by a rephasing of external states.

\subsection{The logarithmic terms of the one-loop AdS$_5\times$S$^5$ S matrix}

The one-loop amplitudes have the general form \eqref{ansatz_1loop} in which all masses are taken to be the same, {\it i.e.}
${\tilde I}_{s,u} \mapsto I_{s,u}$. Using the tree-level amplitudes \eqref{ourfinalequation} it is not difficult to find all components
of the $C_s$ and $C_u$ coefficients. For example,
\be
\frac{1}{J_s}(C_s){}_{ab}^{cd}&=&
(\Atmatrix^{(0)}{}^2+\Btmatrix^{(0)}{}^2+2\Ctmatrix^{(0)}{}\Ftmatrix^{(0)}{})\delta_a^c\delta_b^d
+2(\Atmatrix^{(0)}{}\Btmatrix^{(0)}{}-\Ctmatrix^{(0)}{}\Ftmatrix^{(0)}{})\delta_a^d\delta_b^c
\\
\frac{1}{J_u}(C_u){}_{ab}^{cd}&=&
\Atmatrix^{(0)}{}^2\delta_a^c\delta_b^d   
+2(\Atmatrix^{(0)}{} \Btmatrix^{(0)}{} + \Btmatrix^{(0)}{}^2 - \Htmatrix^{(0)}{} \Ktmatrix^{(0)}{})\delta_a^d\delta_b^c 
 \ .
\ee

As mentioned in the previous section, the one-loop bubble integrals are such that the difference $C_s/J_s-C_u/J_u$
governs the logarithmic dependence on external momenta.  While the complete expressions for $C_s$ and $C_u$ are not immediately transparent, their difference is simple~-- 
\be
\frac{C_s}{J_s}-\frac{C_u}{J_u} = 
+\frac{p^2p'{}^2(\varepsilon\varepsilon'-pp'{})}{(\varepsilon p'{}-\varepsilon'p)^2} \,\unit \ ,
\label{1loopcoefdiff}
\ee
{\it i.e.} it is proportional to the identity operator in field space; it is also independent of the gauge-choice 
parameter $a$, as expected. 
Using the values of the one-loop bubble integrals it follows that the one-loop worldsheet S matrix in \adss~ is
\be
i\tmatrix^{(1)} = 
i\;\left(
\frac{1}{2}\left(-\frac{1}{\pi}\frac{p^2p'{}^2(\varepsilon\varepsilon'-pp'{})}{(\varepsilon p'{}-\varepsilon'p)^2} 
\,\ln\Big|\frac{p'_{-}}{p_{-}}\Big| \right)\unit + \text{rational} \right)=i\,\left(\frac{1}{2}{\hat\theta}_{12}^{(1)}\,\unit+\text{rational}\right)\ .
\label{final_ads5_1loop}
\ee
We thus recover the general form of the one-loop S matrix \eqref{SmatExpansion} with all the logarithmic 
terms\footnote{We note here that, in line with the fact that the integrals we used are 
Lorentz invariant,  the argument of the logarithm is Lorentz-invariant. The coefficient of the logarithm 
is not, however, and can also be written in terms of Lorentz invariants and the constant time-like vector $n$ related to the 
choice of vacuum state: 
${\hat\theta}_{12}=-1/\pi\,({\bf p}\cdot {\bf p}'-n\cdot{\bf p}\,n\cdot {\bf p}')^2({\bf p}\cdot {\bf p}')/({\bf p}\times {\bf p}')^2
\ln|p_-'/p_-|$. Such a rewriting is possible for all the other models we discuss in later sections.
}
given by the one-loop dressing phase in \adss, ${\hat\theta}_{12}^{(1)}$ in eq.~\eqref{1loopphase}.
As mentioned in sec.~\ref{GeneralExpressions}, we can rewrite the one-loop S matrix and separate the rational part 
while keeping an integral representation for the logarithmic dependence:
\be
i\tmatrix^{(1)}=\frac{1}{2} \frac{C_s}{J_s}\,(J_s I_s+1)+\frac{1}{2}\frac{C_u}{J_u}\, J_u I_u +i{\tilde\tmatrix}^{(1)} \ ,
\ee
where the rational part\footnote{We argued in footnote~\ref{notadpoles} that finiteness of rational terms in an integrable theory
relies on the absence of tadpole integrals on external lines (sometimes known as snail graphs). One can check that the 
one-loop integral identified by the one-particle cut constructed 
from the four-point amplitude vanishes upon integration; this is in line with the expected absence of one-loop 
corrections to the dispersion relation of worldsheet fields in \adss. Thus, while we do not determine ${\tilde \tmatrix}^{(1)}$ 
we nevertheless see that it is finite.} of the one-loop S matrix, ${\tilde\tmatrix}^{(1)}$, was defined in eq.~\eqref{tildeT}.\footnote{We 
note that, since the difference of $C_s$ and $C_u$ is independent of the gauge-choice parameter $a$, only the last term, 
${\tilde \tmatrix}^{(1)}$, can depend on it.}
This decomposition will be useful in the next section where we discuss the construction of the logarithmic terms in 
the two-loop S matrix.

\subsection{The logarithmic terms of the two-loop AdS$_5\times$S$^5$ S matrix}

The general form of the two-loop amplitudes 
in \adss~is given by \eqref{twoloop_tmatrix_improved} with all fields having the same mass; the coefficients 
are given by eqs.~\eqref{general_C2loop} and \eqref{Cextra}.
We illustrate these formulae by writing out explicitly the components contributing to $\Atmatrix^{(2)}$ and  $\Btmatrix^{(2)}$;
they can also be rederived easily by iterating the general expressions for two-particle cuts in 
Appendix~\ref{2pc_general_L}:
\be
\frac{1}{J_s^2}(C_a){}_{ab}^{cd}&=&
(\Atmatrix^{(1)}_{s-\text{cut}}\Atmatrix^{(0)}+\Btmatrix^{(1)}_{s-\text{cut}}\Btmatrix^{(0)}
+2\Ctmatrix^{(1)}_{s-\text{cut}}\Ftmatrix^{(0)})\delta_a^c\delta_b^d
\cr
&&\quad   +(\Atmatrix^{(1)}_{s-\text{cut}}\Btmatrix^{(0)}+\Btmatrix^{(1)}_{s-\text{cut}}\Atmatrix^{(0)}
-2\Ctmatrix^{(1)}_{s-\text{cut}}\Ftmatrix^{(0)})\delta_a^d\delta_b^c \ ,
\cr
&=&
(\Atmatrix^{(0)}\Atmatrix^{(1)}_{s-\text{cut}}+\Btmatrix^{(0)}\Btmatrix^{(1)}_{s-\text{cut}}
+2\Ctmatrix^{(0)}\Ftmatrix^{(1)}_{s-\text{cut}})\delta_a^c\delta_b^d
\cr
&&\quad  +(\Atmatrix^{(0)}\Btmatrix^{(1)}_{s-\text{cut}}+\Btmatrix^{(0)}\Atmatrix^{(1)}_{s-\text{cut}}
-2\Ctmatrix^{(0)}\Ftmatrix^{(1)}_{s-\text{cut}})\delta_a^d\delta_b^c \ ,
\label{Ca}
\\
\frac{1}{J_u^2}(C_d){}_{ab}^{cd}&=&
\Atmatrix^{(1)}_{u-\text{cut}}\Atmatrix^{(0)}\delta_a^c\delta_b^d   
\cr
&&
+(\Atmatrix^{(1)}_{u-\text{cut}} \Btmatrix^{0}+\Btmatrix^{(1)}_{u-\text{cut}}\Atmatrix^{(0)} 
  + 2\Btmatrix^{(1)}_{u-\text{cut}}\Btmatrix^{(0)} - 2\Htmatrix^{(1)}_{u-\text{cut}} \Ktmatrix^{(0)})\delta_a^d\delta_b^c
\cr
&=&
\Atmatrix^{(0)}\Atmatrix^{(1)}_{u-\text{cut}}\delta_a^c\delta_b^d   
\label{Cd}
\\
&&
+(\Atmatrix^{(0)} \Btmatrix^{(1)}_{u-\text{cut}}+\Btmatrix^{(0)}\Atmatrix^{(1)}_{u-\text{cut}} 
  + 2\Btmatrix^{(0)}\Btmatrix^{(1)}_{u-\text{cut}} - 2\Htmatrix^{(0)} \Ktmatrix^{(1)}_{u-\text{cut}})\delta_a^d\delta_b^c \ .
\nonumber \ ,
\\
\frac{1}{J_sJ_u}(C_b){}_{ab}^{cd}&=&
(\Atmatrix^{(1)}_{u-\text{cut}}\Atmatrix^{(0)}+\Btmatrix^{(1)}_{u-\text{cut}}\Btmatrix^{(0)}
+2\Ctmatrix^{(1)}_{u-\text{cut}}\Ftmatrix^{(0)})\delta_a^c\delta_b^d
\\
&&\quad   +(\Atmatrix^{(1)}_{u-\text{cut}}\Btmatrix^{(0)}+\Btmatrix^{(1)}_{u-\text{cut}}\Atmatrix^{(0)}
-2\Ctmatrix^{(1)}_{u-\text{cut}}\Ftmatrix^{(0)})\delta_a^d\delta_b^c \ ,
\cr
\frac{1}{J_sJ_u}(C_e){}_{ab}^{cd}&=&
\Atmatrix^{(1)}_{s-\text{cut}}\Atmatrix^{(0)}\delta_a^c\delta_b^d   
\\
&&
\quad 
+(\Atmatrix^{(1)}_{s-\text{cut}} \Btmatrix^{(0)}+\Btmatrix^{(1)}_{s-\text{cut}}\Atmatrix^{(0)} 
  + 2\Btmatrix^{(1)}_{s-\text{cut}}\Btmatrix^{(0)} - 2\Htmatrix^{(1)}_{s-\text{cut}} \Ktmatrix^{(0)})\delta_a^d\delta_b^c \ ,
 \nonumber
\\
\frac{1}{J_sJ_u}(C_c){}_{ab}^{cd}&=&
(\Atmatrix^{(0)}\Atmatrix^{(1)}_{u-\text{cut}}+\Btmatrix^{(0)}\Btmatrix^{(1)}_{u-\text{cut}}
+2\Ctmatrix^{(0)}\Ftmatrix^{(1)}_{u-\text{cut}})\delta_a^c\delta_b^d
\\
&&\quad   +(\Atmatrix^{(0)}\Btmatrix^{(1)}_{u-\text{cut}}+\Btmatrix^{(0)}\Atmatrix^{(1)}_{u-\text{cut}}
-2\Ctmatrix^{(0)}\Ftmatrix^{(1)}_{u-\text{cut}})\delta_a^d\delta_b^c \ ,
\cr
\frac{1}{J_sJ_u}(C_f){}_{ab}^{cd}&=&
\Atmatrix^{(0)}\Atmatrix^{(1)}_{s-\text{cut}}\delta_a^c\delta_b^d   
\label{Cc}
\\
&&
\quad 
+(\Atmatrix^{(0)} \Btmatrix^{(1)}_{s-\text{cut}}+\Btmatrix^{(0)}\Atmatrix^{(1)}_{s-\text{cut}} 
  + 2\Btmatrix^{(0)}\Btmatrix^{(1)}_{s-\text{cut}} - 2\Htmatrix^{(0)} \Ktmatrix^{(1)}_{s-\text{cut}})\delta_a^d\delta_b^c \ .
\nonumber
\ee
The cuts of the one-loop amplitudes are given in terms of $C_s$ and $C_u$ coefficients, e.g.
\be
i\Atmatrix^{(1)}_{s-\text{cut}} \delta_a^c\delta_b^d= \frac{1}{J_s}(C_s){}_{ab}^{cd}
\qquad
i\Atmatrix^{(1)}_{u-\text{cut}}\delta_a^c\delta_b^d = \frac{1}{J_u} (C_u){}_{ab}^{cd} \ .
\ee

Using the explicit expression for the tree-level S-matrix elements it is  not difficult to check that
\be
-2C_a+C_b+C_c-2C_d+C_e+C_f=0 \ ,
\ee 
as mentioned at the end of sec.~\ref{GeneralExpressions}. This implies  
the cancellation of all the double logarithms.\footnote{Such cancellations 
between terms captured by different cuts is characteristic to un-ordered scattering amplitudes and 
was previously observed in {\it e.g.} higher-dimensional supergravity theories.}
This is consistent with the structure of the two-loop terms in eq.~(\ref{SmatExpansion}), since the two-loop 
correction to the BHL/BES dressing phase, ${\hat\theta}^{(2)}$, does not contain any double logarithms (see 
Appendix~\ref{exp_phase}).

The two remaining coefficients, $C_{s,\text{extra}}$ and $C_{u,\text{extra}}$, are given by eq.~\eqref{Cextra} and are 
determined by comparing the single two-particle cuts of the ansatz with the single two-particle cut of the two-loop amplitude. 
Their contribution to $\Atmatrix^{(2)}$ is 
\be
\frac{C^\Atmatrix_{s,\text{extra}}}{J_s} 
  &=&
  2i(i)^2\big(\Atmatrix^{(0)}(i{\tilde\Atmatrix}^{(1)})
  +\Btmatrix^{(0)}(i{\tilde\Btmatrix}^{(1)})
  +\Ctmatrix^{(0)}(i{\tilde\Ftmatrix}^{(1)})+\Ftmatrix^{(0)}(i{\tilde\Ctmatrix}^{(1)})\big)  
  \cr
&&+\frac{C_a^A}{J_s^2}
 - \frac{1}{2}\left(\frac{C^\Atmatrix_b}{J_s}+\frac{C^\Atmatrix_c}{J_s}\right)I_t \ ,
  \\
\frac{C^\Atmatrix_{u,\text{extra}}}{J_u} &=&2(i)^2i \Atmatrix^{(0)}(i{\tilde \Atmatrix}^{(1)}) 
+\frac{1}{2}\left(\frac{C_e^\Atmatrix}{J_sJ_u}+\frac{C_f^\Atmatrix}{J_sJ_u}\right)
-\frac{1}{2}\left(\frac{C^\Atmatrix_e}{J_s}+\frac{C^\Atmatrix_f}{J_s}\right)I_t \ .
\ee
$C_x^Y$ denotes the contribution of $C_x$ to the coefficient $Y$ in eq.~\eqref{eqn:tmatrix-coeff} and 
${\tilde\Atmatrix}^{(1)}$, {\it etc.} are the entries of ${\tilde\tmatrix}^{(1)}$ defined in eq.~\eqref{tildeT}.

Taking their difference and reconstructing the combination on the second line of eq.~\eqref{simple_logs} we observe the 
cancellation  the diagonal term ${\tilde \Atmatrix}^{(1)}$ which potentially contains rational terms undetermined by symmetries.
After some amount of algebra which makes use of the explicit form of the rational terms\footnote{We notice here that the 
off-diagonal rational terms of the one-loop amplitudes are proportional to the tree-level S-matrix elements.} extracted form the exact S matrix
\be
{\hat\Btmatrix}^{(1)} &=&i(\Atmatrix^{(0)}+\Dtmatrix^{(0)})\Btmatrix^{(0)}+\frac{i}{8} aJ\Btmatrix^{(0)}
 \cr
{\hat\Ctmatrix}^{(1)} &=& \frac{i}{2}(\Atmatrix^{(0)}+\Dtmatrix^{(0)})\Ctmatrix^{(0)} 
+ ib(p+p'{})\Ctmatrix^{(0)} \cr
{\hat\Ftmatrix}^{(1)} &=& \frac{i}{2}(\Atmatrix^{(0)}+\Dtmatrix^{(0)})\Ftmatrix^{(0)}
-ib(p+ p'{})\Ftmatrix^{(0)} 
\ee
we find that the second line of eq.~\eqref{simple_logs} vanishes identically. Repeating the same steps for the coefficients 
of all the other tensor structures one can confirm that
\be
\frac{C_{s,\text{extra}}}{J_s}-\frac{C_{u,\text{extra}}}{J_u}
+ \frac{i}{8\pi J_s}\left( C_b + C_c-C_e - C_f\right) = 0 \ .
\ee
The combination of integral coefficients on first line of eq.~\eqref{simple_logs} contributing to $i\Atmatrix^{(2)}$ can be 
written as
\be
-4\pi i \,{\cal C}^{\Atmatrix}_{\ln^1}=\frac{1}{J^2}(2C_a^\Atmatrix -C^\Atmatrix_b -C^\Atmatrix_c)=-
2(i\Atmatrix^{(0)}) \left(\frac{C_s^\Atmatrix}{J_s} - \frac{C_u^\Atmatrix}{J_u}\right) \ ;
\label{identity_2loopcoefs}
\ee
Similar expressions hold for all the other entries. Combining them and using the value of the parenthesis 
above from eq.~\eqref{1loopcoefdiff} we find that the \adss~two-loop S matrix is 
\be
i\tmatrix^{(2)} = -\frac{1}{2}\left(-\frac{1}{\pi} \frac{p^2p'{}^2(\energy\energy'{}-pp'{})}{(\energy'{}p-\energy p'{})^2}
  \ln \frac{p'_{-}}{p_{-}} \right) {\tmatrix}^{(0)}+\text{rational}
= 
-\frac{1}{2}{\hat\theta}^{(1)}_{12} {\tmatrix}^{(0)}
+\text{rational} \ .
\label{final_ads5_2loop}
\ee
It reproduces the logarithmic terms in eq.~\eqref{SmatExpansion} and thus gives strong support to the exponentiation 
of the one-loop logarithms as in eq.~\ref{def_S_phase}. 

We have therefore demonstrated that the generalized unitarity method carried out in two dimensions, together 
with consequences of symmetries of the theory, can be used to efficiently determine all terms with logarithmic 
dependence on external momenta through at least two loop level. We shall now discuss other 
interesting worldsheet theories related to gauge/string dualities and in some cases find novel results regarding their scattering matrices.

\section{On the S matrix for strings in AdS$_4\times$ CP$^3$ \label{AdS4CP3}}

A worldsheet theory that bears close similarity to the \adss~string is that of type IIA string theory on \adscp. 
This is dual to the ABJM  theory \cite{ABJM} and there is evidence for its integrability. Quantized around 
a BMN-like null geodesics, its worldsheet spectrum 
consists of eight  bosons and eight fermions four of each being light ($m^2=1/4$) and four being heavy ($m^2=1$). 
It was argued from a worldsheet perspective \cite{Zarembo:2009au} that the  heavy excitations 
are unstable and decay into a pair of light excitations. The bosonic worldsheet S matrix was found in 
\cite{Kalousios:2009ey}.
In a spin-chain picture the heavy excitations are interpreted as composite and do not exist as 
asymptotic states; thus, the relevant S matrix scatters only the eight light excitations, organized in two representations
of $PSU(2|2)\times U(1)$, typically called the $A$ and the $B$ particles/multiplets. 
For our discussion we shall use the small momentum limit of this S-matrix. As discussed
in sec.~\ref{worldsheetPT}, the different masses of excitations and the properties of the S matrix guarantee 
that this truncation is perturbatively consistent through at least two loops.

In \cite{Gromov:2008qe} a proposal was made for the Bethe equations to all-loop orders. The dressing 
phase for  \adscp~  was suggested to be similar to the one found for \adss~ with the difference that the full 
$\Smatrix$-matrix contained the squared dressing phase  (with the two factors coming from the two factors in 
\eqref{ws_smatrix_decomposition_txt});  in the former case there is one dressing phase factor for each multiplet.

Based on these Bethe ansatz equations an S matrix was conjectured in \cite{Ahn:2008aa}. The proposed S matrix has four 
sectors corresponding to the four ways one can pick the multiplets of the incoming and outgoing particles and it is 
reflectionless ({\it i.e.} particles of different multiplets will not exchange momentum):
\begin{align}
S^{BB}(p,p')=S^{AA}(p,p')&=S_0(p,p')\widehat{S}(p,p'),\\
S^{AB}(p,p')=S^{BA}(p,p')&=\widetilde{S}_0(p,p')\widehat{S}(p,p') \ .
\end{align}
The factors $S_0(p,p')$ and $\widetilde{S}_0(p,p')$ are two (potentially different) dressing phases and 
$\widehat{S}(p,p')$ is the $SU(2|2)$ invariant S matrix found in \cite{Arutyunov:2006yd}.

The reflectionless property of the S matrix implies that different sectors do not mix in the  $s$- and $u$-channel 
cuts so it is sufficient to consider the cuts of a single $SU(2|2)$-invariant S matrix. Our computations in the previous 
section show that, at least through two loops, the dressing phase of such an S matrix would indeed have half of the 
logarithmic terms of the full BES/BHL phase, consistent with the expectation of \cite{Gromov:2008qe} thus providing 
a non-trivial check of that proposal.

An alternative S matrix was considered in \cite{Ahn:2008tv} and rejected as the resulting Bethe equations did not 
match two-loop gauge theory perturbative calculations \cite {Minahan:2008hf}. It is also based 
on the $SU(2|2)$ invariant S matrix, with the only important difference that it is no longer reflectionless.
We can also see from the perspective of generalized unitarity that such an S matrix is not consistent with worldsheet 
perturbation theory truncated to the fields with $m^2=1/4$.

To this end let us consider the scattering of two $A$-type scalars into two $A$-type fermions. Since this is an off-diagonal
matrix element and since (assuming integrability and $SU(2|2)$ symmetry) quantum corrections can yield logarithmic 
terms only in the dressing phase, eq.~\eqref{SmatExpansion} implies that at one-loop level this matrix element should have 
no logarithms. For a reflectionless S matrix this is realized by the $s$- and $u$-channel cuts being equal.
Allowing reflections in the S matrix changes the number of particles crossing the $u$-channel cut (from only $A$-type 
particles to both $A$- and $B$-type particles) while not affecting the $s$-channel cut.
Thus, a reflection-containing S matrix is not consistent with  worldsheet perturbation theory truncated to 
the light fields.

\section{The S matrix for strings in  AdS$_3\times$S$^3\times$S$^3\times$S$^1$ \label{adss}}

The Green-Schwarz string in \adssss~\cite{Pesando:1998wm} was shown in \cite{StefanskyZarembo} to be related to 
the ${\bf Z}_4$-graded supercoset  $D(2,1;\alpha) \otimes D(2,1;\alpha)/(SO(1,2)\times SO(3)\times SO(3))$ and an 
extra free boson, where  $\alpha$ is related to the radii of the various factors by
\be
\alpha=\frac{R^2_\text{AdS}}{R^2_{S^3_1}}=1-\frac{R^2_\text{AdS}}{R^2_{S^3_2}} \ .
\ee 
The BMN limit was studied in \cite{Rughoonauth:2012qd} and it was found that
the perturbative worldsheet spectrum consists of 

\smallskip

\centerline{
\begin{tabular}{lcl}
two bosons and two fermions of $m=1$
&~~&
two bosons and two fermions of $m=0$
\\
two bosons and two fermions of $m=\alpha$
&~~&
two bosons and two fermions of $m=1-\alpha\,$.
\end{tabular}
}

\noindent One massless mode corresponds to excitations on S$^1$ and the other to an excitation shared between 
the two three-spheres. 
Certain entries of the tree-level S matrix for particles with $\alpha$-dependent masses  were found in \cite{Sundin:2013ypa}.

The symmetry group of the light-cone gauge-fixed worldsheet theory is $PSU(1|1)^2$. As discussed in \cite{Borsato:2012ud}, 
the states with $\alpha$-dependent masses  may be written as a tensor product of left excitations ($L$), $|\phi\rangle$ 
and $|\psi\rangle$, and the conjugate right excitations ($R$) , $|\bar{\phi}\rangle$ and $|\bar{\psi}\rangle$; they 
transform (in conjugate representations) under one or the other factors of the symmetry group.\footnote{
The $L$ and $R$ excitations have been denoted by $+$ and $-$ in related a context  \cite{Hoare:2013pma}. 
We mention here again that these excitations are not directly related to left- and right- worldsheet motion. 
} 
The two representations 
are no longer decoupled in the centrally-extended $PSU(1|1)^2$ and, while  $PSU(1|1)^2$-invariant, the S matrix 
describes nontrivial $LL$, $RR$, $LR$ and $RL$ scattering. 
This sector of massive states was discussed from a spin-chain perspective in 
\cite{Borsato:2012ud,  Borsato:2012ss} and independently in \cite{Ahn:2012hw} where the 
symmetry-determined parts of the S-matrices were proposed. 
The direct tree-level calculation of \cite{Sundin:2013ypa} favors the S matrix proposed in  
\cite{Borsato:2012ud,  Borsato:2012ss}.
The one-loop correction to the dressing phase was  found indirectly in \cite{Beccaria:2012kb}, by comparing 
one-loop corrections to energies of semiclassical states to the energy predictions of the Asymptotic Bethe Ansatz based on 
the exact S matrix.

As described in \cite{Borsato:2012ss}, the proposed S-matrices scatter only states with 
normalized masses $m=\alpha$ and $m=(1-\alpha)$. 
One of their characteristic feature is that individual masses are conserved in a scattering process.
Let us consider some generic scattering process:
\begin{align}
|\mathcal{X}^{(in)}_{p_{in}}\mathcal{X}^{(in)}_{p'_{in}}\rangle&\to|\mathcal{X}^{(out)}_{p_{out}}\mathcal{X}^{(out)}_{p'_{out}}\rangle,
\end{align}
\noindent where we also assume that 
\footnote{
This condition is equivalent to assuming a definite sign for the Jacobian $J=\varepsilon'p-p'\varepsilon$.
} 
$\frac{p_{in}}{m_{in}}>\frac{p'_{in}}{m'_{in}}$, $\frac{p_{out}}{m_{out}}>\frac{p'_{out}}{m'_{out}}$, 
where $m$ denote the normalized masses of particles. 
When all the masses are the same momentum and energy conservation leads to the solution $p_{in}=p_{out}$, 
$p'_{in}=p'_{out}$ plus another solution that does not satisfy the assumed ordering.

When masses are not the same one can still have a similar solution to the conservation equations, 
$p_{in}=p_{out}$, $p'_{in}=p'_{out}$, $m_{in}=m_{out}\neq m'_{in}=m'_{out}$ but there is also another 
solution where the outgoing momenta are not equal to the incoming momenta. The proposed 
S-matrices  are reflectionless, {\it i.e.} they forbid this second possibility. 
The scattering of states in different multiplets of the $PSU(1|1)^2$  is also reflectionless
even if they have the same mass.

The proposed S matrices to not describe the scattering of states with $m=0$ and $m=1$. The calculation of 
\cite{Sundin:2013ypa} does not shed light on the properties of the relevant S-matrix elements as the relevant 
matrix elements involving $m=1$ states were evaluated only for $\alpha=1$ ({\it i.e.} they are S-matrix elements 
for strings on \adsst) and the $m=0$ states were not considered\footnote{A proposal for the inclusion of the 
massless degrees of freedom in the spin chain Asymptotic Bethe Ansatz for M$=$S$^3\times$S$^1$ and 
M$=$T$^4$ was discussed in \cite{Sax:2012jv}.}. In the following we shall assume that the scattering of the
$m=0$ and $m=1$ states off states with $m=\alpha$ and $m=(1-\alpha)$ is also reflectionless and thus that 
they do not contribute to the one- and two-loop logarithmic terms. This is justified {\it a posteriori}, as our results 
are consistent with those of~\cite{Beccaria:2012kb}.


\subsection{The tree-level S matrix }

Entries of the tree-level worldsheet S matrix were constructed in \cite{Sundin:2013ypa} for general $\alpha$. The result 
reproduces the small momentum limit of the full S-matrices of \cite{Borsato:2012ud}; an alternative S matrix 
was proposed in \cite{Ahn:2012hw}; both the S matrix of Borsato, Ohlsson Sax and Sfondrini (BOSS) \cite{Borsato:2012ud} 
and that of Ahn and Bombardelli (AB) \cite{Ahn:2012hw}  
are included in Appendix~\ref{ads3s3s3s1}. In a notation close to that used in \adss, we parametrize the $\tmatrix$ 
as follows:
\begin{align}
\mathrm{T}^{\phi\phi}_{\phi\phi}&=\Atmatrix_{LL},&\mathrm{T}_{\psi\psi}^{\psi\psi}&=\Dtmatrix_{LL},&
\mathrm{T}_{\phi\psi}^{\phi\psi}&=\Gtmatrix_{LL},\nonumber\\
\mathrm{T}_{\phi\psi}^{\psi\phi}&=\Htmatrix_{LL},&\mathrm{T}_{\psi\phi}^{\phi\psi}&=\Ktmatrix_{LL},&
\mathrm{T}_{\psi\phi}^{\psi\phi}&=\Ltmatrix_{LL},\nonumber\\
\mathrm{T}_{\phi\bar{\phi}}^{\phi\bar{\phi}}&=\Atmatrix_{LR},
&\mathrm{T}_{\phi\bar{\phi}}^{\psi\bar{\psi}}&=\Ctmatrix_{LR},
&\mathrm{T}_{\psi\bar{\psi}}^{\psi\bar{\psi}}&=\Dtmatrix_{LR},\label{T matrix}\\
\mathrm{T}_{\phi\bar{\psi}}^{\phi\bar{\psi}}&=\Gtmatrix_{LR},
&\mathrm{T}_{\psi\bar{\psi}}^{\phi\bar{\phi}}&=\Ftmatrix_{LR},
&\mathrm{T}_{\phi\bar{\psi}}^{\psi\bar{\phi}}&=\Htmatrix_{LR},\nonumber\\
\mathrm{T}_{\psi\bar{\phi}}^{\phi\bar{\psi}}&=\Ktmatrix_{LR},
&
\mathrm{T}_{\psi\bar{\phi}}^{\psi\bar{\phi}}&=\Ltmatrix_{LR} \ .\nonumber
\end{align}
The lower indices denote the incoming state and the upper indices denote the outgoing state.
Here we did not assign an index for the mass of the excitation; such an index can be added 
without difficulty. Apart from fields, the indices $L$ and $R$ should carry the same index related to the mass.
Each component has an expansion in the appropriate worldsheet coupling constant $\hg^{-1}$ 
similar to eq.~\ref{egExpansionA}, {\it e.g.}
\be
\Atmatrix_{LL}=\frac{1}{\hg}\Atmatrix^{(0)}_{LL}+\frac{1}{\hg^2} \Atmatrix^{(1)}_{LL}+\dots \ .
\ee
The  tree-level worldsheet S-matrix elements, $\Atmatrix^{(0)}_{LL}$ {\it etc.}, 
follow from \cite{Sundin:2013ypa} or can be extracted
from the small momentum expansion of the exact S matrix \cite{Borsato:2012ud} with additional minus signs for the scattering 
of two fermions. The latter approach also yields the symmetry-determined rational terms at higher loops.

We will not need the explicit form of the diagonal tree-level matrix elements, so we will not list them here. Their only property 
that is important for our calculation is that they are related by
\begin{align}
\label{identity_adssss}
\Atmatrix^{(0)} + \Dtmatrix^{(0)} - \Gtmatrix^{(0)} - \Ltmatrix^{(0)}&=0 \ ,
\end{align}
for any choice of masses and for any choice of $L$ and $R$ states.

With these preparations, the tree-level entries of the BOSS S-matrix that 
we will need are \cite{Sundin:2013ypa,Borsato:2012ud}:
\begin{align}
\Htmatrix^{(0)}{}^{\rm BOSS}_{LL}=\Ktmatrix^{(0)}{}^{\rm BOSS}_{LL}
&=\frac{1}{2}\frac{pp'}{\varepsilon'p-p'\varepsilon}\frac{(\varepsilon+m)(\varepsilon'+m')-pp'}{\sqrt{(\varepsilon+m)(\varepsilon'+m')}},\\
%
\Ctmatrix^{(0)}{}^{\rm BOSS}_{LR}=\Ftmatrix^{(0)}{}^{\rm BOSS}_{LR}&=\frac{1}{2}\frac{\sqrt{(\varepsilon+m)(\varepsilon'+m')}
(\varepsilon'p-\varepsilon p'+p' m-pm')}{\varepsilon'p-p'\varepsilon},\\
\Htmatrix^{(0)}{}^{\rm BOSS}_{LR}=\Ktmatrix^{(0)}{}^{\rm BOSS}_{LR}&=0 \ .
\label{BOSStree}
\end{align}
Here we combined all different choices of masses in a single expression; different mass sectors correspond to 
different choices of $m, m'=\alpha, 1-\alpha$; also, $\energy$ and $\energy'$ are the standard relativistic energies for
particles of masses $m$ and $m'$, respectively.

Similarly, expanding the non-diagonal entries of the AB S matrix and combining different mass sectors 
we find:
\begin{align}
\Htmatrix^{(0)}{}^{\rm AB}_{LL}=\Htmatrix^{(0)}{}^{\rm AB}_{LR}&=\frac{1}{2}\frac{pp'}{\varepsilon'p-p'\varepsilon}
\frac{(\varepsilon+m)(\varepsilon'+m')-pp'}{(\varepsilon+m)},\\
\Ctmatrix^{(0)}{}^{\rm AB}_{LR}=\Ftmatrix^{(0)}{}^{\rm AB}_{LR}&=0,\\
\Ktmatrix^{(0)}{}^{\rm AB}_{LL}=\Ktmatrix^{(0)}{}^{\rm AB}_{LR}&=\frac{1}{2}\frac{pp'}{\varepsilon'p-p'\varepsilon}
\frac{(\varepsilon+m)(\varepsilon'+m')-pp'}{(\varepsilon+m')} \ .
\label{ABtree}
\end{align}
Notice that although the functions $H_{LL}$ and $K_{LL}$ are different in eqs.~\eqref{BOSStree} 
and \eqref{ABtree}, their product remains the same. This observation will be useful shortly and leads to the equality 
of the one-loop corrections in the  $LL$/$RR$ sectors of the two S-matrices.

\subsection{The logarithmic terms of the one-loop  \adssss~$\smatrix$-matrix}

Let us now illustrate the calculation of the one-loop S matrix following the general discussion in sec.~\ref{GeneralExpressions}.
Each multi-index $A,B,C,\dots$ there stands for the triplet $(\text{field label, mass, sector})$, {\it e.g.} $(\phi, m, L)$; the grade of 
the index $[A]$ is the grade of the field label ($0/1$ for bosons/fermions). 
The general expression for one-loop four-point amplitudes is \eqref{ansatz_1loop} with integrals having different masses.
They obey the relations
\begin{align}
J_s\tilde{I}_s-J_u\tilde{I}_u&=-\frac{2i}{\pi}\left(\mathrm{ln}\left|\frac{p_-'}{p_-}
\right|-\mathrm{ln}\left|\frac{m'}{m}\right|\right)-1\ ,\\
\label{difference JI}
J_s\tilde{I}_s+J_u\tilde{I}_u+1&=0,
\end{align}
which implies that the logarithmic dependence of the S matrix is governed by the difference $C_s/J_s-C_u/J_u$.
It is instructive to consider separately the $LL$ and $LR$ sectors and find the one-loop correction in the various 
possible sectors. In the process we shall also expand the slightly cryptic multi-index form of the one-loop coefficients 
\eqref{generalC1loop}.

\subsubsection{The $LL$ and $RR$ sectors}

The tree-level amplitudes in the BOSS and AB S-matrices have the same vanishing entries; therefore the 
one-loop coefficients \eqref{generalC1loop} have formally similar expressions:
\be
\begin{array}{lcl}
\frac{1}{J_s}(C_{s,LL})\phantom{}^{\phi\phi}_{\phi\phi} =\Atmatrix^{(0)2}_{LL}
&~&
\frac{1}{J_u}(C_{u,LL})\phantom{}^{\phi\phi}_{\phi\phi} =\Atmatrix^{(0)2}_{LL}-\Htmatrix^{(0)}_{LL}\Ktmatrix^{(0)}_{LL}
\\[1pt]
\frac{1}{J_s}(C_{s,LL})\phantom{}^{\psi\psi}_{\psi\psi} =\Dtmatrix^{(0)2}_{LL}
&~&
\frac{1}{J_u}(C_{u,LL})\phantom{}^{\psi\psi}_{\psi\psi} =\Dtmatrix^{(0)2}_{LL}-\Htmatrix^{(0)}_{LL}\Ktmatrix^{(0)}_{LL}
\\[1pt]
\frac{1}{J_s}(C_{s,LL})\phantom{}^{\phi\psi}_{\phi\psi} =\Gtmatrix^{(0)2}_{LL}+\Htmatrix^{(0)}_{LL}\Ktmatrix^{(0)}_{LL}
&~&
\frac{1}{J_u}(C_{u,LL})\phantom{}^{\phi\psi}_{\phi\psi} =\Gtmatrix^{(0)2}_{LL}
\\[1pt]
\frac{1}{J_s}(C_{s,LL})\phantom{}^{\phi\psi}_{\psi\phi} =\Gtmatrix^{(0)}_{LL}\Htmatrix^{(0)}_{LL}
+\Htmatrix^{(0)}_{LL}\Ltmatrix^{(0)}_{LL}
&~&
\frac{1}{J_u}(C_{u,LL})\phantom{}^{\phi\psi}_{\psi\phi} =\Dtmatrix^{(0)}_{LL}\Htmatrix^{(0)}_{LL}
+\Htmatrix^{(0)}_{LL}\Atmatrix^{(0)}_{LL}
\\[1pt]
\frac{1}{J_s}(C_{s,LL})\phantom{}^{\psi\phi}_{\phi\psi} =\Ltmatrix^{(0)}_{LL}\Ktmatrix^{(0)}_{LL}
+\Ktmatrix^{(0)}_{LL}\Gtmatrix^{(0)}_{LL}
&~&
\frac{1}{J_u}(C_{u,LL})\phantom{}^{\psi\phi}_{\phi\psi} =\Atmatrix^{(0)}_{LL}\Ktmatrix^{(0)}_{LL}
+\Ktmatrix^{(0)}_{LL}\Dtmatrix^{(0)}_{LL}
\\[1pt]
\frac{1}{J_s}(C_{s,LL})\phantom{}^{\psi\phi}_{\psi\phi} =\Ltmatrix^{(0)2}_{LL}+\Htmatrix^{(0)}_{LL}\Ktmatrix^{(0)}_{LL}
&~&
\frac{1}{J_u}(C_{u,LL})\phantom{}^{\psi\phi}_{\psi\phi} =\Ltmatrix^{(0)2}_{LL}
\end{array}
\ee
It is easy to see that the difference $(C_s/J_s-C_u/J_u)$, governing the logarithmic dependence on external momenta,
depends only on the product $H_{LL}^{(0)}K_{LL}^{(0)}$ which is the same for both S-matrices \eqref{BOSStree} and
\eqref{ABtree}. We therefore find that
\be
\frac{C_{s,LL}^{\rm BOSS}}{J_s}-\frac{C_{u,LL}^{\rm BOSS}}{J_u} =
\frac{C_{s,LL}^{\rm AB}}{J_s}-\frac{C_{u,LL}^{\rm AB}}{J_u} =
 {}\frac{p^2(p')^2({\bf p}\cdot{\bf p'}+mm')}{2(\varepsilon'p-p'\varepsilon)^2}\,\unit \ ,
\ee
where $\unit$ is the identity matrix in field space. This in turn implies that the one-loop correction to the $LL$ sector of both
S-matrices is
\be
i\tmatrix^{(1)}_{LL}=i\, \left(\frac{1}{2}\left(
{}-\frac{1}{\pi}\frac{p^2(p')^2({\bf p}\cdot{\bf p'}+mm')}{2(\varepsilon'p-p'\varepsilon)^2}
\left(\mathrm{ln}\left|\frac{p_-'}{p_-}\right|-\mathrm{ln}\left|\frac{m'}{m}\right|\right)\right)\,\unit +\text{rational}\right) \ .
\cr
\label{resultat LL}
\ee
This expression reproduces  the small momentum limit of the $LL$ dressing phase of 
\cite{Beccaria:2012kb} for general values of $m$ and $m'$, see eq.~\eqref{fase1}. 
The additional factor of 1/2 is reminiscent 
of \adscp~ as the dressing phase to which we are comparing comes from a factorized S matrix.

The calculation in the $RR$ sector is completely equivalent for both S matrices. While consistent with the BOSS S matrix, 
the presence of logarithms appears to contradict the conjecture \cite{Ahn:2012hw} that in the sector with two 
different masses, $m\ne m'$, there is no dressing phase. This implies that the S matrix proposed 
there is not the S matrix of a quantum field theory.

\subsubsection{The $LR$ and $RL$ sectors}

Due to the structure of the tree-level AB S matrix, in particular the vanishing of $\Ctmatrix{}^{(0)}_{LR}$ and  
$\Ftmatrix{}^{(0)}_{LR}$
as well as the fact that $\Htmatrix^{(0)}_{LR}=\Htmatrix^{(0)}_{LL}$ and $\Ktmatrix^{(0)}_{LR}=\Ktmatrix^{(0)}_{LL}$, 
the calculation of the one-loop S matrix in the $LR/RL$ sectors is identical to the one in the $LL/RR$ sectors in the 
previous section and $i\tmatrix^{(1), AB}_{LR}$ is given by the right-hand side of eq.~\eqref{resultat LL}.

Using the vanishing entries of the $BOSS$ tree-level S matrix, the various components of the coefficients $C_s$ and $C_u$ 
describing its one-loop corrections are given by \eqref{generalC1loop}: 
\be
\begin{array}{lcl}
\frac{1}{J_s}(C^{\rm BOSS}_{s, LR})\phantom{}^{\phi\bar{\phi}}_{\phi\bar{\phi}}=\Atmatrix^{(0)2}_{LR}
+\Ctmatrix^{(0)}_{LR}\Ftmatrix^{(0)}_{LR} &~&
\frac{1}{J_u}(C^{\rm BOSS}_{u, LR})\phantom{}^{\phi\bar{\phi}}_{\phi\bar{\phi}}=\Atmatrix^{(0)2}_{LR}
\\[1pt]
\frac{1}{J_s}(C^{\rm BOSS}_{s, LR})\phantom{}^{\psi\bar{\psi}}_{\phi\bar{\phi}}=\Atmatrix^{(0)}_{LR}\Ctmatrix^{(0)}_{LR}
+\Ctmatrix^{(0)}_{LR}\Dtmatrix^{(0)}_{LR} &~&
\frac{1}{J_u}(C^{\rm BOSS}_{u, LR})\phantom{}^{\psi\bar{\psi}}_{\phi\bar{\phi}}=\Gtmatrix^{(0)}_{LR}\Ctmatrix^{(0)}_{LR}
+\Ctmatrix^{(0)}_{LR}\Ltmatrix^{(0)}_{LR}
\\[1pt]
\frac{1}{J_s}(C^{\rm BOSS}_{s, LR})\phantom{}^{\psi\bar{\psi}}_{\psi\bar{\psi}}=\Dtmatrix^{(0)2}_{LR}
+\Ctmatrix^{(0)}_{LR}\Ftmatrix^{(0)}_{LR}&~&
\frac{1}{J_u}(C^{\rm BOSS}_{u, LR})\phantom{}^{\psi\bar{\psi}}_{\psi\bar{\psi}}=\Dtmatrix^{(0)2}_{LR}
\\[1pt]
\frac{1}{J_s}(C^{\rm BOSS}_{s, LR})\phantom{}^{\phi\bar{\phi}}_{\psi\bar{\psi}}=\Dtmatrix^{(0)}_{LR}\Ftmatrix^{(0)}_{LR}
+\Ftmatrix^{(0)}_{LR}\Atmatrix^{(0)}_{LR} &~&
\frac{1}{J_u}(C^{\rm BOSS}_{u, LR})\phantom{}^{\phi\bar{\phi}}_{\psi\bar{\psi}}=\Ltmatrix^{(0)}_{LR}\Ftmatrix^{(0)}_{LR}
+\Ftmatrix^{(0)}_{LR}\Gtmatrix^{(0)}_{LR}
\\[1pt]
\frac{1}{J_s}(C^{\rm BOSS}_{s, LR})\phantom{}^{\phi\bar{\psi}}_{\phi\bar{\psi}}=\Gtmatrix^{(0)2}_{LR}&~&
\frac{1}{J_u}(C^{\rm BOSS}_{u, LR})\phantom{}^{\phi\bar{\psi}}_{\phi\bar{\psi}}=\Gtmatrix^{(0)2}_{LR}
-\Ctmatrix^{(0)}_{LR}\Ftmatrix^{(0)}_{LR}
\\[1pt]
\frac{1}{J_s}(C^{\rm BOSS}_{s, LR})\phantom{}^{\psi\bar{\phi}}_{\psi\bar{\phi}}=\Ltmatrix^{(0)2}_{LR}&~&
\frac{1}{J_u}(C^{\rm BOSS}_{u, LR})\phantom{}^{\psi\bar{\phi}}_{\psi\bar{\phi}}=\Ltmatrix^{(0)2}_{LR}
-\Ctmatrix^{(0)}_{LR}\Ftmatrix^{(0)}_{LR}
\end{array}
\ee
From here the difference $C_s/J_s-C_u/J_u$ that governs the S matrix' logarithmic dependence on external momenta 
is
\begin{align}
\frac{C^{\rm BOSS}_s}{J_s}-\frac{C^{\rm BOSS}_u}{J_u}=
{}\frac{p^2(p')^2({\bf p}\cdot{\bf p'}-mm')}{2(\varepsilon'p-p'\varepsilon)^2} \,\unit \ .
\end{align}
where $\unit$ is the identity matrix in field space. In turn this implies that the one-loop correction to the $BOSS$ S matrix
is
\begin{align}
i\tmatrix^{(1),BOSS}_{LR}=i\left(\frac{1}{2}\left(
-\frac{1}{\pi}\frac{p^2(p')^2({\bf p}\cdot{\bf p'}-m m')}{2(\varepsilon'p-p'\varepsilon)^2}
\left(\mathrm{ln}\left|\frac{p_-'}{p_-}\right|-\mathrm{ln}\left|\frac{m'}{m}\right|\right)\right)\,\unit+\text{rational}\right) \ .
\label{resultat LR}
\end{align}
This expression reproduces the small momentum limit of the $LR$ dressing phase of 
\cite{Beccaria:2012kb} for general values of $m$ and $m'$, see eq.~\eqref{fase2}. 
The additional factor of 1/2 has the same origin as in the $LL$ and $RR$ sectors.

\section{The S matrix for strings in AdS$_3\times$S$^3\times$T$^4$ \label{adst}}

\adsst~can be sourced by a mixture of RR and NSNS fluxes \cite{Maldacena:1997re, Giveon:1998ns, Berkovits:1999im}. 
The Green-Schwarz string in such a background
was constructed in \cite{Pesando:1998wm}; the AdS$_3\times$S$^3$ part is described by the supercoset 
$PSU(1,1|2)\times PSU(1,1|2)/(SU(1,1)\times SU(2))$ and it is classically integrable \cite{Cagnazzo:2012se}. 
This model was further studied in \cite{Hoare:2013pma} where the tree-level four-point S matrix was found for a generic 
mixture of both types of fluxes.

In this section we will compute the logarithmic part of the one-loop S matrix in the presence of both 
NSNS and RR fluxes.
For a pure RR flux background we extract the off-diagonal rational terms from the 
all-loop symmetry-determined S matrix found in \cite{Borsato:2013qpa} 
and compute the two-loop logarithmic terms.

The massive worldsheet spectrum in \adsst~ consists of eight modes (four bosons and 
four fermions) which are organized in two bifundamental representations of the light-cone 
gauge symmetry group $SU(1|1)^2\times SU(1|1)^2$; they are denoted by  \cite{Hoare:2013pma}
\be
&&|y_+ \rangle = |\phi\rangle \otimes |\phi\rangle
\qquad  
|z_+ \rangle = |\psi \rangle \otimes |\psi \rangle
\qquad
|\zeta_+ \rangle = |\phi \rangle \otimes |\psi \rangle
\qquad  
|\chi_+  \rangle = |\psi \rangle \otimes |\phi \rangle \ ,
\cr
&&|y_- \rangle = |{\bar \phi}\rangle \otimes |{\bar \phi}\rangle
\qquad  
|z_- \rangle = |{\bar \psi} \rangle \otimes |{\bar \psi} \rangle
\qquad
|\zeta_- \rangle = |{\bar \phi} \rangle \otimes |{\bar \psi} \rangle
\qquad  
|\chi_-  \rangle = |{\bar \psi} \rangle \otimes |{\bar \phi} \rangle \ .
\ee
We slightly adjusted the notation, in particular $\phi_+\rightarrow\phi$, $\phi_-\rightarrow{\bar\phi}$, to make 
it match the pure RR flux S matrix of \cite{Borsato:2013qpa} where $(\phi, \psi)$ are the excitations in the 
$L$ sector and  $(\bar\phi, \bar\psi)$ are the excitations in the $R$ sector. 
The symmetry group implies that the S matrix has the usual factorized structure \eqref{ws_smatrix_decomposition_txt}
\be
\Smatrix=\smatrix_{su(1|1)^2}\otimes \smatrix_{su(1|1)^2} \ .
\ee
As in \adssss, each of the two $\smatrix$-matrix factors has  four sectors: $LL$, $LR$, $RL$ and $RR$ 
(or $(++), (+-), (-+), (--)$ in the notation of \cite{Hoare:2013pma}). 

\subsection{String theory tree-level S-matrices}

The $\tmatrix$ matrix is parameterized in the same way as in \adssss, {\it i.e.}
\begin{align}
\mathrm{T}^{\phi\phi}_{\phi\phi}&=\Atmatrix_{LL},&\mathrm{T}_{\psi\psi}^{\psi\psi}&=\Dtmatrix_{LL},&
\mathrm{T}_{\phi\psi}^{\phi\psi}&=\Gtmatrix_{LL},\nonumber\\
\mathrm{T}_{\phi\psi}^{\psi\phi}&=\Htmatrix_{LL},&\mathrm{T}_{\psi\phi}^{\phi\psi}&=\Ktmatrix_{LL},&
\mathrm{T}_{\psi\phi}^{\psi\phi}&=\Ltmatrix_{LL},\nonumber\\
\mathrm{T}_{\phi\bar{\phi}}^{\phi\bar{\phi}}&=\Atmatrix_{LR},
&\mathrm{T}_{\phi\bar{\phi}}^{\psi\bar{\psi}}&=\Ctmatrix_{LR},
&\mathrm{T}_{\psi\bar{\psi}}^{\psi\bar{\psi}}&=\Dtmatrix_{LR},\label{T matrix_adsst}\\
\mathrm{T}_{\phi\bar{\psi}}^{\phi\bar{\psi}}&=\Gtmatrix_{LR},
&\mathrm{T}_{\psi\bar{\psi}}^{\phi\bar{\phi}}&=\Ftmatrix_{LR},&
\mathrm{T}_{\psi\bar{\phi}}^{\psi\bar{\phi}}&=\Ltmatrix_{LR},\nonumber
\end{align}
with each coefficient having an expansion in the inverse string tension. For the background supported by 
a mixture of RR and NSNS flux the $LL$ sector of the tree-level S matrix is \cite{Hoare:2013pma}
\begin{align}
\mathrm{T}^{(0)}_{LL}|\phi\phi'\rangle&=\frac{1}{2}(l_1+c)|\phi\phi'\rangle,&
\mathrm{T}^{(0)}_{LL}|\phi\psi'\rangle&=\frac{1}{2}(l_3+c)|\phi\psi'\rangle-l_5|\psi\phi'\rangle,\\
\mathrm{T}^{(0)}_{LL}|\psi\psi'\rangle&=\frac{1}{2}(-l_1+c)|\psi\psi'\rangle\ ,&
\mathrm{T}^{(0)}_{LL}|\psi\phi'\rangle&=\frac{1}{2}(-l_3+c)|\psi\phi'\rangle-l_5|\phi\psi'\rangle \ ,
\end{align}
and the LR sector is
\begin{align}
\mathrm{T}^{(0)}_{LR}|\phi{\bar \psi}'\rangle&=\frac{1}{2}(l_3+c)|\phi{\bar \psi}'\rangle,&
\mathrm{T}^{(0)}_{LR}|\phi{\bar \phi}'\rangle&=\frac{1}{2}(l_2+c)|\phi{\bar \phi}'\rangle+l_4|\psi{\bar \psi}'\rangle,\\
\mathrm{T}^{(0)}_{LR}|\psi{\bar \phi}'\rangle&=\frac{1}{2}(-l_3+c)|\psi{\bar \phi}'\rangle \ ,&
\mathrm{T}^{(0)}_{LR}|\psi{\bar \psi}'\rangle&=\frac{1}{2}(-l_2+c)|\psi{\bar \psi}'\rangle+l_4|\phi{\bar \phi}'\rangle \ .
\end{align}
In the following we shall need only the expressions for $l_4$ and $l_5$:
\begin{align}
l_4&=-\frac{pp'}{2(p+p')}\left(\sqrt{(\varepsilon_++(p+q))(\varepsilon'_-+(p'-q))}-\sqrt{(\varepsilon_+-(p+q))(\varepsilon'_--(p'-q))}\right),\\
l_5&=-\frac{pp'}{2(p-p')}\left(\sqrt{(\varepsilon_++(p+q))(\varepsilon'_++(p'+q))}+\sqrt{(\varepsilon_+-(p+q))(\varepsilon'_+-(p'+q))}\right),\\
\varepsilon_\pm&=\sqrt{(p\pm q)^2+1-q^2} \ .
\label{disp_rel_new}
\end{align}
Here $q$ is a measure of the ratio between the NSNS and RR fluxes; $q\rightarrow 0$ yields the pure RR flux theory. 
The other two sectors of the S matrix can be found by exchanging $\varepsilon_+$ and $\varepsilon_-$ as well as 
barred and un-barred fields. We note that for $q\rightarrow 0$ the form of the S matrix in the $LL$ and $LR$ sectors 
is the same as the $\alpha\to 1$ limit of the $LL$ and $LR$ sectors of the BOSS S matrix \cite{Borsato:2012ud, Sundin:2013ypa}.
As before, the tree-level S-matrix coefficients are related by:
\begin{align}
\Atmatrix^{(0)} +\Dtmatrix^{(0)} -\Gtmatrix^{(0)} -\Ltmatrix^{(0)}  &=0 \ ,
\end{align}
in all sectors. This identity will be useful for the consistency of the construction of loop amplitudes we now discuss.

\subsection{One-loop logarithmic terms for mixed RR/NSNS \adsst}

The one-loop amplitudes 
take again the general form \eqref{ansatz_1loop}; however, because the dispersion relations in the presence 
of a non vanishing $q$ are not the standard relativistic ones, the integrals ${\tilde I}$ have a slightly different interpretation: 
the space-like 
component of the momentum of a propagator is shifted by $\pm q$ for a field in the $L/R$ sector. Moreover, the masses are 
$(1-q^2)$. These integrals may be interpreted as regular integrals with equal masses evaluated at shifted external 
momenta\footnote{To derive these expressions it is necessary to parametrize the integrals to respect the $p\leftrightarrow p'$ symmetry
of the graph.}:
\be
{\tilde I}(p, p')_{LL} &=& 
I(p+q, p'+q)
\cr
{\tilde I}(p, p')_{LR} &=& 
I(p+q, p'-q) \ .
\label{shifted_ints}
\ee
Their expressions follow immediately from Appendix~\ref{one_n_twoloop_ints}.
Due to the modified dispersion relations, the Jacobian factors \eqref{Jacobian} are modified from their usual form 
$J=4(p\energy' - p'\energy)$ to $J_{LL}$ and $J_{LR}$ following from \eqref{Jac0} for  the dispersion relation in 
eq.~\eqref{disp_rel_new}:
\be
J_{LL} = 4((p+q)\energy'_+-(p'+q)\energy_+)
\qquad
J_{LR} = 4((p+q)\energy'_--(p'-q)\energy_+) \ .
\ee
Since all masses are equal, the multi-indices used in sec.~\ref{GeneralExpressions} 
now represent the pair $(\text{field label}, \text{sector})$.

The non-zero integral coefficients in the $LL$ sector are given by \eqref{generalC1loop}
\be
\begin{array}{lcl}
\frac{1}{J_{s,LL}}(C_{s,LL})\phantom{}^{\phi\phi}_{\phi\phi} =\Atmatrix^{(0)2}_{LL}
&~&
\frac{1}{J_{u,LL}}(C_{u,LL})\phantom{}^{\phi\phi}_{\phi\phi} =\Atmatrix^{(0)2}_{LL}-\Htmatrix^{(0)}_{LL}\Ktmatrix^{(0)}_{LL}
\\[1pt]
\frac{1}{J_{s,LL}}(C_{s,LL})\phantom{}^{\psi\psi}_{\psi\psi} =\Dtmatrix^{(0)2}_{LL}
&~&
\frac{1}{J_{u,LL}}(C_{u,LL})\phantom{}^{\psi\psi}_{\psi\psi} =\Dtmatrix^{(0)2}_{LL}-\Htmatrix^{(0)}_{LL}\Ktmatrix^{(0)}_{LL}
\\[1pt]
\frac{1}{J_{s,LL}}(C_{s,LL})\phantom{}^{\phi\psi}_{\phi\psi} =\Gtmatrix^{(0)2}_{LL}+\Htmatrix^{(0)}_{LL}\Ktmatrix^{(0)}_{LL}
&~&
\frac{1}{J_{u,LL}}(C_{u,LL})\phantom{}^{\phi\psi}_{\phi\psi} =\Gtmatrix^{(0)2}_{LL}
\\[1pt]
\frac{1}{J_{s,LL}}(C_{s,LL})\phantom{}^{\phi\psi}_{\psi\phi} =\Gtmatrix^{(0)}_{LL}\Htmatrix^{(0)}_{LL}+\Htmatrix^{(0)}_{LL}\Ltmatrix^{(0)}_{LL}
&~&
\frac{1}{J_{u,LL}}(C_{u,LL})\phantom{}^{\phi\psi}_{\psi\phi} =\Dtmatrix^{(0)}_{LL}\Htmatrix^{(0)}_{LL}+\Htmatrix^{(0)}_{LL}\Atmatrix^{(0)}_{LL}
\\[1pt]
\frac{1}{J_{s,LL}}(C_{s,LL})\phantom{}^{\psi\phi}_{\phi\psi} =\Ltmatrix^{(0)}_{LL}\Ktmatrix^{(0)}_{LL}+\Ktmatrix^{(0)}_{LL}\Gtmatrix^{(0)}_{LL}
&~&
\frac{1}{J_{u,LL}}(C_{u,LL})\phantom{}^{\psi\phi}_{\phi\psi} =\Atmatrix^{(0)}_{LL}\Ktmatrix^{(0)}_{LL}+\Ktmatrix^{(0)}_{LL}
\Dtmatrix^{(0)}_{LL}
\\[1pt]
\frac{1}{J_{s,LL}}(C_{s,LL})\phantom{}^{\psi\phi}_{\psi\phi} =\Ltmatrix^{(0)2}_{LL}+\Htmatrix^{(0)}_{LL}\Ktmatrix^{(0)}_{LL}
&~&
\frac{1}{J_{u,LL}}(C_{u,LL})\phantom{}^{\psi\phi}_{\psi\phi} =\Ltmatrix^{(0)2}_{LL}
\end{array}
\label{adsstqne01loopcoefs}
\ee

From here follows that the difference of $C_s$ and $C_u$ coefficients relevant for the logarithmic terms 
in \eqref{difference} in the $LL$ sector is  
\begin{align}
\frac{C_{s,LL}}{J_s}-\frac{C_{u,LL}}{J_u}
=\frac{p^2(p')^2}{2(p-p')^2}
\left(\varepsilon_+\varepsilon_+'+(p+q)(p'+q)+(1-q^2)\right)
\,\unit  \,
\label{difference1loop_ADST_LL}
\end{align}
which depends only on the product  $\Htmatrix^{(0)}\Ktmatrix^{(0)}$.

The nonzero integral coefficients in the $LR$ sector are given by 
\be
\begin{array}{lcl}
\frac{1}{J_{s, LR}}(C_{s,LR})\phantom{}^{\phi\bar{\phi}}_{\phi\bar{\phi}} =
\Atmatrix^{(0)2}_{LR}+\Ctmatrix^{(0)}_{LR}\Ftmatrix^{(0)}_{LR} &~&
\frac{1}{J_{u, LR}}(C_{u,LR})\phantom{}^{\phi\bar{\phi}}_{\phi\bar{\phi}} =
\Atmatrix^{(0)2}_{LR}
\\[1pt]
\frac{1}{J_{s, LR}}(C_{s,LR})\phantom{}^{\psi\bar{\psi}}_{\phi\bar{\phi}} =
\Atmatrix^{(0)}_{LR}\Ctmatrix^{(0)}_{LR}+\Ctmatrix^{(0)}_{LR}\Dtmatrix^{(0)}_{LR} &~&
\frac{1}{J_{u, LR}}(C_{u,LR})\phantom{}^{\psi\bar{\psi}}_{\phi\bar{\phi}} =
\Gtmatrix^{(0)}_{LR}\Ctmatrix^{(0)}_{LR}+\Ctmatrix^{(0)}_{LR}\Ltmatrix^{(0)}_{LR}
\\[1pt]
\frac{1}{J_{s, LR}}(C_{s,LR})\phantom{}^{\psi\bar{\psi}}_{\psi\bar{\psi}} =
\Dtmatrix^{(0)2}_{LR}+\Ctmatrix^{(0)}_{LR}\Ftmatrix^{(0)}_{LR}&~&
\frac{1}{J_{u, LR}}(C_{u,LR})\phantom{}^{\psi\bar{\psi}}_{\psi\bar{\psi}} =
\Dtmatrix^{(0)2}_{LR}
\\[1pt]
\frac{1}{J_{s, LR}}(C_{s,LR})\phantom{}^{\phi\bar{\phi}}_{\psi\bar{\psi}} =
\Dtmatrix^{(0)}_{LR}\Ftmatrix^{(0)}_{LR}+\Ftmatrix^{(0)}_{LR}\Atmatrix^{(0)}_{LR} &~&
\frac{1}{J_{u, LR}}(C_{u,LR})\phantom{}^{\phi\bar{\phi}}_{\psi\bar{\psi}} =
\Ltmatrix^{(0)}_{LR}\Ftmatrix^{(0)}_{LR}+\Ftmatrix^{(0)}_{LR}\Gtmatrix^{(0)}_{LR}
\\[1pt]
\frac{1}{J_{s, LR}}(C_{s,LR})\phantom{}^{\phi\bar{\psi}}_{\phi\bar{\psi}} =
\Gtmatrix^{(0)2}_{LR}&~&
\frac{1}{J_{u, LR}}(C_{u,LR})\phantom{}^{\phi\bar{\psi}}_{\phi\bar{\psi}} =
\Gtmatrix^{(0)2}_{LR}-\Ctmatrix^{(0)}_{LR}\Ftmatrix^{(0)}_{LR}
\\[1pt]
\frac{1}{J_{s, LR}}(C_{s,LR})\phantom{}^{\psi\bar{\phi}}_{\psi\bar{\phi}} =
\Ltmatrix^{(0)2}_{LR}&~&
\frac{1}{J_{u, LR}}(C_{u,LR})\phantom{}^{\psi\bar{\phi}}_{\psi\bar{\phi}} =
\Ltmatrix^{(0)2}_{LR}-\Ctmatrix^{(0)}_{LR}\Ftmatrix^{(0)}_{LR} \ .
\end{array}
\ee
The corresponding difference \eqref{difference} relevant for the logarithmic terms 
in this sector is then 
\begin{align}
\frac{C_{s,LR}}{J_{s,LR}}-\frac{C_{u,LR}}{J_{u,LR}}
=\frac{p^2(p')^2}{2(p+p')^2}
\left(\varepsilon_+\varepsilon_-'+(p+q)(p'-q)-(1-q^2)\right)
\,\unit \ ,
\label{difference1loop_ADST_LR}
\end{align}
and depends only on $\Ctmatrix^{(0)}\Ftmatrix^{(0)}$.

Using the integrals \eqref{shifted_ints} to reconstruct the one-loop S matrix we find:
\begin{align}
&i\tmatrix^{(1)}_{LL}=i\Big(\frac{1}{2}\Big(-\frac{1}{\pi}\frac{p^2(p')^2
\left(\varepsilon_+\varepsilon_+'+(p+q)(p'+q)+(1-q^2)\right)
}{2(p-p')^2}\,\ln\left|\frac{\energy'_+ -p'-q}{\energy_+ - p-q}\right|\Big)\,\unit+\text{rat.}\Big)\ ,
\label{++phase}
\\
&i\tmatrix^{(1)}_{LR}=i\Big(\frac{1}{2}\Big(-\frac{1}{\pi}\frac{p^2(p')^2
\left(\varepsilon_+\varepsilon_-'+(p+q)(p'-q)-(1-q^2)\right)
}{2(p+p')^2}\,\ln\left|\frac{\energy'_- -p'+q}{\energy_+ - p-q}\right|\Big)\,\unit+\text{rat.}\Big) \ ,
\label{+-phase}
\end{align}
i.e. only the diagonal entries are corrected. This result is in line with the expectation \eqref{SmatExpansion} 
that at one loop all logarithmic corrections are proportional to the identity matrix. Comparison with that equation 
identifies the coefficients of $(i/2\, \unit)$ with the one-loop dressing phases $\theta^{(1)}_{LL}$ and $\theta^{(1)}_{LR}$ in 
the $LL$ and  $LR$ sectors.

In the limit $q\rightarrow 0$ the momentum dependence of eqs.~\eqref{++phase} and \eqref{+-phase}
becomes the same as that of the small momentum limit of the phase factors found in \cite{Beccaria:2012kb}, 
eqs.~\eqref{fase1} and \eqref{fase2}.\footnote{The simplest way to see this is to solve the on shell conditions and write 
both expressions in terms of $p_-$ and $p'_-$.} The near-flat space limit of these phases was also found through 
a Feynman graph calculation in \cite{Sundin:2013ypa}.
An additional factor of $1/2$ relates to the fact that the phase factor of \cite{Beccaria:2012kb} receives contributions 
from both factors of the factorized S matrix \eqref{ws_smatrix_decomposition_txt}. This pattern is the same as in \adss.

\subsection{Two-loop logarithmic terms in \adsst~ with RR flux}


The two-loop logarithmic terms for mixed NRNR and RR flux background can be found as 
soon as the symmetry-determined part of the S matrix becomes available~\cite{HTexactNSNSRR}.
For the $q=0$ case the relevant S matrix was recently
suggested in \cite{Borsato:2013qpa}; up to the choice of mass scale it is the same as the S matrix in \adssss~ 
which is included in Appendix~\ref{ads3s3s3s1}. 

The two-loop amplitude has the general form \eqref{twoloop_tmatrix_improved} with coefficients given by 
\eqref{general_C2loop} and \eqref{Cextra}. As for $q\ne 0$, the multi-indices are pairs $(\text{field label}, \text{sector})$.
We discuss separately the $LL$ and $LR$ sectors, 
focusing on the S-matrix element $\Atmatrix^{(2)}$. The contribution of the $C$ coefficients to it will be denoted 
by $C^{\Atmatrix_{LL}}$ and $C^{\Atmatrix_{LR}}$.

\subsubsection{The $LL$ and $RR$ sectors}

Expressed in terms of cuts of the one-loop amplitudes, the six coefficients $C_a^{\Atmatrix_{LL}}$  \eqref{general_C2loop} are
\be
\begin{array}{l}
\frac{1}{J^2}{C^{\Atmatrix_{LL}}_a}={}\Atmatrix_{LL\ s-\text{cut}}^{(1)}\Atmatrix^{(0)}_{LL},\\[1pt]
\frac{1}{J^2}{C^{\Atmatrix_{LL}}_d}={}\Atmatrix_{LL\ u-\text{cut}}^{(1)}\Atmatrix^{(0)}_{LL}
-\frac{1}{2}\Htmatrix_{LL\ u-\text{cut}}^{(1)}\Ktmatrix^{(0)}_{LL}
-\frac{1}{2}H^{(0)}_{LL}K_{LL\ u-\text{cut}}^{(1)},\\[1pt]
\frac{1}{J^2}{C^{\Atmatrix_{LL}}_b}={}\Atmatrix_{LL\ u-\text{cut}}^{(1)}\Atmatrix^{(0)}_{LL},\\[1pt]
\frac{1}{J^2}{C^{\Atmatrix_{LL}}_e}={}\Atmatrix_{LL\ s-\text{cut}}^{(1)}\Atmatrix^{(0)}_{LL}
-\Htmatrix_{LL\ s-\text{cut}}^{(1)}\Ktmatrix^{(0)}_{LL},\\[1pt]
\frac{1}{J^2}{C^{\Atmatrix_{LL}}_c}={}\Atmatrix_{LL\ u-\text{cut}}^{(1)}\Atmatrix^{(0)}_{LL},\\[1pt]
\frac{1}{J^2}{C^{\Atmatrix_{LL}}_f}={}\Atmatrix_{LL\ s-\text{cut}}^{(1)}\Atmatrix^{(0)}_{LL}
-\Htmatrix^{(0)}_{LL}\Ktmatrix_{LL\ s-\text{cut}}^{(1)} \ .
\end{array}
\label{adsst2loop}
\ee
We recall that the cuts of the one-loop coefficients are given by the relevant components of 
the $C_s$ and $C_u$ coefficients \eqref{adsstqne01loopcoefs}. It is trivial to check that 
\be
\label{no_squared_log}
{\cal C}^{\Atmatrix_{LL}}_{\ln^2}\propto -2C^{\Atmatrix_{LL}}_a+C^{\Atmatrix_{LL}}_b+C^{\Atmatrix_{LL}}_c
-2C^{\Atmatrix_{LL}}_d+C^{\Atmatrix_{LL}}_e+C^{\Atmatrix_{LL}}_f =0 \ ,
\ee
which implies that the double logarithms cancel out of the $\Atmatrix^{(2)}_{LL}$. The same holds for all 
other components of the two-loop S matrix. 

We note here that eqs.~\eqref{adsst2loop} hold for $q\ne 0$ as well; the resulting two-loop integral coefficients 
$C_a\dots C_f$ also obey the relation \eqref{no_squared_log}, guaranteeing the absence of double logarithms
in the dressing phases of the S matrix in the mixed NSNS/RR-flux background as well.

The single-log terms in $\Atmatrix_{LL}$ have the same form in terms of the $C$ coefficients as 
given in eq.~\eqref{simple_logs}. 
The remaining coefficients $C_{s,\text{extra}}$ and $C_{u,\text{extra}}$, obtained by matching the 
single two-particle cuts of the ansatz onto the single two-particle cuts of the four-point amplitudes, 
are given by \eqref{Cextra}
\begin{align}
\frac{C_{s,\text{extra}}^{\Atmatrix_{LL}}}{J}=&{}-i\Atmatrix^{(0)}_{LL}\frac{C_s^{\Atmatrix_{LL}}}{J}
-\frac{1}{2}\left(\frac{C_b^{\Atmatrix_{LL}}}{J^2}
+\frac{C_c^{\Atmatrix_{LL}}}{J^2}\right)J_uI_t
+2\Atmatrix^{(0)}_{LL}{\tilde \Atmatrix}^{(1)}_{LL} \ ,\nonumber
\\
\frac{C_{u,\text{extra}}^{\Atmatrix_{LL}}}{J}=&{}-i\Atmatrix^{(0)}_{LL}\frac{C_s^{\Atmatrix_{LL}}}{J}
+\frac{1}{2}i\Htmatrix^{(0)}_{LL}\frac{C_s^{\Ktmatrix_{LL}}}{J}+\frac{1}{2}i\Ktmatrix^{(0)}_{LL}\frac{C_s^{\Htmatrix_{LL}}}{J}
  -\frac{1}{2}\left(\frac{C_e^{\Atmatrix_{LL}}}{J^2}
+\frac{C_f^{\Atmatrix_{LL}}}{J^2}\right)J_uI_t\nonumber \\
&+2\Atmatrix^{(0)}_{LL}{\tilde\Atmatrix}^{(1)}_{LL}-\Htmatrix^{(0)}_{LL}{\tilde\Ktmatrix}^{(1)}_{LL}
-\Ktmatrix^{(0)}_{LL}{\tilde\Htmatrix}^{(1)}_{LL} \ ,
\end{align}
with the coefficients with tilde being the entries of the rational part, ${\tilde\tmatrix}^{(1)}$, of the one-loop S matrix 
introduced in eq.~\eqref{tildeT}.
The difference of the $C_\text{extra}$ coefficients that enters eq.~\eqref{simple_logs} is then
\begin{align}
\frac{C_{s,\text{extra}}^{\Atmatrix_{LL}}}{J}-\frac{C_{u,\text{extra}}^{\Atmatrix_{LL}}}{J}=&
{}\frac{1}{2}\left(\frac{C_e^{\Atmatrix_{LL}}}{J^2}
+\frac{C_f^{\Atmatrix_{LL}}}{J^2}-\frac{C_b^{\Atmatrix_{LL}}}{J^2}-\frac{C_c^{\Atmatrix_{LL}}}{J^2}\right)J_u\,I_t\\
&-\frac{1}{2}\Htmatrix^{(0)}_{LL}(2{\tilde\Ktmatrix}_{LL}^{(1)}-(i)^2\Ktmatrix^{(1)}_{LL\ s-\text{cut}})\nonumber\\
&-\frac{1}{2}\Ktmatrix^{(0)}_{LL}(2{\tilde\Htmatrix}_{LL}^{(1)}-(i)^2\Htmatrix^{(1)}_{LL\ s-\text{cut}})\nonumber \ ,
\end{align}
where we used $J_s=J_u$. We notice the cancellation of the diagonal rational term ${\tilde\Atmatrix}^{(1)}$ which 
carries the only dependence on the undetermined rational function $\Phi$.
The relevant one-loop rational terms are\footnote{We notice here that, similarly to the \adss~S~matrix, the one-loop 
off-diagonal rational terms are proportional to the tree-level S-matrix elements.}
\begin{align}
{\tilde\Htmatrix}^{(1)}_{LL}={\hat\Htmatrix}^{(1)}_{LL}&=\frac{i}{2}\left(\Atmatrix^{(0)}_{LL}
+\Dtmatrix^{(0)}_{LL}\right)\Htmatrix^{(0)}_{LL}
+\frac{i}{4}(1+4b)(p-p')\Htmatrix^{(0)}_{LL},\\
{\tilde\Ktmatrix}^{(1)}_{LL}={\hat\Ktmatrix}^{(1)}_{LL}&=\frac{i}{2}\left(\Atmatrix^{(0)}_{LL}
+\Dtmatrix^{(0)}_{LL}\right)\Ktmatrix^{(0)}_{LL}
-\frac{i}{4}(1+4b)(p-p')\Ktmatrix^{(0)}_{LL} \ .
\end{align}
Together with the one-loop coefficients $\Htmatrix^{(1)}$ and $\Ktmatrix^{(1)}$ they imply that 
the terms on the second line of \eqref{simple_logs} vanish identically and that the only contribution 
to the single-log terms in $\Atmatrix_{LL}^{(2)}$ comes from the first line of that equation:
\begin{align}
-4\pi i \, {\cal C}^{\Atmatrix_{LL}}_{\ln^1}=\frac{1}{J^2}(2C^{\Atmatrix_{LL}}_a-C^{\Atmatrix_{LL}}_b-C^{\Atmatrix_{LL}}_c)&=-2(i\Atmatrix^{(0)}_{LL})
\left(\frac{C^{\Atmatrix_{LL}}_s}{J_s}-\frac{C^{\Atmatrix_{LL}}_u}{J_u}\right) \ ;
\label{simple_log_1st_term}
\end{align}
the expression inside the parentheses can be read off of eq.~\eqref{difference1loop_ADST_LL} in the limit $q\rightarrow 0$.

Repeating the calculation for other entries of the two-loop S-matrix 
\footnote{Both here and in the $LR$ sector, the identity 
$$
{\hat\Atmatrix}^{(1)}+{\hat\Dtmatrix}^{(1)}-{\hat\Gtmatrix}^{(1)}-{\hat\Ltmatrix}^{(1)}
+\frac{1}{2}\Atmatrix_{s-\text{cut}}^{(1)}+\frac{1}{2}\Dtmatrix_{s-\text{cut}}^{(1)}-\frac{1}{2}
\Gtmatrix_{s-\text{cut}}^{(1)}-\frac{1}{2}\Ltmatrix_{s-\text{cut}}^{(1)}=0 \ .
$$
is necessary for finding the off-diagonal entries of the S matrix.
}
we find similar results except that 
$(i\Atmatrix^{(0)}_{LL})$ is replaced with the tree-level value of that entry and the parenthesis 
is replaced with some combination of the entries of $C_s/J_s-C_u/J_u$ whose value equals that of
$C^{\Atmatrix_{LL}}_s/J_s-C^{\Atmatrix_{LL}}_u/J_u$. Thus, the two-loop S matrix in the LL sector is
\be
i\tmatrix_{LL}^{(2)} = -\frac{1}{2}\,{\tmatrix}^{(0)}\,\theta^{(1)}_{LL}\,+\text{rational}
\label{ADST_LL_final}
\ee
The $RR$ sector S matrix is completely identical.

\subsubsection{The $LR$ and $RL$ sectors}

The construction of the two-loop S matrix in the $LR$ sector is very similar except for the specifics related to the 
tree-level S matrix in this sector. The six coefficients determined by maximal cuts are  \eqref{general_C2loop}
\be
\begin{array}{l}
\frac{1}{J^2}{C_a^{\Atmatrix_{LR}}}={}\Atmatrix^{(1)}_{LR\ s-\text{cut}}\Atmatrix^{(0)}_{LR}
+\frac{1}{2}\Ctmatrix_{LR}^{(0)}\Ftmatrix^{(1)}_{LR\ s-\text{cut}}
+\frac{1}{2}\Ctmatrix^{(1)}_{LR\ s-\text{cut}}\Ftmatrix_{LR}^{(0)} \ ,\\[1pt]
\frac{1}{J^2}{C_d^{\Atmatrix_{LR}}}={}\Atmatrix^{(1)}_{LR\ s-\text{cut}}\Atmatrix^{(0)}_{LR},\\[1pt]
\frac{1}{J^2}{C_b^{\Atmatrix_{LR}}}={}\Atmatrix^{(1)}_{LR\ u-\text{cut}}\Atmatrix^{(0)}_{LR}
+\Ctmatrix^{(1)}_{LR\ u-\text{cut}}\Ftmatrix_{LR}^{(0)},\\[1pt]
\frac{1}{J^2}{C_e^{\Atmatrix_{LR}}}={}\Atmatrix^{(1)}_{LR\ s-\text{cut}}\Atmatrix^{(0)}_{LR},\\[1pt]
\frac{1}{J^2}{C_c^{\Atmatrix_{LR}}}={}\Atmatrix^{(1)}_{LR\ u-\text{cut}}\Atmatrix^{(0)}_{LR}
+\Ctmatrix_{LR}^{(0)}\Ftmatrix^{(1)}_{LR\ u-\text{cut}},\\[1pt]
\frac{1}{J^2}{C_f^{\Atmatrix_{LR}}}={}\Atmatrix^{(1)}_{LR\ s-\text{cut}}\Atmatrix^{(0)}_{LR} \ .
\end{array}
\ee
As before, the cuts of the one-loop S-matrix elements are given in terms of the appropriate components of $C_s$ and $C_u$.
It is easy to check that 
\be
{\cal C}^{\Atmatrix_{LR}}_{\ln^2}=-2C^{\Atmatrix_{LR}}_a+C^{\Atmatrix_{LR}}_b+C^{\Atmatrix_{LR}}_c
-2C^{\Atmatrix_{LR}}_d+C^{\Atmatrix_{LR}}_e+C^{\Atmatrix_{LR}}_f =0
\label{no_2logs_LR}
\ee
and thus the double logarithms cancel out. The same holds for all other components of the S matrix.

As in the $LL$ sector, the integral coefficients $C_a\dots C_f$ can also be found in a mixed RR/NSNS background 
and they obey the condition \eqref{no_2logs_LR}. Thus, double logarithms are also absent from the two-loop S matrix
in this more general case.

The remaining two coefficients,  $C_{s,\text{extra}}^{\Atmatrix_{LR}}$ and $C_{u,\text{extra}}^{\Atmatrix_{LR}}$, 
follow from \eqref{Cextra}:
\begin{align}
\frac{C_{s,\text{extra}}^{\Atmatrix_{LR}}}{J}=&{}-i\Atmatrix_{LR}^{(0)}\frac{C^{\Atmatrix_{LR}}_s}{J}
-\frac{1}{2}i\Ctmatrix^{(0)}\frac{C_s^{\Ftmatrix_{LR}}}{J}-\frac{1}{2}i\Ftmatrix^{(0)}\frac{C_s^{\Ctmatrix_{LR}}}{J}
-\frac{1}{2}\left(\frac{C_b^{\Atmatrix_{LR}}}{J}+
\frac{C_c^{\Atmatrix_{LR}}}{J}\right)\,I_t\\
&+2\Atmatrix_{LR}^{(0)}{\tilde\Atmatrix}_{LR}^{(1)}+\Ctmatrix_{LR}^{(0)}{\tilde\Ftmatrix}_{LR}^{(1)}
+{\tilde\Ctmatrix}_{LR}^{(1)}\Ftmatrix_{LR}^{(0)}\nonumber
\\
\frac{C_{u,\text{extra}}^{\Atmatrix_{LR}}}{J}=&{}-i\Atmatrix_{LR}^{(0)}\frac{C^{\Atmatrix_{LR}}_s}{J}
-\frac{1}{2}\left(\frac{C_e^{\Atmatrix_{LR}}}{J}+
\frac{C_f^{\Atmatrix_{LR}}}{J}\right)\,I_t
+2\Atmatrix_{LR}^{(0)}{\tilde\Atmatrix}_{LR}^{(1)} \ .
\end{align}
In their difference, which enters eq.~\eqref{simple_logs}, we notice again the cancellation of 
the diagonal one-loop rational term ${\tilde\Atmatrix}_{LR}^{(1)} $, which guarantees that 
all the single-logarithms are independent of the rational part $\Phi$ of the dressing phase, see eq~\eqref{tildeT}.

With the off-diagonal rational terms extracted from the symmetry-determined one-loop  S matrix\footnote{As in the $LL$ sector
and in \adss, the rational terms are proportional to the tree-level S-matrix elements.}
\begin{align}
{\tilde \Ctmatrix}^{(1)}_{LR}={\hat \Ctmatrix}^{(1)}_{LR}&=
\frac{i}{2}\left(A^{(0)}_{LR}+D^{(0)}_{LR}\right)C^{(0)}_{LR}+\frac{i}{4}(1+4b)(p+p')C^{(0)}_{LR},\\
{\tilde\Ftmatrix}^{(1)}_{LR}={\hat\Ftmatrix}^{(1)}_{LR}&=
\frac{i}{2}\left(A^{(0)}_{LR}+D^{(0)}_{LR}\right)F^{(0)}_{LR}-\frac{i}{4}(1+4b)(p+p')F^{(0)}_{LR} \ ,
\end{align}
we find that the second line on the right-hand side of eq.~\eqref{simple_logs} vanishes identically and that
\be
-4\pi i\,{\cal C}^{\Atmatrix_{LR}}_{\ln^1} = \frac{1}{2J^2}(2C^{\Atmatrix_{LR}}_a-C^{\Atmatrix_{LR}}_b-C^{\Atmatrix_{LR}}_c) 
= -2(i\Atmatrix^{(0)}_{LR})\left(\frac{C^{\Atmatrix_{LR}}_s}{J_s}-\frac{C^{\Atmatrix_{LR}}_u}{J_u}\right) \ .
\ee 
The value of the parenthesis can be read off eq.~\eqref{difference1loop_ADST_LR} in the limit $q\rightarrow 0$.

Repeating the calculation for other entries of the two-loop S matrix we find similar results upon using all 
relations between the two-loop integral coefficients.
Thus, the two-loop S matrix in the $LR$ sector is
\be
i\tmatrix_{LR}^{(2)} = -\frac{1}{2}\,{\tmatrix}^{(0)}\,\theta^{(1)}_{LR}\,+\text{rational}
\label{ADST_LR_final}
\ee
This result, as well as eq.~\eqref{ADST_LL_final}, supports the exponentiation of the one-loop dressing phase 
in all sectors of the theory and thus provide support for the quantum integrability of the theory through two loops.
The $RL$ sector S matrix is completely identical.

\section{Concluding remarks \label{close}}

In this paper we discussed the calculation of the logarithmic terms in the S-matrices of two-dimensional
integrable quantum field theories using the generalized unitarity method and its refinements. The calculation 
of unitarity cuts was carried out in two dimensions and thus it potentially drops terms with completely rational 
dependence on external momenta. By supplying the off-diagonal rational lower-loop terms determined by 
symmetries we can recover all the higher-loop logarithms.
We illustrated this approach with one- and two-loop calculations in worldsheet theories relevant to gauge/string dualities -- 
string theory in \adss, \adscp, \adssss~and \adsst. 

Using this approach we successfully recovered  the known logarithmic terms in the \adss~
S matrix and thus provided evidence that, in this approach, the structure of the S matrix is that implied by integrability.
We also computed the logarithmic terms of the one-loop S matrix for strings in \adssss~for a general value of $\alpha$
and reproduced the dressing phase obtained by matching the one-loop energy calculation of semiclassical states with 
the predictions of the Asymptotic Bethe Ansatz. Our result supports the assumption that the 
scattering of the $m=1$ and $m=0$ states off states with $m=\alpha$ and $m=1-\alpha$ is reflectionless.\footnote{It 
would be interesting to check whether similarly with strings in \adscp~\cite{Zarembo:2009au}, the $m=1$ states can 
be thought of as bound states of  lighter states.}
We also discussed string theory in \adsst~sourced by a mixture of RR and NSNS fluxes and found the logarithmic terms 
in the dressing phase. In the limit of vanishing NSNS flux and in the near-flat space limit 
we recovered the result of \cite{Sundin:2013ypa}.  
For vanishing NSNS flux we have also computed the logarithmic terms in the two-loop S matrix and showed 
that they all come from the exponentiation of the dressing phases.
For the mixed case one can see that the double logarithms cancel out; it should not be difficult to test this structure 
once a symmetry-determined S matrix becomes available.

The calculations described in this paper can be extended to higher loops. 
While for strings in \adss~
all  necessary information is available, potential subtleties arise for other backgrounds. We argued 
in sec.~\ref{worldsheetPT} that through two loops, unitarity-based quantum calculations in subsectors 
of theories with reflectionless S-matrices capture all logarithmic terms. A more detailed analysis is necessary to 
ascertain whether this continues to hold at higher loops. An important part  will likely be played by the
structure of higher-loop integrals.
Absence of additional logarithms in the dressing phase at two loops suggests that the pole structure of 
the S matrix is the expected one through this order and that, to this order, the spectrum of bound states is 
the known one. Higher-loop calculations would provide further tests in this direction.

In our calculations we relied on symmetries to determine the off-diagonal rational parts of one-loop 
S-matrix elements. One may continue to do so at higher loops as well. It would of course be desirable
 to have an independent derivation of both diagonal and off-diagonal rational terms, the former 
being unconstrained by symmetries. We expect that they can be found through use of dimensional 
regularization, albeit in that case one would need to supply local counterterms to restore {\it e.g.}
integrability and perhaps also other symmetries.

A direct determination of all rational terms 
would presumably clarify the interplay between regularization and symmetries in the Green-Schwarz string
and it may provide a means to discuss from an S matrix perspective the existence of integrability anomalies such 
as those in the bosonic CP$^{n-1}$ model~\cite{Abdalla:1981yt}.
It would also be an important ingredient in the unitarity-based construction of other interesting worldsheet quantities, 
such as worldsheet form factors \cite{Klose:2012ju} or correlation functions of operators, perhaps along the lines 
of \cite{Engelund:2012re}.
Similar methods can be used to construct the two-point off-shell Green's function of worldsheet fields and thus find 
the loop corrections to the dispersion relation as well as the corrected propagator residue needed for the determination 
of complete higher-loop scattering matrices.

\bigskip

\noindent{\bf Note added:} While the writing of this paper was being completed we received
\cite{Bianchi:2013nra}~by L.~Bianchi, V.~Forini, B.~Hoare where the idea of using generalized unitarity for the 
calculation of worldsheet S-matrices 
was proposed independently and applied to one-loop calculations in several integrable relativistic theories and 
in string theory in \adss. A natural prescription for the coefficient of the $t$-channel bubble integral was shown to lead 
to  complete one-loop S-matrices. 
In contrast, we determine off-diagonal rational terms from symmetry considerations. We also discussed the 
application of generalized unitarity 
in worldsheet theories truncated to a subset of fields,  obtained general expressions for amplitude through two loops 
and discussed in detail string theory in \adss, \adscp, \adssss~and \adsst.

\bigskip

\section*{Acknowledgments }

We would like to thank Z.~Bern, J.J. Carrasco, L.~Dixon, H.~Johansson, A.~Tseytlin and C.~Vergu 
for useful discussions and L.~Bianchi, V.~Forini, B.~Hoare and A.~Tseytlin for comments on the draft.  
We especially  thank D.~Kosower for collaboration at initial stages of this project and on related subjects and for detailed comments on the draft.
We also thank V.~Forini, B.~Hoare and L.~Bianchi for pre-publication copy of \cite{Bianchi:2013nra}.
This work is supported by the US Department of Energy under contract DE-SC0008745. 

\newpage
\begin{appendices}

\section{On the definition of the tree-level $\smatrix$-matrix elements \label{recall_trees}}

The worldsheet tree-level S-matrix elements are computed in a slightly different normalization than in usual 
four-dimensional  calculations. In this appendix we summarize the relevant definitions, which hopefully will 
make transparent the translation to four-dimensional conventions.

\begin{itemize}

\item Fields are expanded in creation and annihilation operators as, ($\phi$ is a generic real scalar field)
\be
\phi = \int \frac{dk_1}{(2\pi)\sqrt{2 k_0}}\Big(a_{k_1} e^{ik\cdot x}+a^\dagger_{k_1} e^{-ik\cdot x}\Big)
\label{mode_expansion}
\ee
i.e. with the measure missing a factor of $(\sqrt{2 k_0})^{-1}$ compared to the standard relativistic normalization.

\item  The commutation relations are missing a factor of $2k_0$ compared to the standard relativistic normalization
\be
[a_k, a^\dagger_{k'}]=2\pi\delta(k-k') 
\ee
This is of course a consequence of the previous item.

\item States are defined in the usual way, e.g.
\be
a^\dagger_{k_{1,1}}a^\dagger_{k_{1,2}}|0\rangle
\ee

\item Momentum conservation $\delta^{(2)}(\sum_kp_k)$ was solved as \cite{KZ} in the presence of the on-shell 
condition for external states
\be
&&\delta(\energy_1+\energy_2-\energy_3-\energy_4)\delta(p_1+p_2-p_3-p_4) \cr
&&\qquad\qquad\qquad
=\frac{\energy_1 \energy_2}{\energy_2 p_1-\energy_1p_2}
(\delta(p_1-p_3)\delta(p_2-p_4)+\delta(p_1-p_4)\delta(p_2-p_3))
\label{resolve_mom_cons}
\ee
This expression assumes that $p_1>p_2$; otherwise the Jacobian factor is negative.

\end{itemize}

\section{Beisert's SU$(2|2)$ spin-chain S matrix \label{sec:results}}

Beisert's spin-chain S matrix \cite{BeSmatrix} is defined by its action on two-particle states:
\begin{align}
\smatrix^B\ket{\phi_a\phi'_b} &=
 \Asmatrix^B\ket{\phi'_{\{a}\phi_{b\}}}
+\Bsmatrix^B\ket{\phi'_{[a}\phi_{b]}}
+\half
\Csmatrix^B\epsilon_{ab}\epsilon^{\alpha\beta}\ket{{\cal
Z}^{-}\psi'_\alpha\psi_\beta} \; ,
\\
\smatrix^B\ket{\psi_\alpha\psi'_\beta} &=
 \Dsmatrix^B\ket{\psi'_{\{\alpha}\psi_{\beta\}}}
+\Esmatrix^B\ket{\psi'_{[\alpha}\psi_{\beta]}}
+\half
\Fsmatrix^B\epsilon_{\alpha\beta}\epsilon^{ab}\ket{{\cal
Z}^{+}\phi'_a\phi_b} \; ,
\\
\smatrix^B\ket{\phi_a\psi'_\beta} &=
 \Gsmatrix^B\ket{\psi'_\beta\phi_{a}}
+\Hsmatrix^B\ket{\phi'_{a}\psi_\beta} \; ,
\\
\smatrix^B\ket{\psi_\alpha\phi'_b} &=
 \Ksmatrix^B\ket{\psi'_\alpha\phi_{b}}
+\Lsmatrix^B\ket{\phi'_{b}\psi_\alpha} \; .
\label{SU22_S_matrix}
\end{align}
Its coefficients are fixed by the requirement that it exhibits $PSU(2|2)$ symmetry; they are
\begin{align}\label{allfromsu22}
 \Asmatrix^B&= S^0_{pp'}\,\frac{x_{p'}^+-x_{p}^-}{x_{p'}^--x_{p}^+}\; , \nn \\
 \Bsmatrix^B&= S^0_{pp'}\,\frac{x_{p'}^+-x_{p}^-}{x_{p'}^--x_{p}^+}\, \left(1
             -2 \,\frac{1-\frac{1}{x^+_{p}x^-_{p'}}}{1-\frac{1}{x^+_{p}x^+_{p'}}}\,
             \frac{x_{p'}^--x_{p}^-}{x_{p'}^+-x_{p}^-}\
             \right) \; , \nn \\
 \Csmatrix^B&= S^0_{pp'}\,\frac{ 2\gamma_p\gamma_{p'}}{x^+_{p}x^+_{p'}}
                \,\frac{1}{1-\frac{1}{x^+_{p}x^+_{p'}}}
                \,\frac{x^-_{p'}-x^-_{p}}{x_{p'}^--x_{p}^+}\; , \nn \\
 \Dsmatrix^B&= -S^0_{pp'},\nn \\
 \Esmatrix^B&= -S^0_{pp'}\, \left(1
-2\,\frac{1-\frac{1}{x^-_{p}x^+_{p'}}}{1-\frac{1}{x^-_{p}x^-_{p'}}}
                              \, \frac{x_{p'}^+-x_{p}^+}{x_{p'}^--x_{p}^+}\ \right) \; , \nn \\
 \Fsmatrix^B&= -S^0_{pp'}\,\frac{2}{\gamma_{p}\gamma_{p'}x^-_{p}x^-_{p'}}
                 \,\frac{(x_{p}^+-x_{p}^-)(x_{p'}^+-x_{p'}^-)}{1-\frac{1}{x^-_{p}x^-_{p'}}}
                 \,\frac{x^+_{p'}-x^+_{p}} {x_{p'}^--x_{p}^+} \; , \nn \\
 \Gsmatrix^B&= S^0_{pp'} \,\frac{x_{p'}^+-x_{p}^+}{x_{p'}^--x_{p}^+} \; ,
 ~~~~~~~~~~~~~~
 \Hsmatrix^B = S^0_{pp'}
\,\frac{\gamma_p}{\gamma_{p'}}\,\frac{x_{p'}^+-x_{p'}^-}{x_{p'}^--x_{p}^+}
\; ,
 \nn \\
 \Ksmatrix^B&= S^0_{pp'}
\,\frac{\gamma_{p'}}{\gamma_p}\,\frac{x_{p}^+-x_{p}^-}{x_{p'}^--x_{p}^+}
\; ,
 ~~~~~~~~~~
 \Lsmatrix^B = S^0_{pp'} \,\frac{x_{p'}^--x_{p}^-}{x_{p'}^--x_{p}^+} \; ,
\end{align}
where
\be
\gamma_p=\abs{x_p^- - x_p^+}^{1/2}
\ee
and
\be
\label{x+-}
x_p^\pm=\frac{\pi e^{\pm \ihalf p}}{\sqrt{\lambda }\sin\frac{p}{2}}
\left(1+\sqrt{1+\frac{\lambda }{\pi ^2}\,\sin^2\frac{p}{2}}\right) \; .
\ee
are the Zhukowsky variables.

The overall phase factor $S^0$ is undetermined by the
algebraic construction. The one that correctly reproduces the
semiclassical string spectrum via Bethe equations is
\be\label{snol}
S^0_{pp'}=\frac{1-\frac{1}{x^+_{p'}x^-_{p}}}{1-\frac{1}{x^-_{p'}x^+_{p}}}
\,e^{i\theta\left(p,p'\right)}
\ee
with the dressing phase $\theta$ given to the leading order in $1/\sqrt{\lambda }$ by \cite{Arutyunov:2004vx}
\begin{eqnarray}\label{afsphase}
 \theta(p,p')&=&\frac{\sqrt{\lambda }}{2\pi
 }\sum_{m, m'=\pm}^{}\,m\,m'\,{\hat \chi} (x_{p}^m,x_{p'}^{m'}),
 \nn \\
{\hat \chi}
 (x,y)&=&(x-y)\left[\frac{1}{xy}+\left(1-\frac{1}{xy}\right)
 \ln\left(1-\frac{1}{xy}\right)\right] \; .
\end{eqnarray}

\section{Strong coupling expansion of the \adss~ dressing phase \label{exp_phase}}

The general form of the dressing phase in terms of higher local conserved charges is \cite{Beisert:2005cw}:
\be
\theta_{12}&=&\sum_{r=2}^\infty\sum_{s=r+1}^\infty \; c_{r,s}(\sql)(q_r(x_1^\pm)q_s(x_2^\pm)-q_r(x_2^\pm)q_s(x_1^\pm))
\\
q_r(x^\pm)&=&\frac{i}{r-1}\left(\frac{1}{(x^+)^{r-1}}-\frac{1}{(x^-)^{r-1}}\right)
\ee
This may be rewritten as  \cite{Arutyunov:2006iu}
\be
\theta_{12}&=&\chi(x_1^+,x_2^+)-\chi(x_1^+,x_2^-)-\chi(x_1^-,x_2^+)+\chi(x_1^-,x_2^-)
\cr
&-&\chi(x_2^+,x_1^+)+\chi(x_2^+,x_1^-)+\chi(x_2^-,x_1^+)-\chi(x_2^-,x_1^-)
\label{theta_ito_chi}
\\
\chi(x_1,x_2)&=&\sum_{r=2}^\infty\sum_{s=r+1}^\infty \frac{-c_{r,s}}{(r-1)(s-1)}\,\frac{1}{x_1^{r-1}x_2^{s-1}} \ .
\label{chifcts}
\ee
The coefficients $c_{r,s}$ depend on the coupling constant $\hg$ as
\be
c_{rs} = \sum_{n=0}^\infty \hg^{1-n}c_{rs}^{(n)}
&&
\hspace{-6truemm}
\rightarrow
\chi(x_1, x_2)=\sum_{n=0}^\infty\,\hg^{1-n}\chi^{(n)}(x_1,x_2) 
\quad
\rightarrow
\quad
\theta_{12}=\sum_{n=0}^\infty\,\hg^{1-n}\theta_{12}^{(n)}
\\
\theta_{12}^{(n)} &=& \chi^{(n)}(x_1^+,x_2^+)-\chi^{(n)}(x_1^+,x_2^-)-\chi^{(n)}(x_1^-,x_2^+)+\chi^{(n)}(x_1^-,x_2^-)
\cr
&-&\chi^{(n)}(x_2^+,x_1^+)+\chi^{(n)}(x_2^+,x_1^-)+\chi^{(n)}(x_2^-,x_1^+)-\chi^{(n)}(x_2^-,x_1^-)
\label{theta12n}
\ .
\ee
For string theory in \adss~the coefficients $c_{rs}^{(0)}$ were found in \cite{AFS} and $c_{rs}^{(1)}$ in \cite{HL}.
An all-loop proposal was put forward in  \cite{BeSt}.

The functions $\chi^{(n)}$ to various loop orders are:
\be
\chi^{(0)}(x_1, x_2)&=&-\frac{1}{x_2}+\left(\frac{1}{x_2}-x_1\right)\ln\left(1-\frac{1}{x_1x_2}\right)
\\
\chi^{(1)}(x_1, x_2)&=&-\frac{1}{2\pi}{\rm Li}_2\frac{\sqrt{x_1}-1/\sqrt{x_2}}{\sqrt{x_1}-\sqrt{x_2}} 
			         -\frac{1}{2\pi}{\rm Li}_2\frac{\sqrt{x_1}+1/\sqrt{x_2}}{\sqrt{x_1}+\sqrt{x_2}} 
			         \cr
			        && +\frac{1}{2\pi}{\rm Li}_2\frac{\sqrt{x_1}+1/\sqrt{x_2}}{\sqrt{x_1}-\sqrt{x_2}} 
			         +\frac{1}{2\pi}{\rm Li}_2\frac{\sqrt{x_1}-1/\sqrt{x_2}}{\sqrt{x_1}+\sqrt{x_2}} 
\\
\chi^{(2)}(x_1, x_2)&=&-\frac{x_2}{24(x_1x_2-1)(x_2^2-1)}
\\
\chi^{(3)}(x_1, x_2)&=&0
\\
\chi^{(4)}(x_1, x_2)&=&-\frac{x_2^3+4x_2^5-9 x_1x_2^6+x_2^7 +3 x_1^2 x_2^7-3x_1x_2^8+3x_1^2x_2^9}{720(x_1x_2-1)^3(x_2^2-1)^5}
\\
\chi^{(2k+1)}(x_1, x_2)&=&0
\ee
An all-order integral representation of the dressing phase was put forward in \cite{Dorey:2007xn}.

We emphasize here that the only logarithmic terms in the dressing phase have a one-loop origin. This will 
nevertheless lead to logarithmic terms at higher loops in worldsheet perturbation theory. The relation between
the spin chain strong coupling expansion and worldsheet perturbation theory was mentioned in sec.~\ref{worldsheetPT} 
will be reviewed in Appendix~\ref{ws_from_sc}.

\section{The AdS$_5\times$S$^5$ worldsheet S matrix from the spin-chain S matrix \label{ws_from_sc}}

The  symmetry of the world sheet theory is $PSU(2|2)^2$; together with the expected integrability they 
imply that the world sheet $S$-matrix factorizes as 
\be
 \Smatrix=\smatrix\otimes \smatrix
 \label{ws_smatrix_decomposition}
\ee
where each factor transforms under a copy of $PSU(2|2)$; neither factor is the S matrix of any obvious excitations
of the worldsheet theory.
The four-point tree-level worldsheet S matrix in AdS$_5\times$S$^5$ was computed in \cite{KMRZ} and this factorization 
was verified.

Since only $\grSU(2)\times\grSU(2)\subset\grPSU(2|2)$ is a manifest symmetry of the gauge-fixed worldsheet theory,
$\smatrix$ may be parametrized in terms of ten unknown functions of the momenta $p$ and $p'$ of the two incoming 
particles:%
\footnote{These definitions are similar but not identical to those
of \cite{BeSmatrix}. The relationship between the two
definitions is given in equation \ref{relation} below.}.
\begin{align}
\quad\smatrix_{\lAA\lBB}^{\lCC\lDD} & = \Asmatrix \,\delta_\lAA^\lCC \delta_\lBB^\lDD + \Bsmatrix \,\delta_\lAA^\lDD \delta_\lBB^\lCC &&
 \;\; , \qquad\quad &
\smatrix_{\lAA\lBB}^{\lcc\ldd} & = \Csmatrix \,\levi_{\lAA\lBB} \levi^{\lcc\ldd} &&
\;\; , \quad \nn \\
\smatrix_{\laa\lbb}^{\lcc\ldd} & = \Dsmatrix \,\delta_\laa^\lcc \delta_\lbb^\ldd + \Esmatrix \,\delta_\laa^\ldd \delta_\lbb^\lcc &&
 \;\; , &
\smatrix_{\laa\lbb}^{\lCC\lDD} & = \Fsmatrix \,\levi_{\laa\lbb} \levi^{\lCC\lDD} &&
 \;\; , \label{smat0} \\
\smatrix_{\lAA\lbb}^{\lCC\ldd} & = \Gsmatrix \,\delta_\lAA^\lCC \delta_\lbb^\ldd &&
 \;\; , &
\smatrix_{\laa\lBB}^{\lcc\lDD} & = \Lsmatrix \,\delta_\laa^\lcc \delta_\lBB^\lDD &&
\;\; , \nn \\
\smatrix_{\lAA\lbb}^{\lcc\lDD} & = \Hsmatrix \,\delta_\lAA^\lDD \delta_\lbb^\lcc &&
 \;\; , &
\smatrix_{\laa\lBB}^{\lCC\ldd} & = \Ksmatrix \,\delta_\laa^\ldd \delta_\lBB^\lCC &&
\;\; . \nn
\end{align}

As described in \cite{KMRZ}, in the comparison between the worldsheet and the spin chain we are 
interested in the coefficients of ${\cal P}_{{\rm g}}P^u_{pp'}S^B$, where ${\cal P}_{{\rm g}}$ is the 
graded permutation operator and $P^u_{pp'}$ exchanges the excitation momenta. 
Furthermore, to find the S matrix for the full $\grPSU(2,2|4)$ theory we use the relation%
\be
 \Smatrix=\frac{1}{\Asmatrix^B}\smatrix^B\otimes \smatrix^B 
 \quad
 \Smatrix_{A\dot{A}B\dot{B}}^{C\dot{C}D\dot{D}}(p,p')=\frac{1}{\Asmatrix^B}
(\smatrix^B)_{\rm AB}^{CD}(p,p') (\smatrix^B)_{\dot{A}\dot{B}}^{\dot{C}\dot{D}}(p,p') \; .
\ee
This is because the
$PSU(2|2)$ S matrix was defined in \cite{BeSmatrix} as the physical
scattering matrix of the fields $\Phi _{A\dot{1}}$ ; in addition to
$\smatrix$ for the left $PSU(2|2)$ indices, the scattering of
this field receives contribution from
$\smatrix_{\dot{1}\dot{1}}^{\dot{1}\dot{1}}=A^B$.

Consequently we can relate the above coefficients to those of
$\smatrix$ used in Beisert's S matrix
\begin{align}
\Asmatrix&= \tfrac{1}{2\sqrt{\Asmatrix^B}}\brk{\Asmatrix^B-\Bsmatrix^B} \; , &
\Bsmatrix&= \tfrac{1}{2\sqrt{\Asmatrix^B}}\brk{\Asmatrix^B+\Bsmatrix^B} \; , &
\Csmatrix&= \tfrac{i}{2 \sqrt{\Asmatrix^B}}\Csmatrix^B\; , \nn \\
\Dsmatrix&= \tfrac{1}{2\sqrt{\Asmatrix^B}}\brk{-\Dsmatrix^B+\Esmatrix^B} \; , &
\Esmatrix&= \tfrac{1}{2\sqrt{\Asmatrix^B}}\brk{-\Dsmatrix^B-\Esmatrix^B} \; , &
\Fsmatrix&=\tfrac{i}{2 \sqrt{\Asmatrix^B}}\Fsmatrix^B\; , \nn\\
\Gsmatrix&= \tfrac{1}{ \sqrt{\Asmatrix^B}}\Gsmatrix^B\; , &
\Hsmatrix&= \tfrac{1}{ \sqrt{\Asmatrix^B}}\Hsmatrix^B\; , \nn \\
\Lsmatrix&= \tfrac{1}{ \sqrt{\Asmatrix^B}}\Lsmatrix^B\; , &
\Ksmatrix&= \tfrac{1}{ \sqrt{\Asmatrix^B}}\Ksmatrix^B\; .
\label{relation}
\end{align}

The worldsheet perturbative expansion of the S matrix is obtained from the spin-chain S matrix in the 
small momentum expansion \cite{Roiban:2006yc, KMRZ}, defined as
the large-$\lambda$ expansion at fixed string moment $p_\text{string}\sim \sqrt{\lambda}p_\text{chain}$:
\be
 p\,\,\longrightarrow\,\, \frac{2\pi p}{\sqrt{\lambda }}
~~~~~~~~
p_{\rm chain}=\frac{2\pi}{\sqrt{\lambda }}p_{\rm string}=\frac{1}{\hg}p_\text{string}\;.
\ee
The matrix elements in (\ref{allfromsu22}) depend on $\hg\propto1/\sqrt{\lambda }$ only
through $x_p^\pm$. In the small momentum expansion these variables (\ref{x+-}) become ($\varepsilon=\sqrt{1+p^2}$):
\begin{equation}
\label{expanded_x+-}
 x_p^\pm=\frac{1+\energy }{p}
 \left(1
 \pm\frac{i p}{2\hg}
 -\frac{p^2(2+2\energy + 3 p^2)}{24(1+\energy+p^2)\hg^2}
 \mp \frac{i p^5}{48(1+\energy+p^2)\hg^3}
+ {\cal O}\left(\frac{1}{\hg^4}\right)\right)~~.
\end{equation}

Plugging these expressions into eq.~\eqref{theta12n} we notice that the term independent of the upper index $\pm$ of the 
Zhukowsky variables  cancels out. Terms linear in $1/\hg$ have a similar fate and the first nonzero terms are proportional to 
$1/\hg^2$ leading to eq.~\eqref{theta12_general_ws_expansion}
\be
\theta_{12}&=&\frac{1}{\hg}\sum_{n=0}^\infty\,\frac{1}{\hg^{n}}{\hat \theta}_{12}^{(n)} \ .
\ee
Each term ${\hat\theta}^{(n)}_{12}$ is given in terms of the derivatives of the functions $\chi^{(n)}$:
\be
{\hat\theta}^{(n)}_{12}=-(1+\energy)(1+\energy'{})
\big(
\partial_x\partial_y\chi^{(n)}(x,y)-\partial_x\partial_y\chi^{(n)}(y,x)
\big)\Big|_{x=\frac{1+\energy}{p}, y=\frac{1+\energy'{}}{p'{}}} + \chi^{(x\le n-1)}\text{-contrib's} \ ,
\label{theta_ito_chi_ws}
\ee
where the contributions of $\chi^{(x\le n-1)}$ involve three or more derivatives. We notice here that since 
the coefficients of the transcendental functions in $\chi^{(0)}$ and $\chi^{(1)}$  depend on at most one of the two 
arguments $x_1$ and $x_2$ the degree of transcendentality of $\theta^{(n)}$ is lower than that of $\chi^{(0)}$ 
and $\chi^{(1)}$ by (at least) one  unit.  In particular,  $\theta^{(0)}$ 
receives only in rational contributions in worldsheet perturbation theory while $\theta^{(1)}$ contributes 
only logarithmic contributions (cf.~\eqref{1loopphase}). Since $\chi^{(n\ge 2)}$ are themselves rational, they also 
receive only rational contributions in worldsheet perturbation theory. 

An alternative parametrization of $\Smatrix$ is obtained by extracting a phase factor form $\smatrix$, that is 
$\smatrix\mapsto{\tilde S}^{1/2}_0{\tilde\smatrix}$. Then $\Smatrix$ becomes
\be
 \Smatrix=  {\tilde S}_0 \, {\tilde \smatrix}\otimes {\tilde \smatrix} \ ;
 \label{ws_smatrix_decomposition_V1}
\ee
The coefficients entering the tensor decomposition (\ref{smat0}) are
\begin{align}
\label{sma}
 &\tilde \Asmatrix=\frac{x'_--x_-}{x'_--x_+}\,\,
 \frac{1-\frac{1}{x'_-x_+}}{1-\frac{1}{x'_+x_+}}
 \; , \nonumber \\
 &\tilde \Bsmatrix=\frac{x'_+-x_-}{x'_--x_+}\left(
 1-\frac{x'_--x_-}{x'_+-x_-}\,\,
 \frac{1-\frac{1}{x'_-x_+}}{1-\frac{1}{x'_+x_+}}
 \right)
  \; , \nonumber \\
 &\tilde \Csmatrix=\frac{i\gamma_{p} \gamma_{p'}}{x_+x'_+}\,\,
 \frac{1}{1-\frac{1}{x'_+x_+}}\,\,
 \frac{x'_--x_-}{x'_--x_+}\,\,{\rm e}\,^{\frac{i p'}{2}}
 \; , \nonumber\\
  &\tilde \Dsmatrix=\frac{x'_+-x_+}{x'_--x_+}\,\,
 \frac{1-\frac{1}{x'_+x_-}}{1-\frac{1}{x'_-x_-}}\,\,{\rm e}\,^{\frac{i (p'-p)}{2}}
 \; , \nonumber \\ 
 &\tilde \Esmatrix = 1-\frac{x'_+-x_+}{x'_--x_+}\,\,
 \frac{1-\frac{1}{x'_+x_-}}{1-\frac{1}{x'_-x_-}}\,\,{\rm e}\,^{\frac{i(p'-p)}{2}}
 \; , \nonumber 
 \end{align}
 \begin{align}
 &\tilde \Fsmatrix=-\frac{i(x_+-x_-)(x'_+-x'_-)}{\gamma_{p} \gamma_{p'} x_-x'_-}\,\,
 \frac{1}{1-\frac{1}{x'_-x_-}}\,\,
 \frac{x'_+-x_+}{x'_--x_+}\,\,{\rm e}\,^{-\frac{ip}{2}}
 \; , \nonumber \\
 &\tilde \Gsmatrix=\frac{x'_+-x_+}{x'_--x_+}\,\,{\rm e}\,^{-\frac{i p}{2}}
  \; , \qquad \quad
 \tilde \Hsmatrix=\frac{\gamma_{p} }{\gamma_{p'}}\,\,\frac{x'_+-x'_-}{x'_--x_+}\,\,{\rm e}\,^{\frac{i (p'-p)}{2}}
 \; , \nonumber \\
 &\tilde \Lsmatrix=\frac{x'_--x_-}{x'_--x_+}\,\,{\rm e}\,^{\frac{ip'}{2}}
 \; , \qquad \qquad
 \tilde \Ksmatrix=\frac{\gamma_{p'}}{\gamma_p}\,\,\frac{x_+-x_-}{x'_--x_+}
 \; ,
\end{align}
with $x_\pm\equiv x_\pm(p)$, $x'_\pm\equiv x_\pm(p')$, 
\[
\gamma_{p}=\abs{x_--x_+}^{1/2},\qquad \gamma_{p'}=\abs{x'_--x'_+}^{1/2} \; .
\]
and
\be
{\tilde S}_0=\frac{1-\frac{1}{x'_+x_-}}{1-\frac{1}{x'_-x_+}}\,\,
 \frac{x'_--x_+}{x'_+-x_-}\,
 \,{\rm e}\,^{i\theta (p,p')} \ .
\ee
As before, $p, p'$ stand for the spin-chain momenta.

\section{Two-particle cut at $L$-loops for string theory in AdS$_5\times$S$^5$ \label{2pc_general_L}}

We collect here the explicit forms $s$- and $u$-channel cuts of the $L$-loop four-point 
$\smatrix$-matrix of the \adss~string.  Here $L=L_1+L_2+1$; the S-matrix  factor at $L_1$-loops -- {\it i.e.} the 
left factor in each term -- carries the incoming particles).

\noindent
$\bullet$ The $s$-channel two-particle cuts are:
\be
\label{schannelL}
i\tmatrix^{(L)}{}_{ab}^{cd}(p,p')\big|_{s-\text{cut}}&=&
(\Atmatrix^{(L_1)}\Atmatrix^{(L_2)}+\Btmatrix^{(L_1)}\Btmatrix^{(L_2)}+2\Ctmatrix^{(L_1)}\Ftmatrix^{(L_2)})\delta_a^c\delta_b^d
\cr
&&   +(\Atmatrix^{(L_1)}\Btmatrix^{(L_2)}+\Btmatrix^{(L_1)}\Atmatrix^{(L_2)}-2\Ctmatrix^{(L_1)}\Ftmatrix^{(L_2)})\delta_a^d\delta_b^c
    \\
i\tmatrix^{(L)}{}_{\alpha\beta}^{\gamma\delta}(p,p')\big|_{s-\text{cut}}&=&    
(\Dtmatrix^{(L_1)}\Dtmatrix^{(L_2)}+\Etmatrix^{(L_1)}\Etmatrix^{(L_2)}+2\Ftmatrix^{(L_1)}\Ctmatrix^{(L_2)})\delta_\alpha^\gamma\delta_\beta^\delta
\cr
&&+ (\Dtmatrix^{(L_1)}\Etmatrix^{(L_1)}+\Etmatrix^{(L_1)} \Dtmatrix^{(L_2)}-2\Ftmatrix^{(L_1)} \Ctmatrix^{(L_2)})\delta_\alpha^\delta\delta_\beta^\gamma
    \\
i\tmatrix^{(L)}{}_{a \beta}^{c\delta}(p,p')\big|_{s-\text{cut}}&=&(\Htmatrix^{(L_1)}\Ktmatrix^{(L_2)}+\Gtmatrix^{(L_1)}\Gtmatrix^{(L_2)})\delta_\beta^\delta\delta_a^c  
   \\
i\tmatrix^{(L)}{}_{a \beta}^{\gamma d}(p,p')\big|_{s-\text{cut}}&=& (\Gtmatrix^{(L_1)}\Htmatrix^{(L_2)}+\Htmatrix^{(L_1)} \Ltmatrix^{(L_2)})\delta_a^d\delta_\beta^\gamma
   \\
i\tmatrix^{(L)}{}_{ab}^{\gamma\delta}(p,p')\big|_{s-\text{cut}}&=&((\Atmatrix^{(L_1)}-\Btmatrix^{(L_1)})\Ctmatrix^{(L_2)}+\Ctmatrix^{(L_1)}(\Dtmatrix^{(L_2)}-\Etmatrix^{(L_2)}))\epsilon_{ab}\epsilon^{\gamma\delta}
   \\
i\tmatrix^{(L)}{}_{\alpha\beta}^{cd}(p,p')\big|_{s-\text{cut}}&=& (\Ftmatrix^{(L_1)}(\Atmatrix^{(L_2)}-\Btmatrix^{(L_2)})
+(\Dtmatrix^{(L_1)}-\Etmatrix^{(L_1)})\Ftmatrix^{(L_2)})\epsilon_{\alpha\beta}\epsilon^{cd}
   \\   
i\tmatrix^{(L)}{}_{\alpha b}^{\gamma d}(p,p')\big|_{s-\text{cut}}&=&(\Ktmatrix^{(L_1)}\Htmatrix^{(L_2)}+\Ltmatrix^{(L_1)}\Ltmatrix^{(L_2)})\delta_\alpha^\gamma\delta_b^d
   \\
i\tmatrix^{(L)}{}_{\alpha b}^{c\delta}(p,p')\big|_{s-\text{cut}}&=&(\Ktmatrix^{(L_1)} \Gtmatrix^{(L_2)}+\Ltmatrix^{(L_1)}\Ktmatrix^{(L_2)})\delta_\alpha^\delta\delta_b^c      
\ee
A factor of $(i)^4$ was set to unity; the first $(i)^2$ is from the two cut propagators and a second factor of 
$(i)^2$ is due to the fact that scattering amplitudes are $i\tmatrix$ while the coefficients $\Atmatrix, \Btmatrix...$ 
parametrize $\tmatrix$, cf. eq.~(\ref{eqn:tmatrix-coeff}).

\noindent
$\bullet$ The $u$-channel two-particle cuts of an $L$-loop amplitude: as before, $L=L_1+L_2+1$; the S-matrix factor at 
$L_1$-loop order -- {\it i.e.} the left factor in each term -- carries the first lower index and the second upper index: 
\be
\label{uchannelL}
i\tmatrix^{(L)}{}_{ab}^{cd}(p,p')\big|_{u-\text{cut}}&=&
\Atmatrix^{(L_1)}\Atmatrix^{(L_2)}\delta_a^c\delta_b^d   
\cr
&&
+(\Atmatrix^{(L_1)} \Btmatrix^{(L_2)}+\Btmatrix^{(L_1)}\Atmatrix^{(L_2)} + 2\Btmatrix^{(L_1)}\Btmatrix^{(L_2)} - 2\Htmatrix^{(L_1)} \Ktmatrix^{(L_2)})\delta_a^d\delta_b^c
    \\
i\tmatrix^{(L)}{}_{\alpha\beta}^{\gamma\delta}(p,p')\big|_{u-\text{cut}}&=&    
\Dtmatrix^{(L_1)}\Dtmatrix^{(L_2)} \delta_\alpha^\gamma\delta_\beta^\delta
\cr
&&+(\Dtmatrix^{(L_1)}\Etmatrix^{(L_2)}+\Etmatrix^{(L_1)}\Dtmatrix^{(L_2)}+ 2\Etmatrix^{(L_1)}\Etmatrix^{(L_2)} - 2 \Ktmatrix^{(L_1)} \Htmatrix^{(L_2)})\delta_\alpha^\delta\delta_\beta^\gamma
    \\
i\tmatrix^{(L)}{}_{a \beta}^{c\delta}(p,p')\big|_{u-\text{cut}}&=&(-\Ctmatrix^{(L_1)}\Ftmatrix^{(L_2)} + \Gtmatrix^{(L_1)}\Gtmatrix^{(L_2)})\delta_\beta^\delta\delta_a^c  
   \\
i\tmatrix^{(L)}{}_{a \beta}^{\gamma d}(p,p')\big|_{u-\text{cut}}&=& (\Htmatrix^{(L_1)}\Dtmatrix^{(L_2)}+\Atmatrix^{(L_1)}\Htmatrix^{(L_1)}+2\Btmatrix^{(L_1)}\Htmatrix^{(L_2)} +2\Htmatrix^{(L_1)}\Etmatrix^{(L_2)})  \delta_a^d\delta_\beta^\gamma
   \\
i\tmatrix^{(L)}{}_{ab}^{\gamma\delta}(p,p')\big|_{u-\text{cut}}&=&(\Gtmatrix ^{(L_1)} \Ctmatrix^{(L_2)}+ \Ctmatrix^{(L_1)}\Ltmatrix^{(L_2)}) \epsilon_{ab}\epsilon^{\gamma\delta}
   \\
i\tmatrix^{(L)}{}_{\alpha\beta}^{cd}(p,p')\big|_{u-\text{cut}}&=& (\Ftmatrix^{(L_1)} \Gtmatrix^{(L_2)} + \Ltmatrix^{(L_1)}\Ftmatrix^{(L_2)}) \epsilon_{\alpha\beta}\epsilon^{cd}
   \\   
i\tmatrix^{(L)}{}_{\alpha b}^{\gamma d}(p,p')\big|_{u-\text{cut}}&=&(-\Ftmatrix^{(L_1)} \Ctmatrix^{(L_2)} + \Ltmatrix^{(L_1)}\Ltmatrix^{(L_2)}) \delta_\alpha^\gamma\delta_b^d
   \\
i\tmatrix^{(L)}{}_{\alpha b}^{c\delta}(p,p')\big|_{u-\text{cut}}&=&(\Ktmatrix^{(L_1)} \Atmatrix^{(L_2)}+\Dtmatrix^{(L_1)}\Ktmatrix^{(L_2)} + 2 \Ktmatrix^{(L_1)} \Btmatrix^{(L_2)}  + 2 \Etmatrix^{(L_1)} \Ktmatrix^{(L_2)}) \delta_\alpha^\delta\delta_b^c      
\ee
As in the $s$-channel cuts, a factor of $(i)^4$ of the same origin as there was set to unity.

Clearly, to reconstruct the S-matrix element one is to consider all choices of $L_1$ and $L_2$ such that 
$L_1+L_2=L-1$. Generalized cuts can be constructed iteratively, by using (generalized) cuts in place of 
$\Atmatrix^{L_i}, \Btmatrix^{L_i}, \dots$.


\section{One- and two-loop integrals \label{one_n_twoloop_ints}  \label{1loopint}  \label{multimass}}

The one-loop massive bubble integral in two dimensions shown in fig.~\ref{1loop_integrals}$(a)$ and \ref{1loop_integrals}$(b)$ were computed previously in \cite{Klose:2012ju} using the techniques of 
\cite{KahllenToll}; the third integral carried no momentum dependence. The first two integrals may be 
compactly written as
\be
I(p, p'{}) &=&\int\frac{d^2q}{(2\pi)^2}  \frac{1}{(q^2+1+i\varepsilon)((q+p+p'{})^2+1+i\varepsilon)}
\\
&=& \frac{i}{2\pi m^2}\frac{p_{-}p'_{-}}{p_{-}^2-p'_{-}{}^2}\ln\Big|\frac{p'_{-}}{p_{-}}\Big|
-\frac{p_{-}p'_{-}}{4m^2(p_{-}+p'_{-})|p_{-}-p'_{-}|}\left(\frac{p_{-}}{|p_{-}|}+\frac{p'_{-}}{|p'_{-}|}\right)
\ee
or, explicitly,
\be
I(p, p'{}) = \frac{i}{2\pi m^2}\frac{p_{-}p'_{-}}{p_{-}^2-p'_{-}{}^2}
\left\{
\begin{array}{ll}
\ln\left(\frac{p'_{-}}{p_{-}} \right) - i \pi & \text{ for } 0<p_{-}<p'_{-} \text{ or } p'_{-}<p_{-}<0
\cr
\ln\left(-\frac{p'_{-}}{p_{-}}\right)  & \text{ for } p_{-}<0<p'_{-} \text{ or } p'_{-}<0<p_{-}
\cr
\ln\left(\frac{p'_{-}}{p_{-}} \right) + i \pi & \text{ for } p_{-}<p'_{-}<0 \text{ or } 0< p'_{-}<p_{-}
\end{array}
\right.
\ee
where 
\be
p_\pm = \frac{1}{2}(\energy\pm p)\quad\text{and}\qquad
p_+p_-=\frac{1}{4} \ .
\ee

Assuming that  $p_{\pm}, p'_{\pm}>0$ and $p_{-}<p_{-}'$
the integrals that enter all amplitudes one- and two-loop amplitudes in worldsheet theories with 
excitations of {\em equal} masses are \cite{Klose:2007rz}:
\be
I_s&=&\frac{1}{J_s}\left(-\frac{i}{\pi}\ln \frac{p'_{-}}{p_{-}}  -   1\right)
\cr
I_u&=&\frac{1}{J_u}\left(+\frac{i}{\pi}\ln \frac{p'_{-}}{p_{-}} +0\right)
\cr
I_t&=&\frac{i}{4\pi}
\cr
I_a&=&\left(\frac{1}{J_s}\left(-\frac{i}{\pi}\ln \frac{p'_{-}}{p_{-}}  -   1\right)\right)^2
\cr
I_d&=&\left(\frac{1}{J_u}\left(+\frac{i}{\pi}\ln \frac{p'_{-}}{p_{-}}  +0\right)\right)^2
\cr
&&
\cr
I_b&=&\frac{1}{16\pi^2}\left(\frac{4}{J_u^2}\ln^2\frac{p'_{-}}{p_{-}}
+\left(-  \frac{8i\pi}{J_u^2}+\frac{2}{J_u}\right)\ln\frac{p'_{-}}{p_{-}}+\text{rational}\right)
\cr
I_c&=&\frac{1}{16\pi^2}\left(\frac{4}{J_u^2}\ln^2\frac{p'_{-}}{p_{-}}
+\left(-  \frac{8i\pi}{J_u^2}+\frac{2}{J_u}\right)\ln\frac{p'_{-}}{p_{-}}+\text{rational}\right)
\cr
I_e&=&\frac{1}{16\pi^2}\left(\frac{4}{J_u^2}\ln^2\frac{p'_{-}}{p_{-}}-\frac{2}{J_u}\ln\frac{p'_{-}}{p_{-}}+\text{rational}\right)
\cr
I_f&=&\frac{1}{16\pi^2}\left(\frac{4}{J_u^2}\ln^2\frac{p'_{-}}{p_{-}}-\frac{2}{J_u}\ln\frac{p'_{-}}{p_{-}}+\text{rational}\right)
\ee

Integrals with {\em different} masses assuming that $\frac{p}{m}>\frac{p'}{m'}$:
\be
{\tilde I}_s=\frac{-i}{4\pi(p\energy' - p'\energy)}\left( \ln\Big|\frac{p'_-}{p_-}\Big| - \ln\Big|\frac{m'}{m}\Big| -i\pi\right)
\ee
\be
{\tilde I}_u=\frac{+i}{4\pi(p\energy' - p'\energy)}\left( \ln\Big|\frac{p'_-}{p_-}\Big| - \ln\Big|\frac{m'}{m}\Big| \right)
\ee
where 
\be
\energy=\sqrt{p^2+m^2} \ .
\ee

\section{$s$- and $u$-channel cuts of the one- and two-loop integrals \label{one_n_twoloop_cuts}}

In the construction of the two-loop $C_{s,\text{extra}}$ and $C_{u,\text{extra}}$ integral coefficients it is 
necessary to take of the ansatz \eqref{twoloop_tmatrix_improved}. We list here the two-particle cuts 
of the integrals that appear in the text.
\be
\begin{array}{lcl}
I_s\big|_{s-\text{cut}}={\displaystyle{\frac{2}{4(p\energy'{} -p'{}\energy)}}}
&~~&
I_s\big|_{u-\text{cut}}=0
\cr
I_u\big|_{s-\text{cut}}=0
&~~&
I_u\big|_{u-\text{cut}}={\displaystyle{\frac{2}{4(p\energy'{} -p'{}\energy)} }}
\cr
I_a\big|_{s-\text{cut}} = {\displaystyle{\frac{2\times 2}{4(p\energy'{} -p'{}\energy)}I_s }}
&~~&
I_a\big|_{u-\text{cut}} =0
\cr
I_b\big|_{s-\text{cut}} ={\displaystyle{\frac{1}{4(p\energy'{} -p'{}\energy)}(I_u+  I_t) }}
&~~&
I_b\big|_{u-\text{cut}} =0
\cr
I_c\big|_{s-\text{cut}} ={\displaystyle{\frac{1}{4(p\energy'{} -p'{}\energy)}(I_u+  I_t) }}
&~~&
I_c\big|_{u-\text{cut}} =0
\cr
I_d\big|_{s-\text{cut}} =0
&~~&
I_d\big|_{u-\text{cut}} = {\displaystyle{\frac{2\times 2}{4(p\energy'{} -p'{}\energy)} I_u }}
\cr
I_e\big|_{s-\text{cut}} =0
&~~&
I_e\big|_{u-\text{cut}} ={\displaystyle{\frac{1}{2(p\energy'{} -p'{}\energy)}(I_s+  I_t) }}
\cr
I_f\big|_{s-\text{cut}} =0
&~~&
I_f\big|_{u-\text{cut}} ={\displaystyle{\frac{1}{2(p\energy'{} -p'{}\energy)}(I_s+  I_t) }}
\end{array}
\ee
In $I_a$ and $I_d$ one factor of 2 comes from the two chained bubbles and the second from the two 
solutions to the cut conditions.

\section{AdS$_3\times$S$^3\times$S$^3\times$S$^1$ S-matrices \label{ads3s3s3s1}}

\noindent{$\bullet$ {\bf The S matrix of Borsato, Ohlson Sax and Sfondrini}}

\medskip

The S matrix proposed by Borsato, Ohlson Sax and Sfondrini assigns different amplitudes depending on 
whether the scattered states are a left-mover and a right-mover or are two of the same kind (where the left- and 
right-movers are representations of one of the other $PSU(1|1)$ factors of the $PSU(1|1)^2$ symmetry group of 
the gauge-fixed theory). There are also 
differences depending on the masses of the two states but for the most part this is contained within the Zhukowsky 
variables and we can write down the S matrix in terms of general coefficients without having to specify what 
the masses are. For the $LL$ sectors it is defined as:
\begin{align}
{\bf S}^{\rm BOSS}|\phi\phi'\rangle&={\bf A}^{\rm BOSS}_{LL}|\phi'\phi\rangle,&
{\bf S}^{\rm BOSS}|\phi\psi'\rangle&={\bf G}^{\rm BOSS}_{LL}|\psi'\phi\rangle+{\bf H}^{\rm BOSS}_{LL}|\phi'\psi\rangle,\\
{\bf S}^{\rm BOSS}|\psi\psi'\rangle&={\bf D}^{\rm BOSS}_{LL}|\psi'\psi\rangle,&
{\bf S}^{\rm BOSS}|\psi\phi'\rangle&={\bf K}^{\rm BOSS}_{LL}|\psi'\phi\rangle+{\bf L}^{\rm BOSS}_{LL}|\phi'\psi\rangle,
\end{align}
\noindent with the $RR$ sectors behaving in a completely equivalent way. For the $LR$ sectors the S matrix is defined as:
\begin{align}
{\bf S}^{\rm BOSS}|\phi\bar{\phi}'\rangle&={\bf A}^{\rm BOSS}_{LR}|\bar{\phi}'\phi\rangle+{\bf C}^{\rm BOSS}_{LR}|\bar{\psi}'\psi Z^-\rangle,&
{\bf S}^{\rm BOSS}|\phi\bar{\psi}'\rangle&={\bf G}^{\rm BOSS}_{LR}|\bar{\psi}'\phi\rangle,\\
{\bf S}^{\rm BOSS}|\psi\bar{\psi}'\rangle&={\bf D}^{\rm BOSS}_{LR}|\bar{\psi}'\psi\rangle+{\bf F}^{\rm BOSS}_{LR}|\bar{\phi}'\phi Z^+\rangle,&
{\bf S}^{\rm BOSS}|\psi\bar{\phi}'\rangle&={\bf L}^{\rm BOSS}_{LR}|\bar{\phi}'\psi\rangle,
\end{align}
\noindent and again the $RL$ sectors are similar to this.

For the $L_1L_1$ sector -- {\it i.e.} the $LL$ scattering of excitation of mass $m_1=\alpha$ --  the coefficients are given by:
\begin{align}
{\bf A}^{\rm BOSS}_{L_1L_1}&=S_{L_1L_1}\frac{x_{p'}^+-x_p^-}{x_{p'}^--x_p^+},&
{\bf D}^{\rm BOSS}_{L_1L_1}&=-S_{L_1L_1},\\
{\bf G}^{\rm BOSS}_{L_1L_1}&=S_{L_1L_1}\frac{x_{p'}^+-x_p^+}{x_{p'}^--x_p^+},&
{\bf H}^{\rm BOSS}_{L_1L_1}&=S_{L_1L_1}\frac{x_{p'}^+-x_{p'}^-}{x_{p'}^--x_p^+}\frac{\eta_p}{\eta_{p'}},\\
{\bf K}^{\rm BOSS}_{L_1L_1}&=S_{L_1L_1}\frac{x_{p}^+-x_p^-}{x_{p'}^--x_p^+}\frac{\eta_{p'}}{\eta_p},&
{\bf L}^{\rm BOSS}_{L_1L_1}&=S_{L_1L_1}\frac{x_{p'}^--x_p^-}{x_{p'}^--x_p^+}.
\end{align}
The $R_1R_1$ sector is exactly the same with $S_{L_1L_1}=S_{R_1R_1}$ and so are the $L_2L_2$/$R_2R_2$ sectors except 
there the mass appearing in Zhukowsky variables are different $m_2=1-\alpha$ and the dressing phase factor could be 
different as well.

The coefficients of the $L_1L_2$ sectors S matrix -- {\it i.e.} the $LL$ scattering of excitation of different masses -- are given by:
\begin{align}
{\bf A}^{\rm BOSS}_{L_1L_2}&=S_{L_1L_2},&
{\bf D}^{\rm BOSS}_{L_1L_2}&=-S_{L_1L_2}\frac{x_{p'}^--x_p^+}{x_{p'}^+-x_p^-},\\
{\bf G}^{\rm BOSS}_{L_1L_2}&=S_{L_1L_2}\frac{x_{p'}^+-x_p^+}{x_{p'}^+-x_p^-},&
{\bf H}^{\rm BOSS}_{L_1L_2}&=S_{L_1L_2}\frac{x_{p'}^+-x_{p'}^-}{x_{p'}^+-x_p^-}\frac{\eta_p}{\eta_{p'}}
,\\
{\bf K}^{\rm BOSS}_{L_1L_2}&=S_{L_1L_2}\frac{x_{p}^+-x_p^-}{x_{p'}^+-x_p^-}\frac{\eta_{p'}}{\eta_p},&
{\bf L}^{\rm BOSS}_{L_1L_2}&=S_{L_1L_2}\frac{x_{p'}^--x_p^-}{x_{p'}^+-x_p^-}.
\end{align}
Again the $R_1R_2$ sector is the same with $S_{L_1L_2}=S_{R_1R_2}$ and the $L_2L_1$/$R_2R_1$ sectors only differ by 
change of masses and dressing phase factor.

In the $L_1R_1$ sector the coefficients are:
\begin{align}
{\bf A}^{\rm BOSS}_{L_1R_1}&=S_{L_1R_1}\frac{1-\frac{1}{x_p^+x_{p'}^-}}{1-\frac{1}{x_p^-x_{p'}^-}},&
{\bf C}^{\rm BOSS}_{L_1R_1}&=-S_{L_1R_1}\frac{\eta_p\eta_{p'}}{x_p^-x_{p'}^-}\frac{i}{1-\frac{1}{x_p^-x_{p'}^-}},\\
{\bf D}^{\rm BOSS}_{L_1R_1}&=-S_{L_1R_1}\frac{1-\frac{1}{x_p^-x_{p'}^+}}{1-\frac{1}{x_p^-x_{p'}^-}},&
{\bf F}^{\rm BOSS}_{L_1R_1}&=S_{L_1R_1}\frac{\eta_p\eta_{p'}}{x_p^+x_{p'}^+}\frac{i}{1-\frac{1}{x_p^-x_{p'}^-}}
,\\
{\bf G}^{\rm BOSS}_{L_1R_1}&=S_{L_1R_1},&
{\bf L}^{\rm BOSS}_{L_1R_1}&=S_{L_1R_1}\frac{1-\frac{1}{x_p^+x_{p'}^+}}{1-\frac{1}{x_p^-x_{p'}^-}}.
\end{align}
The $L_2R_1$/$L_1R_2$/$L_2R_2$ sectors behave in exactly the same way with the possibility of a different dressing phase.

The coefficients of the $R_1L_1$ sector are given by:
\begin{align}
{\bf A}^{\rm BOSS}_{R_1L_1}&=S_{R_1L_1}\frac{1-\frac{1}{x_p^+x_{p'}^-}}{1-\frac{1}{x_p^+x_{p'}^+}},&
{\bf C}^{\rm BOSS}_{R_1L_1}&=-S_{R_1L_1}\frac{\eta_p\eta_{p'}}{x_p^-x_{p'}^-}\frac{i}{1-\frac{1}{x_p^+x_{p'}^+}},\\
{\bf D}^{\rm BOSS}_{R_1L_1}&=-S_{R_1L_1}\frac{1-\frac{1}{x_p^-x_{p'}^+}}{1-\frac{1}{x_p^+x_{p'}^+}},&
{\bf F}^{\rm BOSS}_{R_1L_1}&=S_{R_1L_1}\frac{\eta_p\eta_{p'}}{x_p^+x_{p'}^+}\frac{i}{1-\frac{1}{x_p^+x_{p'}^+}}
,\\
{\bf G}^{\rm BOSS}_{R_1L_1}&=S_{R_1L_1}\frac{1-\frac{1}{x_p^-x_{p'}^-}}{1-\frac{1}{x_p^+x_{p'}^+}},&
{\bf L}^{\rm BOSS}_{R_1L_1}&=S_{R_1L_1}.
\end{align}
The $R_2L_1$/$R_1L_2$/$R_2L_2$ sectors are the same with the possibility of different dressing phase.

There are further relations between the different scalar factors which are implied by the crossing equation 
\cite{JanikCrossing}; we will not review them since they are not important for our calculation.

Phases can be added to make the S matrix satisfy the untwisted Yang-Baxter equations; it is this S matrix we will be using in the article:

\begin{align}
{\hat \Asmatrix}^{\rm BOSS}&=\Asmatrix^{\rm BOSS} e^{\frac{i}{2}(p-p')}&
{\hat \Csmatrix}^{\rm BOSS}&=\Csmatrix^{\rm BOSS} e^{-\frac{i}{2}p'+ib(p+p')}\nonumber\\
{\hat \Dsmatrix}^{\rm BOSS}&=\Dsmatrix^{\rm BOSS} &
{\hat \Fsmatrix}^{\rm BOSS}&=\Fsmatrix^{\rm BOSS} e^{\frac{i}{2}p-ib(p+p')}\\
{\hat \Gsmatrix}^{\rm BOSS}&=\Gsmatrix^{\rm BOSS} e^{-\frac{i}{2}p'}&
{\hat \Hsmatrix}^{\rm BOSS}&=\Hsmatrix^{\rm BOSS} e^{\frac{i}{2}(p-p')+ib(p-p')}\nonumber\\
{\hat \Ksmatrix}^{\rm BOSS}&=\Ksmatrix^{\rm BOSS} e^{-ib(p-p')}&
{\hat \Lsmatrix}^{\rm BOSS}&=\Lsmatrix^{\rm BOSS} e^{\frac{i}{2}p}\nonumber
\end{align}

The one-loop dressing phases in both mixed and unmixed sectors were extracted in \cite{Beccaria:2012kb} by 
comparing the one-loop corrections to the energy of classical string solutions with the predictions of the 
Asymptotic Bethe Ansatz:

\begin{align}
\theta_{LL}=\theta(x_p,x_{p'})&=-\frac{\hat{\alpha}(x_p)\hat{\alpha}(x_{p'})}{\pi(x_p-x_p')^2}\mathrm{ln}\left(\frac{x_p+1}{x_p-1}\frac{x_{p'}-1}{x_{p'}+1}\right)+\mathrm{rational},\\
\theta_{LR}=\tilde{\theta}(x_p,x_{p'})&=-\frac{\hat{\alpha}(x_p)\hat{\alpha}(x_{p'})}{\pi(1-x_px_p')^2}\mathrm{ln}\left(\frac{x_p+1}{x_p-1}\frac{x_{p'}-1}{x_{p'}+1}\right)+\mathrm{rational}.
\end{align}
The factors ${\hat\alpha}$ are:
\begin{align}
\hat{\alpha}(x_p)&=\frac{2m}{h}\frac{x^2}{x^2-1},&x_p&=\frac{p}{\varepsilon-m}.
\end{align}
Here $h$ is plays the role of coupling constant. Its relation to the gauge theory coupling constant is not fixed by 
integrability or symmetries. At the level of the tree level worldsheet theory it is
\be
h=\frac{\sql}{2\pi} \ ;
\ee
it potentially receives regularization scheme-dependent corrections and $m$ is formally the mass parameter.
Keeping $m$ and $m'$ and expanding in the small momentum limit, the one-loop phase factors are:
\begin{align}
\theta_{LL}=\theta(x_p,x_{p'})&=\frac{-1}{2\pi h^2}\frac{p^2(p')^2[{\bf p}\cdot{\bf p'}+m m']}{(\varepsilon'p-p'\varepsilon)^2}\left(
\mathrm{ln}\left(\frac{p_-'}{p_-}\right)-\mathrm{ln}\left(\frac{m'}{m}\right)\right)+\mathrm{rational},\label{fase1}\\
\theta_{LR}=\tilde{\theta}(x_p,x_{p'})&=\frac{-1}{2\pi h^2}\frac{p^2(p')^2[{\bf p}\cdot{\bf p'}-m m']}{(\varepsilon'p-p'\varepsilon)^2}\left(
\mathrm{ln}\left(\frac{p_-'}{p_-}\right)-\mathrm{ln}\left(\frac{m'}{m}\right)\right)+\mathrm{rational}.\label{fase2}
\end{align}
We notice that these phases correspond to twice the computed expressions in \eqref{resultat LL} and \eqref{resultat LR}.

\

\noindent{$\bullet$ {\bf The S matrix of Ahn and Bombardelli}}

\medskip

The S matrix proposed in \cite{Ahn:2012hw} is somewhat simpler in that it does not depend as much on what representations the scattered states are in.

The $L_1L_1$ sector of the S matrix can be written as

\begin{align}
{\bf S}^{\rm AB}|\phi\phi'\rangle&={\bf A}^{\rm AB}_{LL}|\phi'\phi\rangle,&
{\bf S}^{\rm AB}|\phi\psi'\rangle&={\bf G}^{\rm AB}_{LL}|\psi'\phi\rangle+{\bf H}^{\rm AB}_{LL}|\phi'\psi\rangle,\\
{\bf S}^{\rm AB}|\psi\psi'\rangle&={\bf D}^{\rm AB}_{LL}|\psi'\psi\rangle,&
{\bf S}^{\rm AB}|\psi\phi'\rangle&={\bf K}^{\rm AB}_{LL}|\psi'\phi\rangle+{\bf L}^{\rm AB}_{LL}|\phi'\psi\rangle,
\end{align}
\noindent with the coefficients given by:

\begin{align}
{\bf A}^{\rm AB}_{L_1L_2}&=S_{L_1L_2},&
{\bf D}^{\rm AB}_{L_1L_2}&=-S_{L_1L_2}\frac{x_{p'}^--x_p^+}{x_{p'}^+-x_p^-},\\
{\bf G}^{\rm AB}_{L_1L_2}&=S_{L_1L_2}\frac{x_{p'}^+-x_p^+}{x_{p'}^+-x_p^-},&
{\bf H}^{\rm AB}_{L_1L_2}&=S_{L_1L_2}\frac{x_{p'}^+-x_{p'}^-}{x_{p'}^+-x_p^-}\frac{\omega_p}{\omega_{p'}}
,\\
{\bf K}^{\rm AB}_{L_1L_2}&=S_{L_1L_2}\frac{x_{p}^+-x_p^-}{x_{p'}^+-x_p^-}\frac{\omega_{p'}}{\omega_p},&
{\bf L}^{\rm AB}_{L_1L_2}&=S_{L_1L_2}\frac{x_{p'}^--x_p^-}{x_{p'}^+-x_p^-}.
\end{align}
\noindent where $\omega_p$ and $\omega_{p'}$ are chosen to be 1 and $S_{XY}$ are dressing phases. 
The $L_2L_2$/$R_1R_1$/$R_2R_2$ sectors are 
exactly the same with the appropiate changes of masses. The $L_1R_1$/$L_2R_1$/$R_1L_2$/$R_2L_1$ are also  similar, 
but with a different dressing phase, while the $L_1L_2$/$L_2L_1$/$R_1R_2$/$R_2R_1$ and 
the $L_1R_2$/$L_2R_1$/$R_1L_2$/$R_2L_1$ sectors are also the same except that the dressing phase is set to be 1.

\end{appendices}

\end{document}